\documentclass[11pt]{article}

\usepackage{pdfpages}
\usepackage{amssymb}
\usepackage{amsmath}
\usepackage{graphicx}
\usepackage{epstopdf}
\usepackage[figuresright]{rotating}
\usepackage{verbatim}
\usepackage[section]{placeins}
\usepackage{multirow}
\usepackage{datetime}
\usepackage{cite}
\usepackage[affil-it]{authblk}
\usepackage{setspace}
\usepackage{graphics}
\usepackage{float}
\usepackage{fancyhdr}
\usepackage{color}   
\usepackage[linkbordercolor={0 0 1}]{hyperref}
\hypersetup{
    colorlinks,
    citecolor=blue,
    filecolor=black,
    linkcolor=blue,
    urlcolor=black
}

\hypersetup{linktocpage}

\newdateformat{mydate}{\monthname[\THEMONTH] \THEDAY, \THEYEAR}
\begin{document}




\fancypagestyle{plain}{%
   \fancyhead[R]{\bf P-1063}
   \renewcommand{\headrulewidth}{0pt}
}


\title{LETTER OF INTENT:\\
The Accelerator Neutrino Neutron Interaction Experiment (ANNIE) }


\date{\mydate December 21, 2014}

\newcommand{\ANL}{Argonne National Laboratory; Lemont, IL 60439, USA}
\newcommand{\BNL}{Brookhaven National Laboratory;  Upton, NY 11973, USA }
\newcommand{\FNAL}{Fermi National Accelerator Laboratory; Batavia, IL 60510, USA}
\newcommand{\IC}{Imperial College of London; London SW7 2AZ, UK}
\newcommand{\ISU}{Iowa State University; Ames, IA 50011, USA}
\newcommand{\Hopkins}{Johns Hopkins University; Baltimore, MD 21218, USA}
\newcommand{\MIT}{Massachusetts Institute of Technology; Cambridge, MA 02139, USA}
\newcommand{\NGA}{National Geospatial-Intelligence Agency; Springfield ,VA 22150, USA}
\newcommand{\OSU}{Ohio State University; Columbus, OH 43210, USA}
\newcommand{\Purdue}{Purdue University;  West Lafayette, IN 47907, USA}
\newcommand{\Ultralytics}{Ultralytics, LLC; Arlington, VA 22101, USA}
\newcommand{\Davis}{University of California at Davis; Davis, CA 95817, USA}
\newcommand{\Irvine}{University of California at Irvine; Irvine, CA 92697, USA}
\newcommand{\Chicago}{University of Chicago, Enrico Fermi Institute; Chicago IL 60637, USA}
\newcommand{\Hawaii}{University of Hawaii; Honolulu, HI 96822, USA}
\newcommand{\Queen}{Queen Mary University of London; London E14NS, UK}

\author[1,5]{I.~Anghel}

\author[8]{J.~F.~Beacom}

\author[11]{M.~Bergevin}

\author[9]{C.~Blanco}

\author[5]{E. Catano-Mur}


\author[16]{F.~Di~Lodovico}

\author[14]{A.~Elagin}

\author[14]{H.~Frisch}

\author[12]{J. Griskevich}

\author[14]{R.~Hill}

\author[10]{G.~Jocher}

\author[16]{T.~Katori}

\author[5]{F. Krennrich}

\author[15]{J.~Learned}

\author[4]{M.~Malek}

\author[14]{R.~Northrop}

\author[14]{C.~Pilcher}

\author[3]{E.~Ramberg}

\author[1]{J.~Repond}

\author[16]{R.~Sacco}

\author[1,5]{M.C.~Sanchez~\footnote{Corresponding author: Mayly Sanchez (mayly.sanchez@iastate.edu)}}

\author[12]{M.~Smy}

\author[12]{H.~Sobel}

\author[11]{R.~Svoboda}

\author[6,7]{S.M.~Usman}

\author[12]{M.~Vagins}

\author[15]{G.~Varner}

\author [1] {R.~Wagner} 

\author[5]{A. Weinstein}

\author[14]{M.~Wetstein~\footnote{Corresponding author: Matthew Wetstein (mwetstein@uchicago.edu)}}

\author[13]{L.~Winslow}

\author[1]{L.~Xia}

\author[2]{M.~Yeh}

\affil[1]{\ANL}
\affil[2]{\BNL}
\affil[3]{\FNAL}
\affil[4]{\IC}
\affil[5]{\ISU}
\affil[6]{\Hopkins}
\affil[7]{\NGA}
\affil[8]{\OSU}
\affil[9]{\Purdue}
\affil[10]{\Ultralytics}
\affil[11]{\Davis}
\affil[12]{\Irvine}
\affil[13]{\MIT}
\affil[14]{\Chicago}
\affil[15]{\Hawaii}
\affil[16]{\Queen}

\maketitle



\noindent

\pagebreak


\begin{abstract}
The observation of proton decay (PDK) would rank among the most important discoveries in particle physics to date, confirming a key prediction of Grand Unification Theories and reinforcing the idea that the laws of physics become increasingly symmetric and simple at higher energies. Proposed Water Cherenkov (WCh) detectors, such as Hyper-Kamiokande are within reach of PDK detection according to many general PDK models. However, these experiments will also achieve size scales large enough to see PDK-like backgrounds from atmospheric neutrino interactions at a rate of roughly a few events per year per megaton. Given the rarity of proton decay and significance of the measurement,  the observation of proton-decay should be experimentally unambiguous. 

Neutron tagging in Gadolinium-loaded water may play a significant role in reducing these backgrounds from atmospheric neutrinos in next-generation searches. Neutrino interactions typically produce one or more neutrons in the final state, whereas proton decay events rarely produce any. The ability to tag neutrons in the final state provides discrimination between signal and background. Gadolinium salts dissolved in water have high neutron capture cross-sections and produce  $\sim$8 MeV in gammas, several tens of microseconds after the initial event. This delayed 8 MeV signal is much easier to detect than the 2 MeV gammas from neutron capture in pure water. Nonetheless, even the detection of this signature will not be perfectly efficient in large WCh detectors, especially those with low photodetector coverage.

It is not enough to identify the presence or absence of neutrons in an interaction.  In proton-decay searches, the presence of neutrons can  be used to remove background events. However, the absence of a tagged neutron is insufficient to attribute confidence to the observation of a proton decay event since the absence of a neutron may be explained by detection inefficiencies in WCh detectors. For moderately efficient neutron tagging and backgrounds peaked at higher neutron multiplicity, the absence of {\it any} neutron would increase confidence in the observation of a PDK candidate event. Calculating an exact confidence for discovery will require a detailed picture of the number of neutrons produced by neutrino interactions in water as a function of momentum transfer. Making this measurement in a neutrino test-beam is thus critical to future proton-decay searches. 

The neutron tagging techniques based on such measurement will also be useful to a broader program of physics beyond proton decay. For example, in the detection of diffuse supernova neutrino background, neutron tagging can be used to separate between genuine neutrinos and various radiogenic and spallation backgrounds. In the event of a core collapse supernova, the detection of neutrons can be used to help discriminate among different interactions in the water such as inverse beta decay and neutrino-oxygen scattering. 

Here we propose the Accelerator Neutrino Neutron Interaction Experiment (ANNIE), designed to measure the neutron yield of neutrino interactions in gadolinium-loaded water. While existing experiments such as Super-Kamiokande have attempted {\it in situ} measurements of neutron yield, the analyses were limited by detection inefficiencies and unknowns in the flux and energy of atmospheric neutrinos. ANNIE would represent a small, dedicated experiment designed to make this measurement on a beamline with known characteristics.

An innovative aspect of the ANNIE design is the use of precision timing to localize interaction vertices in the small fiducial volume of the detector. We propose to achieve this by using early prototypes of LAPPDs (Large Area Picosecond Photodetectors). This experiment will be a first application of these devices demonstrating their feasibility for WCh neutrino detectors. The ideas explored by ANNIE could have a transformative impact on water Cherenkov and other photodetection-based neutrino detector technologies, such as the Water-based Liquid Scintillator Concept detector proposed for Homestake.

\end{abstract}

\newpage
\tableofcontents
\setcounter{tocdepth}{3}
\newpage

\section*{List of Abbreviations}
\label{subsec-acronyms}

\begin{tabular}{ll}
ALD & atomic layer deposition \\
ASDC & advanced scintillator detector concept \\
ASIC & application specific integrated circuit \\
BNB & booster neutrino beam \\
CC & charged current \\
CCQE & charged current quasi-elastic \\
CMOS & complementary metal oxide semiconductor \\
DAQ & data acquisition system \\
DCTPC & Double Chooz time projection chamber \\
DIS & deep ineleastic scatter \\
DSNB & diffuse supernova neutrino background \\
FACC & front anti-coincidence counter \\
GUT & grand unified theory \\
LAPPD & large area picosecond photodetector \\
LBNF & Long Baseline Neutrino Facility \\
LS & Liquid Scintillator \\
MC & Monte Carlo \\
MCP & microchannel plates \\
MRD & muon range detector \\
NC & neutral current \\
ND & near detector \\
NDOS & Nova near detector on surface \\
NuMI & Neutrinos to Minnesota beamline \\
PDK & proton decay \\
PMT & photomultiplier tube \\
POT & protons on target \\
QE & quasi-elastic \\
SCA & switched capacitor array \\
SN & super nova \\
TPC & time projection chamber \\
wbLS & water-based Liquid Scintillator \\
WCh & Water Cherenkov \\
\end{tabular}
\newpage

\section{Introduction}
We are presenting a Letter of Intent to carry out the Accelerator Neutrino Neutron Interaction Experiment (ANNIE). We had previously submitted an Expression of Interest to the Fermilab Physics Advisory Committee (PAC) in January 2014~\cite{eoi}. Here we update the physics and design case for the experiment. We present a detailed site study that suggests that the SciBooNE hall on the Booster Neutrino Beam (BNB) is an ideal location to achieve the physics of this experiment. We also describe our efforts to better understand the neutron background at this location and we propose an initial phase of this test experiment to characterize this important background. The goal is to initiate the installation of this phase in the Summer of 2015. We also provide more details on the development of the experiment's design as well as describe the status of several key components already committed to the realization of this experiment. 

\section{The Physics of ANNIE}
\label{sec-physics}

The ability to detect final state neutrons from nuclear interactions would have a transformative impact on a wide variety of physics measurements in very large Water Cherenkov (WCh) and water-based liquid scintillator (wbLS) detectors~\cite{FSneutrons}.  Neutrino interactions in water often produce one or more neutrons in the final-state. Tagging events by the presence and number of final-state neutrons can provide physics analyses with a better handle for signal-background separation and even allow for more subtle discrimination between different varieties of neutrino interactions. For example, the main background on proton decay experiments arises from atmospheric neutrino interactions. These interactions almost always produce at least one final-state neutron, whereas proton decays are expected to produce neutrons less than 10$\%$ of the time~\cite{Ejiri}.

A promising technique for detecting final state neutrons is the search for a delayed signal from their capture on Gadolinium dissolved in water. Even moderately energetic neutrons ranging from tens to hundreds of MeV will quickly lose energy by collisions with free protons and oxygen nuclei in water. Once thermalized, the neutrons undergo radiative capture, combining with a nearby nucleus to produce a more tightly bound final state, with excess energy released in a gamma cascade. Neutron capture in pure water typically produces around 2.2 MeV in gamma particles ($\gamma$)~\cite{Ncapturewater}. However, these low energy photons produce very little optical light and are difficult to detect in large WCh tanks. The introduction of Gadolinium (Gd) salts dissolved in water is proposed as an effective way to improve the detection efficiency of thermal neutrons. With a significantly larger capture cross-section (49,000 barns compared with 0.3 barns on a free proton), Gd-captures happen roughly 10 times faster, on the order of tens of microseconds~\cite{NcaptureGd}. In addition, the Gd-capture produces an 8 MeV cascade of typically 2-3 gammas, producing sufficient optical light to be more reliably detected in large volumes.

A major limitation on the effective execution of neutron tagging techniques comes from large uncertainties on the nuclear mechanisms that produce neutrons and consequently on how many neutrons are produced by high energy (GeV-scale) neutrino interactions. The number of neutrons is expected to depend on the type of neutrino interaction and on the momentum transfer with higher energy interactions producing more than one neutron. However, the exact number of neutrons is determined by a variety of poorly understood nuclear processes and therefore it is not well-known. 

It is not enough to identify the presence or absence of neutrons in an interaction.  While the presence of neutrons can  be used to remove background events, the absence of a tagged neutron is insufficient to attribute confidence to the discovery of a proton decay observation. The absence of a neutron may be explained by detection inefficiencies in the WCh detector for example. On the other hand, if typical backgrounds consistently produce {\it more} than one neutron, the absence of {\it any} neutron would significantly increase the confidence in a PDK-like event. Calculating an exact confidence for discovery will require a detailed picture of the number of neutrons produced by neutrino interactions in water as a function of momentum transferred. 

\begin{figure}
	\begin{center}
		\includegraphics[width=0.5\linewidth]{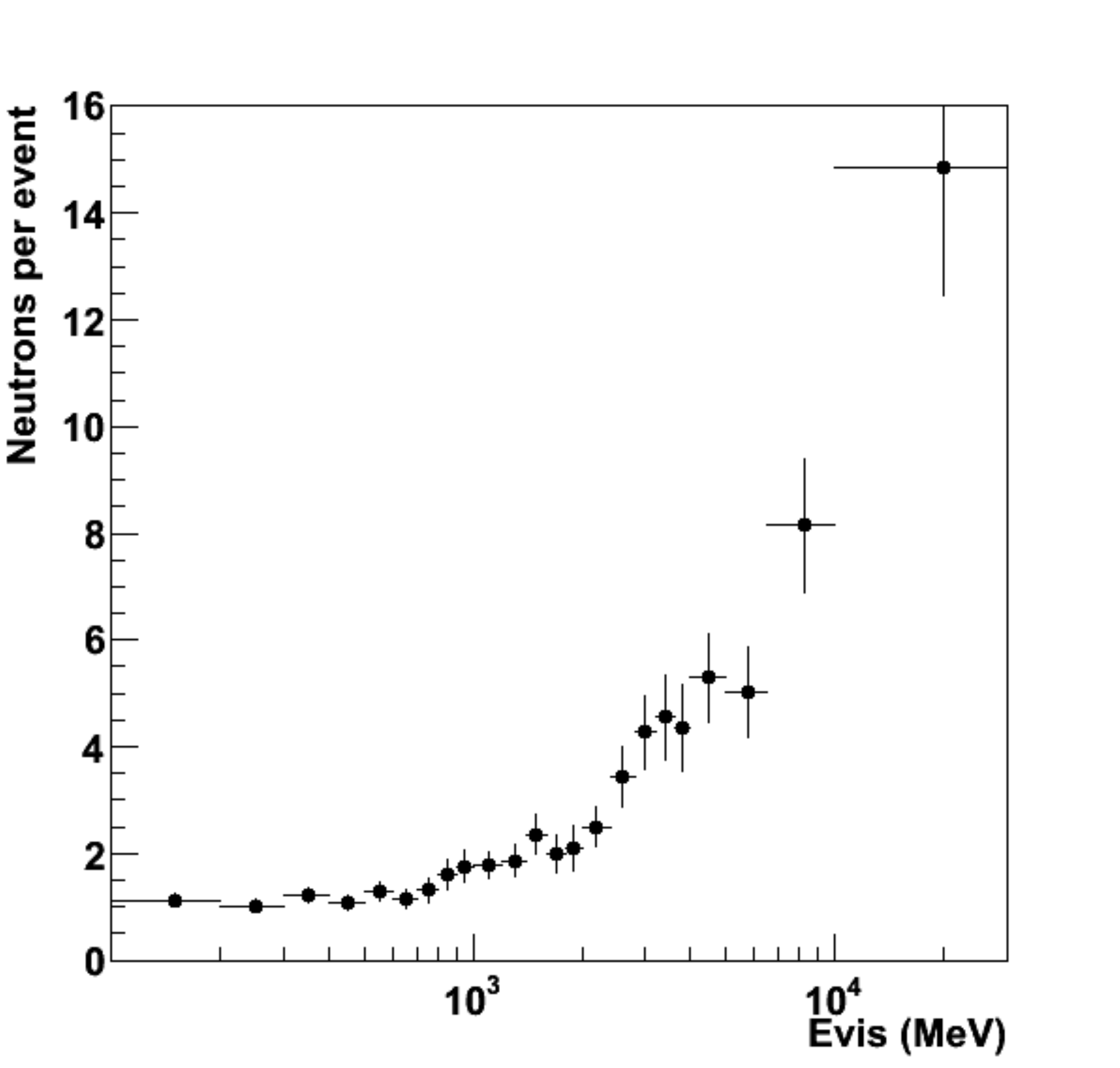}
	\end{center}
	\caption{Measurement of neutron multiplicity in pure water versus visible energy by the Super-K collaboration~\cite{SKneutronyield}.}
	\label{SKneutronmult}
\end{figure}

The Super-Kamiokande (Super-K) collaboration has attempted to measure the final state neutron abundance. Fig~\ref{SKneutronmult} shows the neutron multiplicity as a function of visible energy from atmospheric neutrino interactions in water, as detected by the 2.2~MeV capture gamma in Super-K~\cite{SKneutronyield}. However, the Super-K analysis is limited by uncertainties on the detection efficiencies for the 2.2~MeV gammas and on the flux of atmospheric neutrinos. Additionally, neither the neutrino energy nor the momentum transfer to the nucleus can be measured precisely. Therefore, it is difficult to incorporate these data into background predictions for proton decay.

Therefore, there is a clear need for a dedicated measurement of neutron yield. Such detailed measurement of the neutron multiplicity is possible in a beam with atmospheric neutrino-like energy spectrum. We propose to build such an experiment. The Accelerator Neutrino Neutron Interaction Experiment would consist of a small, economical Water Cherenkov detector deployable on the intense Booster Neutrino Beam (BNB) at Fermilab, and would largely rely on existing infrastructure. The main deliverable from this experiment is a measurement of the final-state neutron abundance as a function of momentum transfer from charged current (CC) neutrino interactions. This measurement is similar to that  shown in Fig~\ref{SKneutronmult}, except that we would reconstruct the total momentum transfer rather than visible energy and the ANNIE detector will be optimized for efficient detection of captured neutrons produced in the fiducial volume. Furthermore, it may be possible to separate the data between a variety of CC event types and possibly neutral current (NC) interactions. These data will provide an essential input to PDK and neutrino-interaction Monte Carlo models to aid in calculating detection efficiencies, expected background rates, accurate limits, and confidence levels. They can also be used to better constrain nuclear models of neutrino interaction physics and are therefore interesting on their own right.

\section{Potential Physics Impact}

\subsection{Understanding a Critical Background in Proton Decay}

One of the ``Big Ideas" in particle physics is the notion that at higher energies, the laws of physics become increasingly symmetric and simple. In the late 1970s it was suggested that, barring perturbations from other processes, the three running coupling constants become similar in strength in the range of $10^{13}-10^{16}$ GeV~\cite{RabyNakamura}. This convergence hints that the electromagnetic, weak, and strong forces may actually be a single force with the differences at low energy being due to the details of the exchange particle properties and the resulting vacuum polarization. This so-called ``Grand Unified Theory" (GUT) is a touchstone of particle physics in the late 20th and early 21st centuries. Theories ranging from supersymmetry (SUSY) to a wide class of string theories all have this basic ``Big Idea". A major challenge for experimental particle physics is how to determine if it is really true.

A convergence of the coupling constants at a very high ``unification energy" implies that there may be a single force that could connect quarks and leptons at that scale. Such reactions would violate baryon (B) and lepton (L) number by the exchange of very heavy bosons with masses in the range of the unification energy scale. Since that scale is far beyond the reach of any conceivable accelerator, they would only manifest themselves at our low energy scale via virtual particle exchange leading to rare reactions that would violate B and L. This would mean that normal matter (e.g., protons, either free or in nuclei) would not be stable but would decay with some very long lifetime. This phenomenon, generically called {\it proton decay} although neutrons in nuclei are included, has been searched for in a series of experiments dating back more than thirty years. Its discovery would be nothing short of revolutionary.

Proton decay final states depend on the details of a given theory. Experimentally, the modes $p \rightarrow e + \pi_0$ and $p \rightarrow K^+ + \nu$ are common benchmarks. The former represents the lightest anti-lepton plus meson final state, typical for the case where the first generation of quarks and leptons are grouped in a single multiplet, as in SU(5). The second is typical of supersymmetric grand unified theories where dimension-5 operators induce decays that span generations, hence requiring a strange quark. Current published limits from SK for these two modes are $8.2\times 10^{33}$ and $5.9\times10^{33}$ years, respectively~\cite{{SuperK2},{SuperK3}}.

\subsubsection{$p \rightarrow e + \pi_0$}

\begin{figure}
	\begin{center}
		\includegraphics[width=0.6\linewidth]{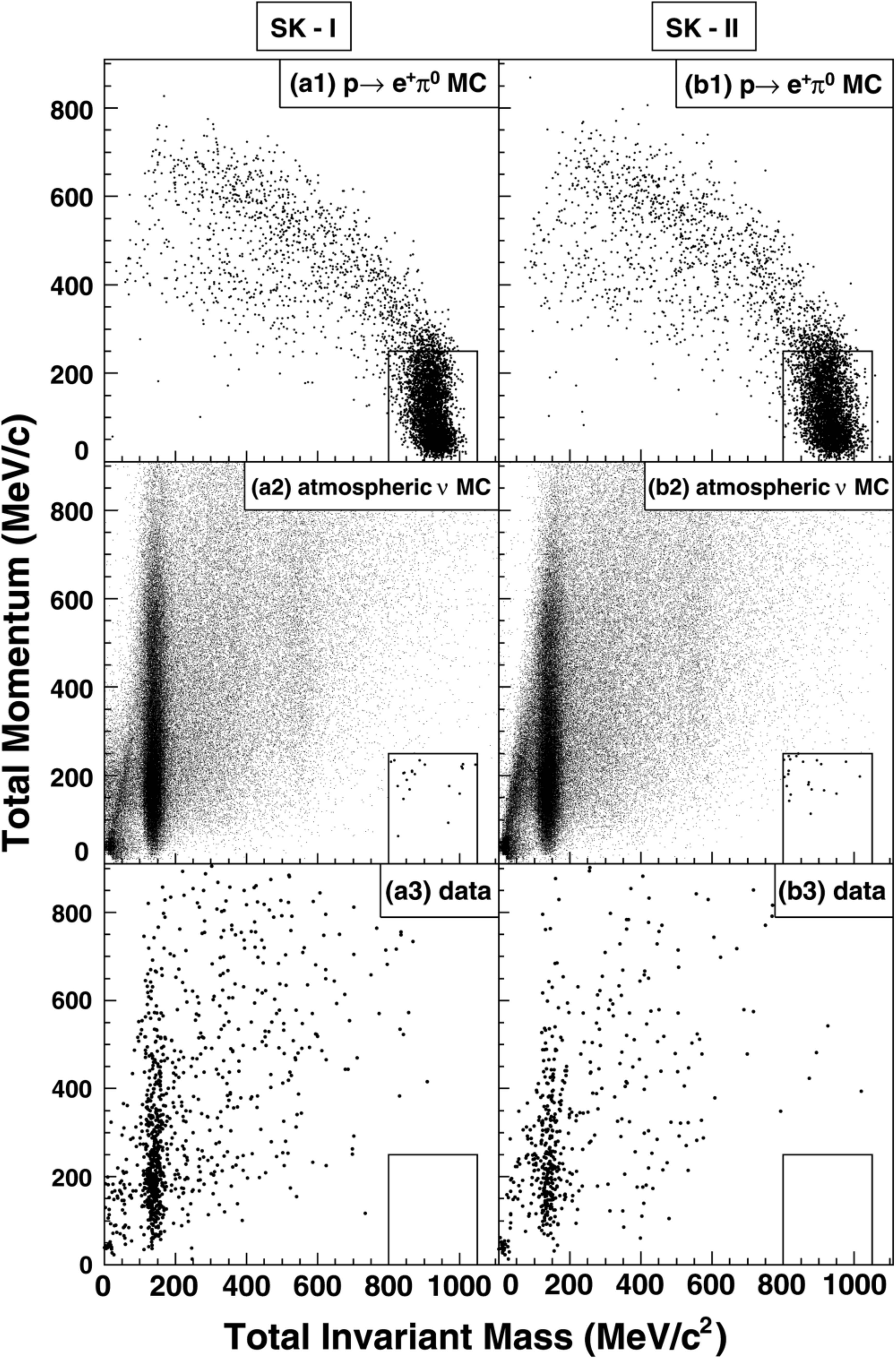}
	\end{center}
	\caption{The reconstructed kinematics of proton decay events in Super-K Monte Carlo (a1,b1), compared with those of atmospheric neutrino Monte Carlo (a2,b2) and data (a3,b3). Atmospheric neutrino events that fall in the signal region of (a2,b2) are enlarged (Ref~\cite{SuperK2}).}
	\label{pdkbkgd}
\end{figure}

It is instructive to describe the analysis currently being used by the Super-K experiment. This analysis consists of (1) selection of events in the detector that have three showering tracks, (2) a requirement that at least one combination of tracks gives an invariant mass close to that of the $\pi^0$ (85-185 MeV), (3) a requirement that there was no follow-on Michel electron (indicating that there was a muon in the event), and (4) that the invariant mass be near that of the proton (800-1050 MeV) and the unbalanced momentum be less than 250 MeV/c. Figure~\ref{pdkbkgd} (reproduced from ~\cite{SuperK2}) shows the invariant mass-unbalanced momentum distributions for two versions of Super-K (the left plots have twice the number of PMTs as the right plots) for the proton decay MC (top), atmospheric neutrino background MC (middle), and data (bottom). At 0.141 Mton-years there are no candidates.

The selection efficiency of the Super-K analysis was estimated to be $45\%$, with an uncertainty of $19\%$ dominated by nuclear effects (mainly pion interactions in the oxygen nucleus). In the center plots, the incursion of background events into the signal region is clearly seen. The MC gives a background estimate of $2.1 \pm 0.9$ events/Mton-year, which is consistent with direct measurements made in the K2K 1-kton near detector ($1.63_{-0.33}^{+0.42}$(stat)$_{-0.51}^{+0.45}$(syst) events/Mton-year)~\cite{SuperK3}. 

According to the Super-K MC, about $81\%$ of the background events are CC, with $47\%$ being events with one or more pions, and $28\%$ being quasi-elastic. In many cases, a $\pi^0$ is produced by an energetic proton scattering in the water. These events could be rejected by means other than invariant mass and unbalanced momentum. Neutron tagging has been proposed as a key method for doing this. Many of these background-producing events should be accompanied by one or more neutrons in the final state. This is because to look like a proton decay, there needs to be significant hadronic activity in the event, and there are many ways to produce secondary neutrons:

\begin{itemize}
\item direct interaction of an anti-neutrino on a proton, converting it into a neutron
\item secondary (p,n) scattering of struck nucleons within the nucleus
\item charge exchange reactions of energetic hadrons in the nucleus (e.g., $\pi^- + p \rightarrow n + \pi^0$)
\item de-excitation by neutron emission of the excited daughter nucleus
\item capture of $\pi^-$ events by protons in the water, or by oxygen nuclei, followed by nuclear breakup
\item secondary neutron production by proton scattering in water
\end{itemize}


Unfortunately, simulations of these processes are not currently data-driven. It is thus not possible to reliably predict the number of neutrons produced following a neutrino interaction. Some experimental data from Super-K supports the idea that atmospheric neutrino interactions in general might have accompanying neutrons but it has not been published and it is thus not definite~\cite{SKneutronyield}. 

This is to be contrasted with signal proton decay events, which are expected to produce very few secondary neutrons. Using general arguments, it is expected that more than $80\%$ of all proton decays should {\it not} have an accompanying neutron: 
\begin{itemize}
\item For water, $20\%$ of all protons are essentially free. If these decay, there is no neutron produced as the $\pi^0$ would decay before scattering in the water, and 400 MeV electrons rarely make hadronic showers that result in free neutrons.
\item Oxygen is a doubly-magic light nucleus, and hence one can use a shell model description with some degree of confidence. Since two protons are therefore in the $p_{1/2}$ valence shell, if they decay to $^{15}N$, the resultant nucleus is bound and no neutron emission occurs except by any final state interactions (FSI) inside the nucleus. 
\item Similarly, if one of the four protons in the $p_{3/2}$ state decays, a proton drops down from the $p_{1/2}$ state emitting a 6 MeV gamma ray, but the nucleus does not break up except by FSI.
\item Finally, if one of the two $s_{1/2}$ protons decays, there {\it is} a chance that the nucleus will de-excite by emission of a neutron from one of the higher shells.
\end{itemize}

Detailed nuclear calculations by Ejiri~\cite{Ejiri} predict that only $8\%$ of proton decays in oxygen will result in neutron emission. This means that only approximately $6\%$ ($8\%$ of $80\%$) of all proton decays in water should result in neutrons (ignoring FSI production by proton decay daughters). Therefore neutron tagging to reject atmospheric neutrino backgrounds incurs a modest loss of signal efficiency. In this proposal our goal is to measure the neutron yield in neutrino interactions as a function of momentum transfer. This will allow us to assess the effectiveness of such strategy.


\begin{figure}
	\begin{center}
		\includegraphics[width=0.65\linewidth]{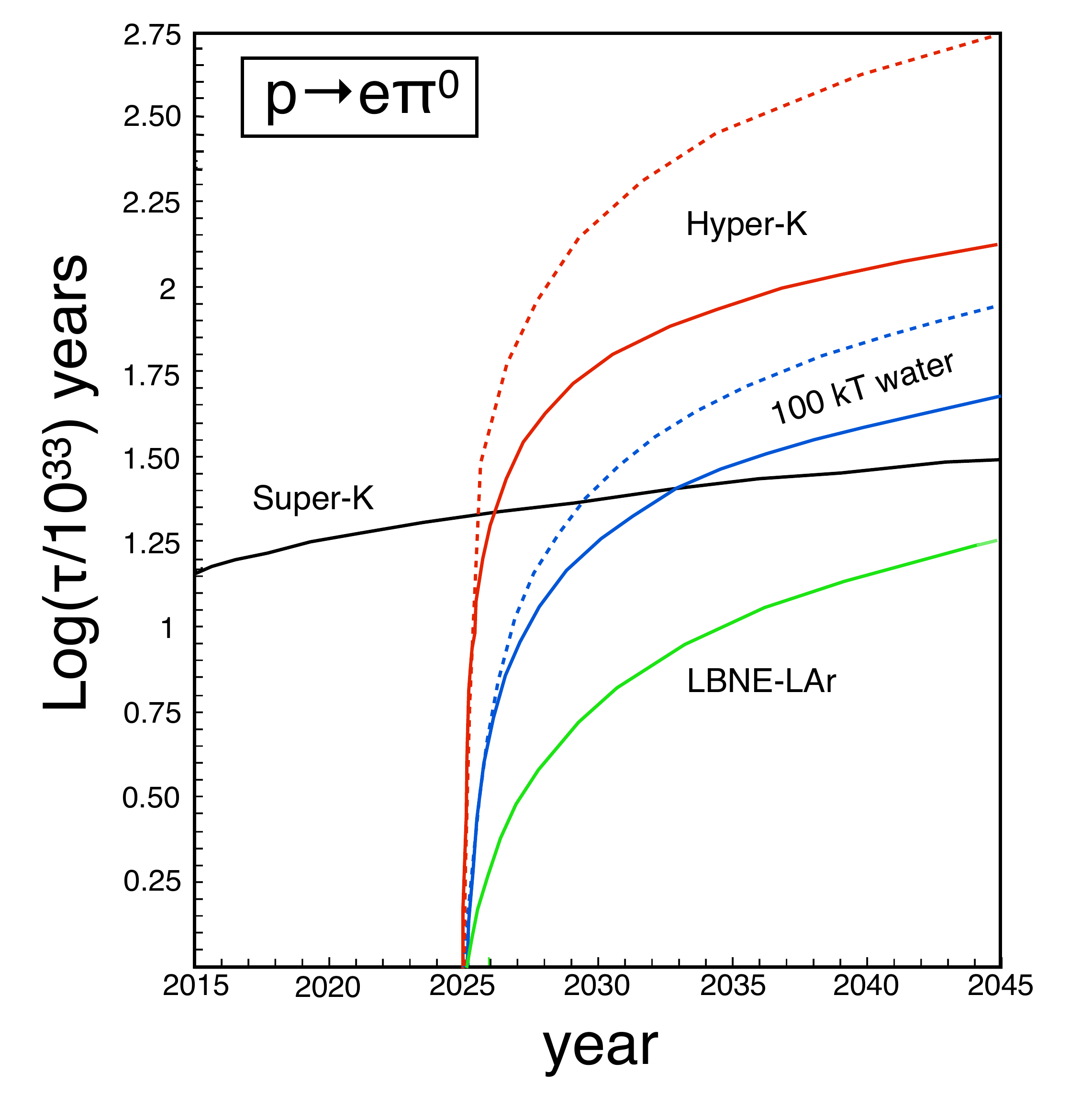}
	\end{center}
	\caption{Proton decay sensitivity in the $p \rightarrow e + \pi_0$ channel for Super-K (black), Hyper-K (red), the LBNF 34-kton LAr detector (green), and a hypothetical 100-kton water volume (blue). For Hyper-K and the 100 kton water volume two scenarios are presented: sensitivity including atmospheric neutrino backgrounds (solid) and limits and sensitivity with 90$\%$ of backgrounds removed (dashed).}
	\label{bkgdlimits}
\end{figure}

As an illustration, Figure~\ref{bkgdlimits} shows the sensitivity of Super-K (green) if it continues to run another 35 years, assuming the expected background rate from atmospheric neutrinos remains unchanged. Uncertainties in the background spectrum are taken into account, and the curves shown are $90\%$ c.l. limits.  Also shown (in blue) is the sensitivity of a 0.5-Mton detector with similar background estimations as Super-K running for a similar amount of time. Substantial background reduction using neutron tagging techniques is expected to significantly improve the sensitivity and discovery potential of very large WCh detectors. For example, the span between the solid and dashed curves highlight the impact of such background reduction. However, the precise background rejection efficiencies have not been demonstrated. ANNIE will accurately evaluate and demonstrate the potential of this method.

\subsubsection{$p \rightarrow K^+ + \nu$ }

As another example, for the $p \rightarrow K^+ + \nu$ mode, the $K^+$ is below the Cherenkov threshold, requiring a search for the decay of a kaon at rest. There is significant atmospheric neutrino background in the dominant ($63\%$) decay mode of $K^+ \rightarrow \mu + \nu_{\mu}$. Super-K uses the prompt nuclear de-excitation gamma ray (6.3 MeV) from the residual $^{15}N$ nucleus to reject background events. Analysis of the hadronic mode, $K^+ \rightarrow \pi^0 \pi^+$ ($21\%$), is hampered by the fact that $\beta_{\mu+}$ = 0.87, so that the amount of Cherenkov light emitted by the decay muon (from the $\pi^+$) is near the detectable threshold. Expectations are that background events will be seen in this mode at a rate of ~8 events/Mton-year. The combined efficiency for the prompt gamma tag of $K^+ \rightarrow \mu + \nu_{\mu}$ plus $K^+ \rightarrow \pi^0 \pi^+$ is $14\% \pm 2\%$ with an expected background of $1.2 \pm 0.4$ events/100-kton/year. Thus even though Super-K does not currently have a candidate, it is expected that this mode will soon start generating background. If a significant fraction of these events could be rejected, sensitivity would continue to rise relatively linearly in a very large detector.

\subsection{Improving  Identification of Supernova Neutrino Interactions}

Supernova explosions throughout the universe left behind a diffuse supernova background of neutrinos that may be detected on Earth~\cite{supernova-beacom}. The flux and spectrum of this background contains information about the rate of supernova explosions as well as their average neutrino temperature.  The main detection channel for supernova relic neutrinos in water Cherenkov detectors comes from positrons emitted by inverse $\beta$ decay reactions. Above $\sim 20$ MeV, the dominant background is due to the decay of sub-Cherenkov threshold muons from atmospheric neutrino interactions. This could be greatly reduced by tagging the neutron that accompanies each inverse $\beta$ reaction. A 200-kton detector loaded with gadolinium and at sufficient depth may be within reach of detecting this neutrino flux~\cite{review-beacom, LBNESN}. In order to achieve this, understanding neutron yields can be used to help statistically discriminate among various radiogenic, spallation and neutrino backgrounds.  

A nearby core collapse supernova will provide a wealth of information via its neutrino signal. The neutrinos are emitted in a burst of a few tens of seconds duration, with about half in the first second. Energies are in the few tens of MeV range, and the luminosity is divided roughly equally among flavors. Neutrino densities in the core are so high that neutrino-neutrino scattering plays a significant role in the dynamics, leading to non-linear oscillation patterns, highly sensitive to fundamental neutrino properties and even new physics. Accurate measurements of the energies, flavors, and time dependent fluxes would also allow one to set limits on coupling to axions, large extra dimensions, and other exotic physics~\cite{Raffelt}. From these details, one could also learn about the explosion mechanism, accretion, neutron star cooling, and possible transitions to quark matter or to a black hole. Neutron tagging would be essential in building a more complete picture of the SN neutrino flux, helping to more efficiently identify neutral current interactions, and separate neutrino-nucleus scattering in the water, which do not produce any neutrons.

\subsection{Testing Nuclear Models of Neutrino Interactions}

There is growing interest among the neutrino cross-section community in better understanding nuclear effects on neutrino interactions. Most of the current and future long-baseline neutrino oscillation experiments are designed to measure neutrinos with energies below 10 GeV. Nuclear effects play a significant role in this regime, as demonstrated by the recent T2K oscillation result~\cite{T2K2}, where the neutrino interaction model is the largest systematic error.

The MiniBooNE experiment has published a first double differential cross section for CC quasi-elastic (CCQE) interactions~\cite{MB1,MB2}. Many aspects of this precision measurement are not understood by traditional nuclear models based on the impulse approximation~\cite{Benhar}. The MiniBooNE data may be better described by models including two-body currents, where low-energy neutrinos scatter off  correlated pairs of nucleons~\cite{Martini,Nieves}. Confirming such processes and incorporating them into oscillation analyses is now a major goal of the cross-section community. A predicted consequence of two-body currents is a higher multiplicity of final-state nucleons~\cite{Martini}. An experiment, like ANNIE, with neutron tagging abilities would provide a unique opportunity to study some of these effects.

Final-state neutrons can also be used used in the statistical separation between NC interactions and CC interactions. In neutrino-mode, neutron multiplicity is expected to be lower for CC interactions. This feature can be used to distinguish $\nu_e$ oscillation candidates from NC backgrounds, such as $\pi^0$ or photon production~\cite{MBp}. ANNIE is in a position to study the feasibility of this technique in water.

In addition, one of the systematics in neutrino oscillation measurements, such as those to be performed by LBNF, is the uncertainty in the reconstructed energy of events identified as being CCQE. One way of understanding and controlling these uncertainties is to look at the multiplicity of final state nucleons, protons {\it and} neutrons. For this reason, one of the key neutrino interaction measurements to be
undertaken by the next generation of liquid argon (LAr) neutrino detectors is the measurement of final states described as: 0$\pi$ + X$_p$ + X$_n$, namely no pions and some number of protons and neutrons. LAr TPCs are well suited to measure the number of final state protons. However the number of final-state neutrons is expected to be difficult. ANNIE will be in a position to measure these states, and thus enhance our knowledge of neutrino interactions as well as complement the LAr short and long baseline programs.

\subsection{Designing a Near Detector for Future Long Baseline Experiments}

Hyper-Kamiokande (Hyper-K) is a potential next generation underground water Cherenkov
detector with a total (fiducial) mass of 0.99 (0.56) Mton
tons, approximately 20 (25) times larger than that of
Super-Kamiokande (Super-K).  It is designed as a detector capable of observing
accelerator, atmospheric and solar neutrinos, proton decays, and
neutrinos from other astrophysical origins, providing a very rich
physics portfolio. One of the main goals of Hyper-K is the
study of the CP asymmetry in the lepton sector using accelerator
neutrino and anti-neutrino beams.

The accelerator neutrino and anti-neutrino event rate observed at Hyper-K depends on the
oscillation probability, neutrino flux, neutrino interaction
cross-section, detection efficiency, and the detector fiducial mass of
the Hyper-K detector.  To extract estimates of the
oscillation parameters from data, one must model the neutrino flux,
cross-section and detection efficiency with sufficient precision.
This is achieved with near detectors. 

The TITUS (Tokai Intermediate Tank for Unoscillated Spectrum) detector
is a proposed near detector for the Hyper-K experiment lead by some of our UK collaborators. The main characteristics of this detector are: a 4$\pi$
phase-space coverage, a water target similarly to the detector at the
far site (Super-K) and the same flux as at the far detector,
being situated at about 2\,km from the beam target.

The main disadvantage of a WCh detector is the inability to
separate positively and negatively charged leptons. This proposed detector aims to overcome
this issue using a Gd-loaded WCh detector by detecting the presence of neutrons in the final state. 
This ability is especially important for a CP violation measurement
where the wrong sign contribution to the neutrino flux should be well
understood. 

%

The TITUS strategy is to take advantage of both the Gd-doping and 
LAPPDs in a similar manner to the ANNIE design.  Thanks to the Gd doping, we will be able to tag the neutrons in the final states.  The neutron multiplicity is measured after the FSI
(Final State Interactions) is taken into account, so it may not be
identical to the neutrino interaction process, as several other
processes can occur before the nucleon leaves the nucleus.  However,
we should be able to relate the final state interaction to the
original process in most cases. 

In TITUS, the very precise timing and spacial resolution of the LAPPDs
will help to both reduce the background and tag the neutron thanks to
a much improved vertex capability that will directly impact the
reconstruction.  ANNIE, as a  Gd-loaded WC detector instrumented with
LAPPDs, allows us to perform studies that are directly relevant to the design and planning of TITUS. It is extremely useful that ANNIE (if located on-axis of the Booster Neutrino beam line) will run at energies similar to those used in T2K and planned for Hyper-K. 

\section{Experimental Overview}
\label{sec-expoverview}

\begin{figure}
	\begin{center}
		\includegraphics[width=0.45 \linewidth]{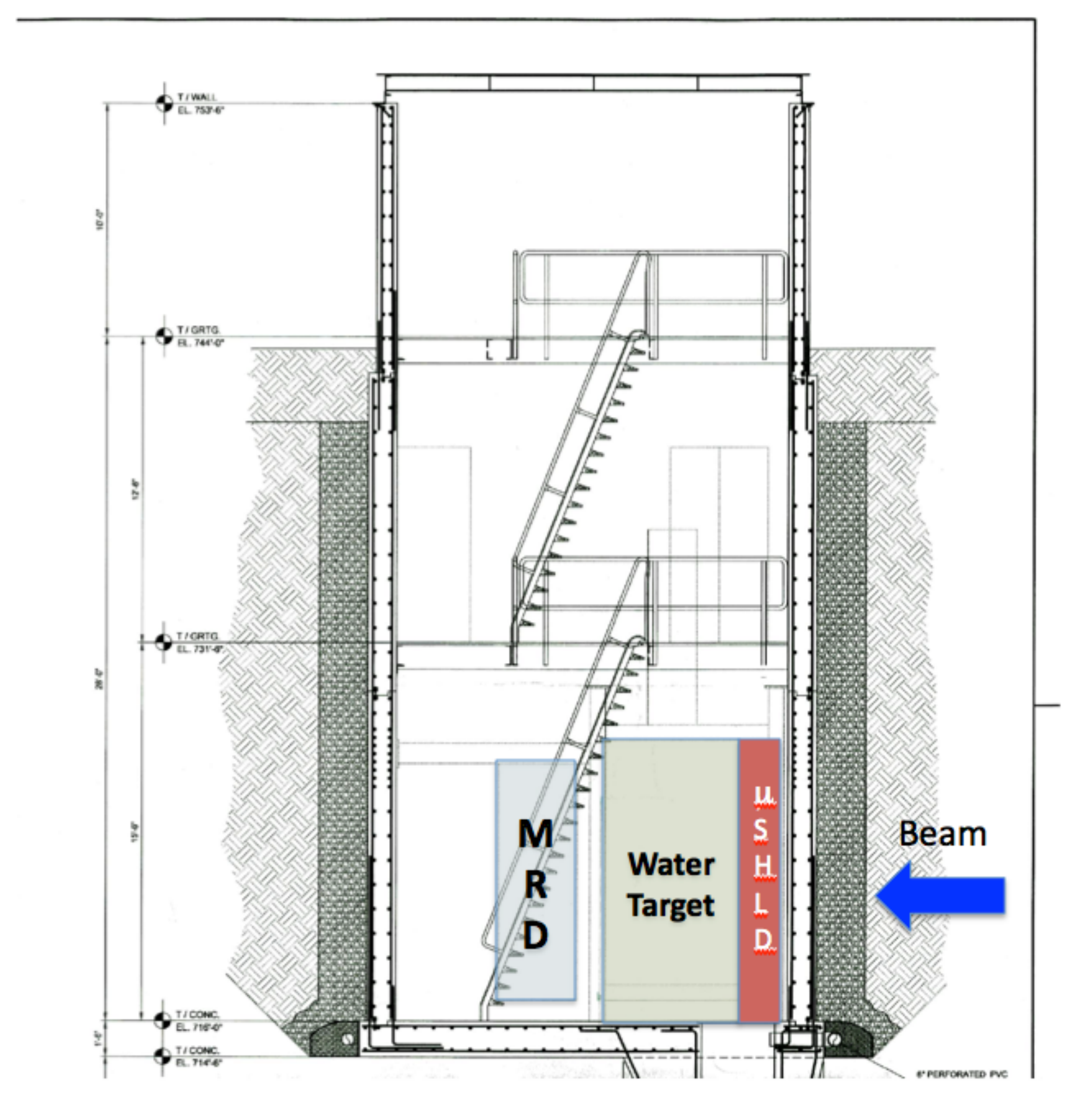}
	\end{center}
	\caption{ANNIE in the SciBooNE Hall.}
	\label{anniehall}
\end{figure}

We propose making a systematic measurement of the neutron yield from neutrino interactions of energies similar to atmospheric neutrinos. We can optimally carry out this measurement by utilizing the existing ``SciBooNE" hall which sits in a prime experimental location on the FNAL booster beamline (Fig~\ref{anniehall}). This hall is currently unused, however it is being investigated as a potential location for the cryogenic system of LAr1ND. The laboratory is currently exploring alternative locations for this system and we have explored alternative locations for ANNIE in Sec.~\ref{sec-beam}. We plan to put a Gd-doped water target sufficiently instrumented in front of a muon range detector to be able to stop and detect the capture gammas from primary and secondary neutrons. We have named this test experiment: the Accelerator Neutrino Neutron Interaction Experiment, or ANNIE.

\subsection{Physics of the Measurement }
 \label{measurementstrategy}
To first order, neutrino interactions with nucleii will predictably yield either 1 or 0 neutrons in the final state: Neutrinos interacting by charged current  (CC) exchange will produce a final-state proton and no additional neutrons, whereas anti-neutrinos produce exactly one final-state neutron. High energy neutral current (NC) interactions tend to produce either protons or neutrons, proportional to the abundance of each nucleon in water. 

%

However, GeV-scale (anti-)neutrinos can produce additional neutrons through the complex interplay of higher-order and multi-scale nuclear physics:

\begin{itemize}
\item secondary (p,n) scattering of struck nucleons within the nucleus
\item charge exchange reactions of energetic hadrons in the nucleus (e.g., $\pi^- + p \rightarrow n + \pi^0$)
\item de-excitation by neutron emission of the excited daughter nucleus
\item capture of $\pi^-$ events by protons in the water, or by oxygen nuclei, followed by nuclear breakup
\item Meson Exchange Currents (MEC), where the neutrino interacts with a correlated pair of nucleons, rather than a single proton or neutron.
\item secondary neutron production by proton or neutron scattering in water
\end{itemize}

Consequently, neutron multiplicity distributions tend to peak at 0 or 1 with long tails. 
Given the highly non-gaussian shape of these distributions, parameters such as the mean neutron yield are not necessarily illuminating. At the simplest level, we want to measure P(N=0), P(N=1), and P(N$>$1) with particular attention to any excesses beyond tree-level expectations. These measurements, binned by interaction type and kinematics, will provide a strong handle to constrain nuclear models, even in the absence of detailed shape information beyond P(N=2). 

When using the presence of final state neutrons to separate experimental backgrounds in various physics analyses, the shape of the far tail becomes increasingly less important with higher N. For example, in the case of proton decay we are interested in the efficiency for detecting any neutrons at all. The rate for atmospheric neutrinos faking a proton decay ($f$) is given by:

\begin{equation}
f = P(0)  + P(1)(1- \epsilon) + P(2)(1- \epsilon)^2 + P(3)(1- \epsilon)^3 + ....
\label{eq:PMTquality}
\end{equation}

where P(N) is the probability of N neutrons given a background event, and $\epsilon$ is the neutron detection efficiency. For high neutron detection efficiencies such as the expected 68$\%$ in a Gd-loaded Super-K fill, higher order terms quickly drop off, and $f$ can be accurately estimated by the integral of P(N$>$2) without any further shape information.

\subsection{Experimental Design and Status}

\begin{figure}
	\begin{center}
		\includegraphics[width=0.60 \linewidth]{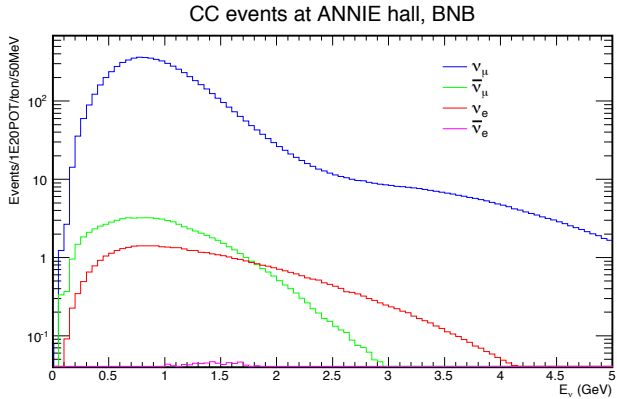}
	\end{center}
	\caption{Neutrino flux spectra expected in the SciBooNE Hall from the BNB.}
	\label{anniefluxes}
\end{figure}

ANNIE would run using the Booster Neutrino Beam (BNB). This beam runs at 7.5 Hz, with roughly 4x$10^{12}$ protons-on-target (POT) per spill. These are delivered in 81 bunches over a $1.6$ $\mu$s spill time to a target and horn combination 100~m upstream of the SciBooNE hall.  This beam is about $93\%$ pure $\nu_{\mu}$ (when running in neutrino mode) and has a spectrum that peaks at about 700 MeV (Fig.~\ref{anniefluxes}). We expect on the order of 7,000 charged current muon neutrino interactions per ton per $10^{20}$ POT over a period of 6 months.  The neutrino rates at various sites at Fermilab are discussed in detail in Sec~\ref{sec-beam}. 

There are several sources of neutron background in ANNIE. These arise from neutrino interactions in the rock and dirt upstream of the detector (dirt neutrons) as well as from ambient neutrons from the beam dump which travel mostly through air and scatter into the hall (sky shine). This background and steps to measure and suppress it are described in Sec~\ref{sec-neutron}. 

The footprint for the water target is essentially that of the SciBooNE detector, a cylindrical volume roughly 3.8 m long and 2.3 m in diameter. The plan is to contain the target volume in a single water tank made of aluminum with a plastic liner. The mechanical design of the water target using an existing tank from UChicago is described in Sec~\ref{sec-mechanical}. The details of the gadolinium loading including a filtration system provided by UC Irvine are discussed in Sec.~\ref{sec-Gd}. The target will be instrumented by 60 to 100 eight-inch PMTs. These PMTs are available from UC Irvine and their status is described in Sec.~\ref{sec-photodetection}.  An iron-scintillator sandwich detector that was used to range out and fit the direction of daughter muons from neutrino interactions in the SciBooNE target is available in the experimental hall~\cite{Sciboone}. Parts of this detector, called the Muon Range Detector (MRD) could be used for ANNIE as discussed in Sec.~\ref{subsec-MRD}. 

In order to select events away from the detector wall, we propose to use vertex reconstruction based on the arrival time of emitted light. This is the simplest option, requiring neither segmentation of the already small target nor the introduction of new materials with unknown neutron capture properties. Given the few-meter length scale of the detector, timing based reconstruction is a challenge. Typical drift times for direct light are below 10 nanoseconds, so it is unlikely that conventional PMTs, with few-nanosecond time resolutions will be good enough to localize the vertex. We intend to use early commercial prototypes of Large Area Picosecond Photodetectors (LAPPDs) with single photoelectron time resolutions below 100 picoseconds. A description of the status of these photodetectors is given in Sec~\ref{sec-photodetection}. 

Neutrino interactions in the water target will produce a flash of light with no signal in a front anti-coincidence counter. Events can be selected so that they are ``CCQE-like", i.e., there is a single muon track in the MRD that points back to the rough position of the vertex in the target. Following a valid CCQE candidate, neutron capture events must be detected in the target for about 100 $\mu$s, or about three capture times. If the vertex is restricted to the central volume of the water target, then there are several hadron scattering lengths in all directions, which should be enough to slow down and stop neutrons in the range of 110~MeV. Higher energy neutrons could require external counters. The readout and electronics required to record both neutrino and neutron capture data are described in Sec~\ref{sec-electronics}. 

The interaction point of each neutrino can be reconstructed using several different approaches, thus providing an effective handle on the location of the interaction point. The ideal method is to fit for the earliest light emitted along the muon track, using the track parameters extracted from the muon range detector. It may be possible to use the isotropic light emitted from the neutron captures themselves to determine where in the volume the captures are happening. By measuring the muon direction to a precision of roughly $10^{\circ}$ and the muon momentum (from a range measurement) to roughly $20\%$, we will be able to accurately reconstruct the multiplicity as a function of the momentum transfer to the nucleus from the neutrino. This is desirable in order to facilitate the incorporation of this measurement into an atmospheric neutrino MC which is relevant to the proton decay searches. 

While basic simulations of the ANNIE detector already exist, we will are in the process of building a fully integrated Geant4 simulation of the experiment. Through these simulations, we can address the technical design issues required for the success of this effort. The current status of ANNIE simulations is described in Sec.~\ref{sec-simulations}.

\section{Beam and Site Requirements}
\label{sec-beam}

In this section we describe the beam and site requirements for the ANNIE experiment. Given ANNIE's physics goals we are interested in studying neutrino interactions with energies comparable to those of atmospheric neutrinos that result in potential background for proton decay. The beam must be sufficiently intense in order to provide the needed statistical power over the lifetime of the experiment. At a given site the beam should have a sufficiently low duty cycle to limit multi-interaction pileup (specially from rock-interactions).

\subsection{Neutrino Beam Spectrum and Intensity}

There are two existing neutrino beams currently running at Fermilab, the Booster Neutrino Beam (BNB) and the Neutrinos from the Main Injector (NuMI) beam. The BNB impinges 8.89 GeV/c protons from the booster on a beryllium target, with $4 \times 10^{12}$ delivered in a spill of approximately 1.6~$\mu$s at a frequency of 7.5 Hz. The NuMI beam throws 120 GeV/c protons from the main injector (MI) on a carbon target. The proton beam contains  $4 \times 10^{13}$ delivered in a spill of approximately 10~$\mu$s at frequency of 0.6 Hz (a 1.67~s cycle). The number of protons incident on the target is defined as protons on target (POT). The projected POT per year for the BNB is about $2 \times 10^{20}$ POT and for NuMI it should ramp up from $3 \times 10^{20}$ to $6 \times 10^{20}$ over the next 4 years. 

The BNB and NuMI neutrino spectra differ significantly on the axis of the beam. For this location the BNB peaks at 0.7 GeV (Fig.~\ref{anniefluxes}) while NuMI peaks at 6 GeV in its medium energy configuration. If an off-axis location is chosen for NuMI the neutrino spectrum can peak as low as 2 GeV (Fig.~\ref{annienovandfluxes}). 

\begin{figure}
	\begin{center}
		\includegraphics[width=0.75 \linewidth]{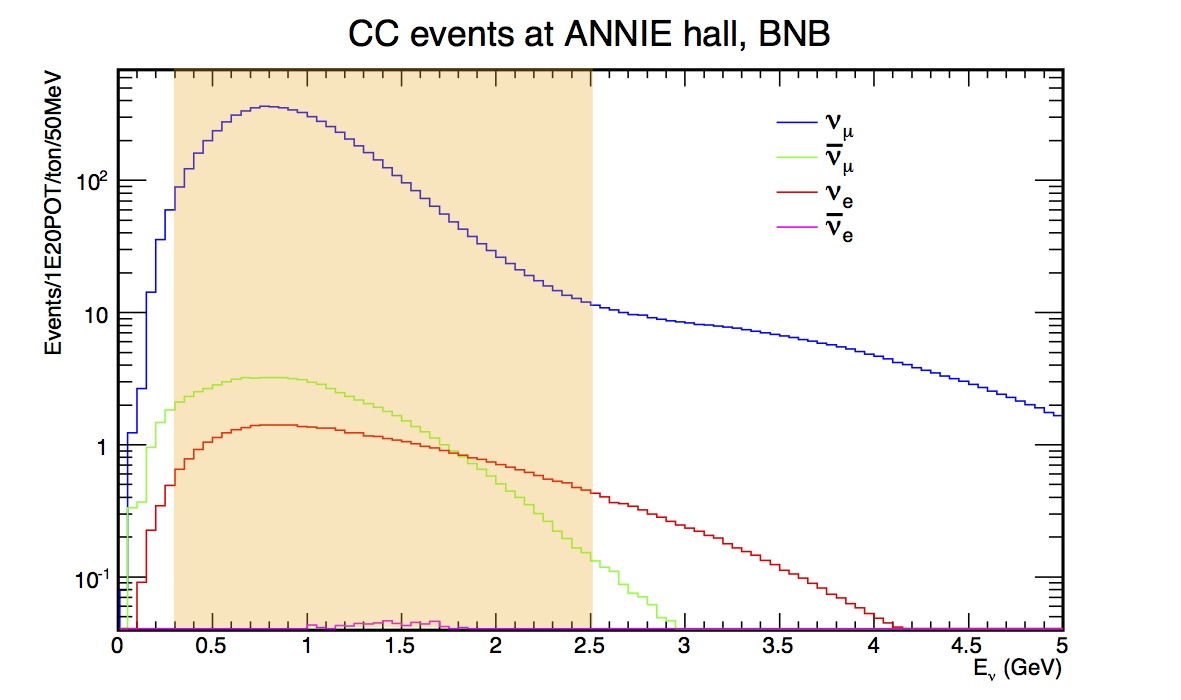}
	\end{center}
	\caption{Neutrino flux spectra expected in the SciBooNE Hall from the BNB. Rates are per ton per $1\times10^{20}$ POT. The shaded region indicates the energy range of the atmospheric neutrino flux that dominates the production of proton decay background events.}
	\label{anniehallfluxwithpdk}
\end{figure}

Figure~\ref{anniehallfluxwithpdk} shows the spectrum of the BNB neutrino spectrum overlaid with the region of interest which represents the portion of the atmospheric neutrino flux that dominates the production of proton decay background events. It is clear from this figure that the the BNB on-axis location possesses  the ideal neutrino spectrum peaked in the region of interest. In the next section we detail the rates in the region of interest for various site locations for these two neutrino beams. 


\subsection{Neutrino Interaction Rates at Different Sites}
\label{subsec-rates}


ANNIE will require collecting data from sufficient neutrino interactions to make accurate statements about neutron yield in an inclusive sample of neutrino interactions. The initial goal is to be able to describe distributions of neutron yield versus various kinematic observables. A more demanding goal is to study neutron yields for specific event classes. For instance studying separately the neutron yield for quasi-elastic and deep inelastic charged current as well as neutral current neutrino interactions. Eventually, with improved detector performance, we will be able to identify explicitly proton decay-like backgrounds at rates that are statistically significant.  It is expected that this would require tens of thousands of neutrino interactions a year. 



When studying the potential neutrino interaction rates and spectra of the existing neutrino beams, the siting of the detector with respect to the beam must be considered. We have studied various detector location with potential for siting the ANNIE detector. While these potential locations might not currently be available, it builds a picture of what flexibility, if any, is available in order to carry out this measurement. 

As potential ANNIE detector locations in the BNB we have considered the SciBooNE hall (on-axis) at 100~m from the target and a location on the surface of the SciBooNE hall (80 mrad or $4.6^\circ$ off-axis). In NuMI we have considered an on-axis location in the MINOS near detector (MINOS ND) hall at 1 km from the target as well as two off-axis locations roughly at the same distance. The first off-axis location for the NuMI beam is in the NOvA near detector (NOvA ND) hall which is at 14 mrad or $0.8^\circ$ from the beam axis. The second location is the NOvA NDOS hall on the surface above the MINOS ND hall at 111 mrad off-axis. The SciBooNE hall, the MINOS ND hall and the NOvA NDOS locations have potential space available for the size of the detector including infrastructure for installing a detector. The NOvA ND hall currently cannot fit the detector and it is occupied for the foreseeable future. The SciBooNE surface is shown as an example of an off-axis location but there is no infrastructure available at this location.  

The NuMI beam simulation data has been obtained from Flugg flux files from 2013 as provided by the NuMI-X group. The files in dk2nu format were processed using software from NOvA experiment designed to propagate  the flux to the designated locations. The BNB flux simulated data has been provided by Zarko Pavlovic (MiniBooNE) appropriately propagated to the SciBoone hall and surface locations. 

The SciBooNE hall as it has been described in Section~\ref{sec-expoverview} is on axis from the BNB  at 8~m below the surface. The spectrum as shown in Figure~\ref{anniefluxes} peaks at 0.7~GeV. The rates expected in 1~ton of water (the approximate usable fiducial volume) per year considering $2\times10^{20}$ POT/year for BNB are about 20K neutrino interactions, 14K of  those would be 
$\nu_\mu$ CC interactions. As mentioned before this spectrum peaks ideally in the region of interest and has the desired rate of neutrino interactions per year. Detailed rates are shown in Table~\ref{expectedratesANNIEhall1} per ton per $1\times10^{20}$ POT . 

%

\begin{table*}
  \begin{center}
    \begin{tabular} {| l | c | c | c |}
     \hline
     $\nu$-type & Total Interactions & Charged Current & Neutral Current\\
     \hline
     \hline
     $\nu_{\mu}$ & 9892 & 6991 & 2900\\
     \hline
     $\bar{\nu}_{\mu}$ & 130 & 83 & 47\\
     \hline
     $\nu_{e}$ & 71 & 51 & 20\\
     \hline
     $\bar{\nu}_{e}$ & 3.0 &2.0 & 1.0\\
     \hline
    \end{tabular}
    
  \end{center}
  \caption{Rates expected in 1 ton of water with 1x$10^{20}$ POT exposure at the SciBooNE Hall.}
  \label{expectedratesANNIEhall1}
\end{table*}

\begin{figure}
	\begin{center}
		\includegraphics[width=0.75 \linewidth]{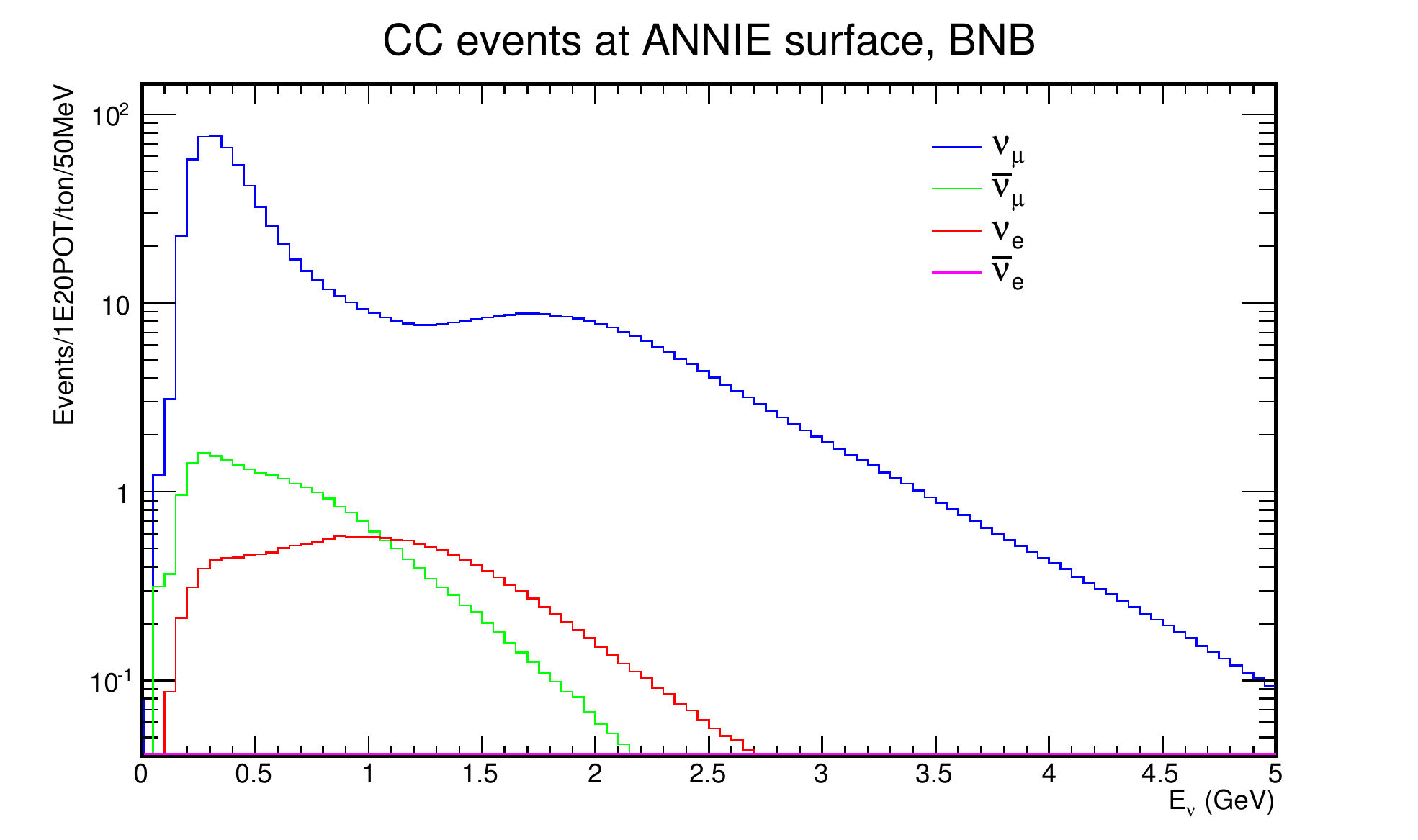}
	\end{center}
	\caption{Neutrino flux spectra expected at the SciBooNE surface from the BNB. Rates are per ton per $1\times10^{20}$ POT.}
	\label{anniesurfacefluxes}
\end{figure}

We have also considered an off-axis location for BNB. Since there are no other facilities or halls built around the BNB, we have used the surface as a point of reference. This is 8~m above the axis of the beam at a distance of 100~m from the target which results in an angle of 80 mrad or $4.6^\circ$ from the axis of the beam. There is no infrastructure at this location in which to build this experiment. The flux at this location drops by a factor of ~8 from the on-axis location as shown in Table~\ref{expectedratesSurface}. The spectrum peaks at a lower energy as shown in Figure~\ref{anniesurfacefluxes}. This is a significant drop that would lengthen the experiment's planned data taking by a large factor. 

\begin{table*}
  \begin{center}
    \begin{tabular} {| l | c | c | c |}
     \hline
     $\nu$-type & Total Interactions & Charged Current & Neutral Current\\
     \hline
     \hline
     $\nu_{\mu}$ & 1251 & 847 & 404 \\
     \hline
     $\bar{\nu}_{\mu}$ & 44 & 26 & 18\\
     \hline
     $\nu_{e}$ & 24 & 17 & 6.9\\
     \hline
     $\bar{\nu}_{e}$ & 1.5 & 1.0 & 0.5\\
     \hline
    \end{tabular}
   
  \end{center}
   \caption{Rates expected in 1 ton of water with 1x$10^{20}$ POT exposure at the surface of the SciBooNE Hall.}
  \label{expectedratesSurface}
\end{table*}

\begin{figure}
	\begin{center}
		\includegraphics[width=0.75 \linewidth]{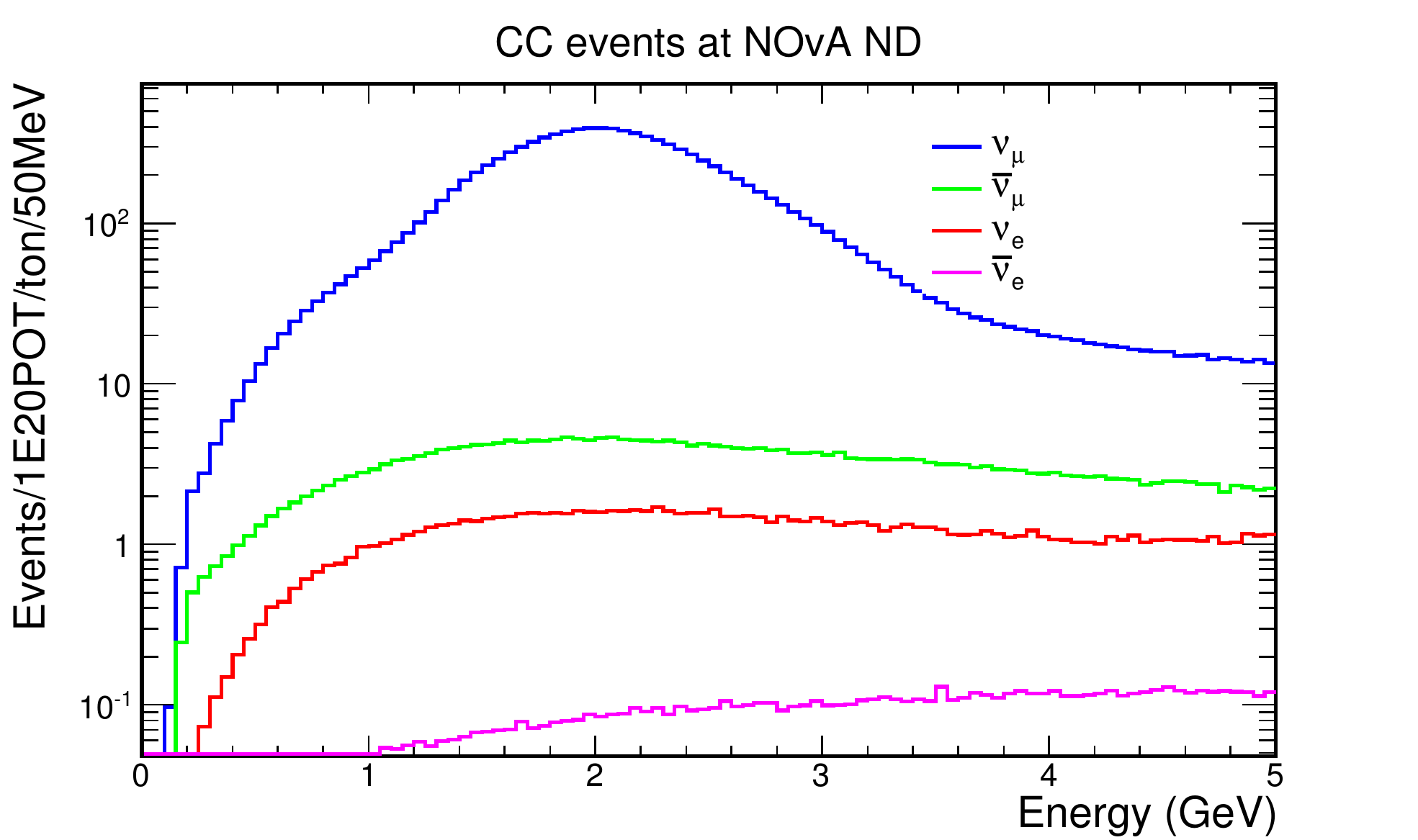}
	\end{center}
	\caption{Neutrino flux spectra expected at the Nova Near Detector Hall from the NuMI beam. Rates are per ton per $1\times10^{20}$ POT.}
	\label{annienovandfluxes}
\end{figure}

In the NuMI beam we have considered the on-axis location in the MINOS ND hall. While the rates are high (over 200k  neutrino interactions), the beam spectra peaks at 6 GeV thus making the rates in the region of interest much lower.  This also will increase event interaction pileup rate from neutrino interactions occurring in the rock. Off-axis locations that shift the peak to the relevant portion of the neutrino spectrum are more interesting. For example the neutrino spectrum at the NOvA ND hall which is 14 mrad off-axis is shown in Figure~\ref{annienovandfluxes} peaking at 2~GeV. The rates expected in this case are around 45K neutrino interactions considering $3 \times 10^{20}$ POT/year for NuMI (see Table~\ref{expectedratesNOvAND}). There is however no space in the NOvA ND detector hall and no other similar off-axis locations are available in the underground tunnels. The hall was in fact specifically excavated for the NOvA ND. This case is shown as an example of what is the minimum distance that we would need to be off-axis from the NuMI beam in its medium energy configuration in order to obtain a neutrino spectrum usable for this experiment.


%
%

\begin{table*}
  \begin{center}
    \begin{tabular} {| l | c | c | c |}
     \hline
     $\nu$-type & Total Interactions & Charged Current & Neutral Current\\
     \hline
     \hline
     $\nu_{\mu}$ & 16563 & 12074 & 4489 \\
     \hline
     $\bar{\nu}_{\mu}$ & 636 & 445 & 191\\
     \hline
     $\nu_{e}$ & 300 & 221 & 79\\
     \hline
     $\bar{\nu}_{e}$ & 28 & 20 & 7.9\\
     \hline
    \end{tabular}
   
  \end{center}
   \caption{Rates expected in 1 ton of water with 1x$10^{20}$ POT exposure at the NOvA Near Detector Hall.}
  \label{expectedratesNOvAND}
\end{table*}

\begin{figure}
	\begin{center}
		\includegraphics[width=0.75 \linewidth]{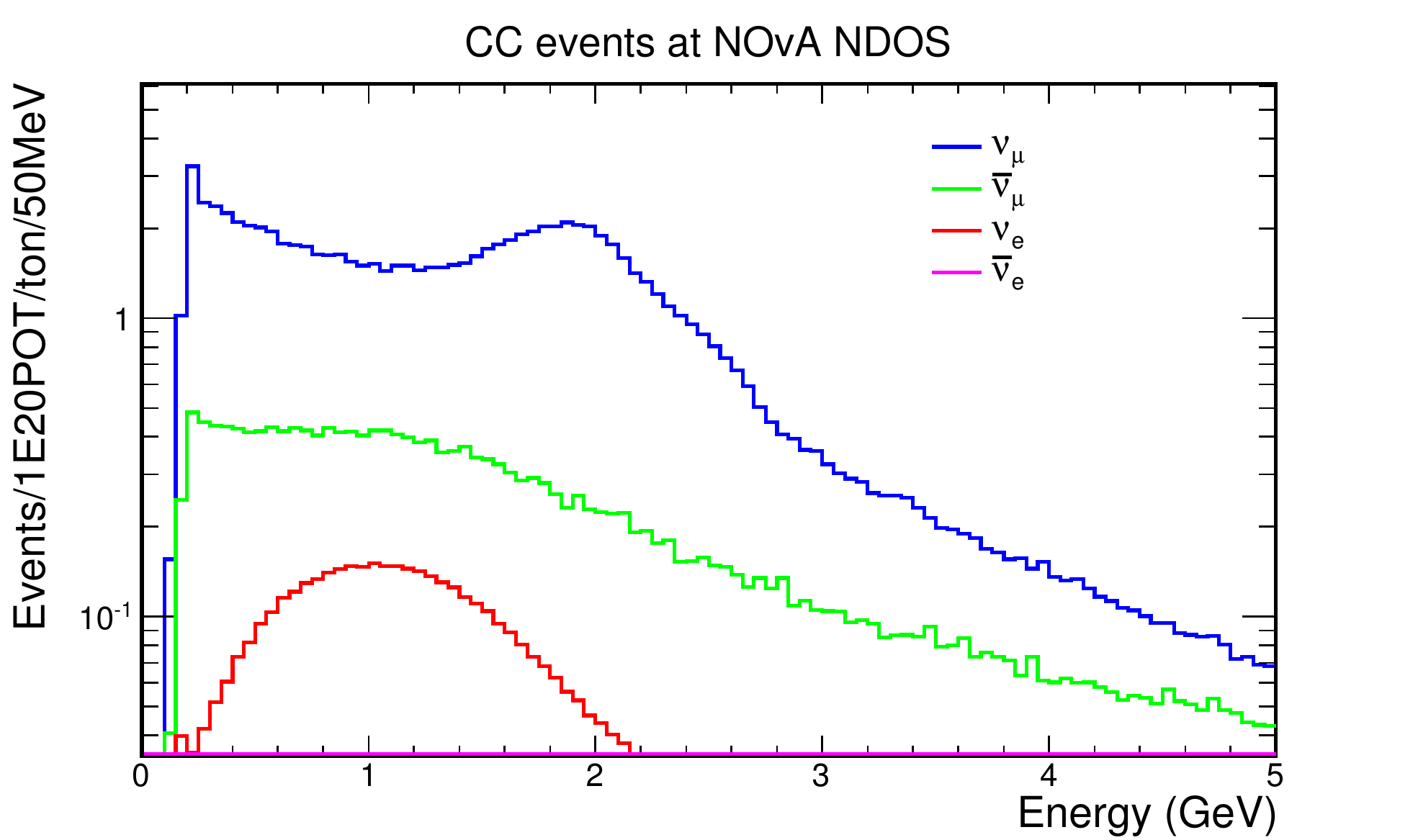}
	\end{center}
	\caption{Neutrino flux spectra expected at the Nova NDOS Hall from the NuMI beam. Rates are per ton per $1\times10^{20}$ POT.}
	\label{annienovandosfluxes}
\end{figure}

The last location considered is the NOvA NDOS hall on the surface of the MINOS and NOvA ND halls. This location is important to consider as there is the infrastructure available to potentially install the ANNIE detector. The neutrino spectra at this location still have a peak around 2 GeV but a significant fraction of the spectrum shifts to energies below 1~GeV (Figure~\ref{annienovandosfluxes}). However, the rates as shown in Table~\ref{expectedratesNOvANDOS} drop by several orders of magnitude to roughly 400 muon neutrino interactions per year in the fiducial volume of the detector. This rate is too low to be able to carry out any of the measurements proposed.

\begin{table*}
  \begin{center}
    \begin{tabular} {| l | c | c | c |}
     \hline
     $\nu$-type & Total Interactions & Charged Current & Neutral Current\\
     \hline
     \hline
     $\nu_{\mu}$ & 130 & 91 & 39 \\
     \hline
     $\bar{\nu}_{\mu}$ & 32 & 21 & 11\\
     \hline
     $\nu_{e}$ & 6.2 & 4.5 & 1.7\\
     \hline
     $\bar{\nu}_{e}$ & 1.5 & 1.0 & 0.5\\
     \hline
    \end{tabular}
   
  \end{center}
   \caption{Rates expected in 1 ton of water with 1x$10^{20}$ POT exposure at the Nova Near Detector On Surface (NDOS) Hall.}
  \label{expectedratesNOvANDOS}
\end{table*}

Finally, we show the rates for all locations considered in Table~\ref{expectedratesALL} which shows the fraction of the spectrum that is in the region of interest between 0.25 to 2.5 GeV. It is of note that the SciBooNE hall and the NuMI off-axis locations have the larger fraction of neutrinos in the interesting region with the NDOS hall showing the lowest rates. From this table we can conclude that the SciBooNE hall represents the optimal solution with the highest rate and potential infrastructure available. An off-axis location on the NuMI beam (at no less than 14 mrad but significantly less than 111 mrad) would also be optimal but such infrastructure does not currently exist. 

\begin{table*}
  \begin{center}
    \begin{tabular} {| l | c | c | c |}
     \hline
     Location & $\nu_{\mu}$ CC [0.25-2.5 GeV] & $\nu_{\mu}$ CC [0-10 GeV] & Percentage\\
     \hline
     \hline
     SciBooNE Hall & 6626 & 6991 & 95\% \\
     \hline
     SciBooNE surface & 708 & 847 & 84\% \\
     \hline
     MINOS ND & 3362 & 168078 & 2\% \\
    \hline
     NOvA ND & 8115 & 12074 & 67\% \\
     \hline
     NDOS & 76 & 91 & 84\% \\
    \hline
    \end{tabular}
   
  \end{center}
   \caption{Rates of $\nu_{\mu}$ CC interactions expected in 1 ton of water with 1x$10^{20}$ POT exposure at two different energy ranges, and the percentage of events between 0.25-2.5 GeV, for different detector locations.}
  \label{expectedratesALL}
\end{table*}

\subsection{Neutrino Event Interactions Pileup Rates}

Another consideration beyond the rates and spectra of in-detector neutrino interactions is the probability of seeing multiple events in one beam spill. Ideally we would want to collect all light from one interaction before a second starts. We have developed a toy Monte Carlo simulation to estimate the event interaction pileup rates of both in-detector and outside of detector (occurring in the rock) neutrino interactions. 

In order to study these rates, we must consider the beam structure. Each beam has a different time structure. For BNB it is one booster batch of protons per beam spill spread over 1.6 $\mu$s in 84 bunches of protons separated by 19 ns from one another. For NuMI we have 5 or 6 batches per spill  spread in 84 bunches each over 10 $\mu$s. 

The relevance of the fine structure depends on the typical time length of the event. We can define a characteristic time $\tau$ as the time that it takes a Cherenkov photon to travel from one corner to the opposite, i.e. the maximum  possible distance, inside the detector. Setting a window of 4-5 $\tau$ is a conservative expectation for the time required to collect all the photons associated with a vertex. For our initial studies, we use a window of  $\approx 100$ ns.

Considering the neutrino flux at a given location, the detector size and cross sections, we can find the corresponding expected number of in-detector events per spill. To account for the rock events we have used a MC based (average) rock to in-detector ratio from complete Monte Carlo simulations of each of the relevant experiments. While the SciBooNE and NOvA ND hall are the most interesting, the MINOS ND location is calculated for comparison of an on-axis high duty-cycle beam location. The ratios are shown in Table~\ref{averageratio}. 

\begin{table} [H] \centering
\begin{tabular}{|c|c|c|} \hline
Location  & Energy Peak & Ratio rock/in-detector\\ \hline \hline
SciBooNE  & 0.6 GeV & 3 \\ \hline
NOvA ND  & 2 GeV & 4\\ \hline
MINOS ND & 6 GeV & 10\\ \hline
\end{tabular}
\caption{Average ratio of neutrino interactions occurring in the detector to neutrino interactions outside of the detector (rock events).}
\label{averageratio}
\end{table}

We then use the resulting expected number of events (in-detector plus rock) per spill as the parameter for a Poisson distribution to get a \emph{simulated} number of events for a given spill, and the time structure of the spill itself as a probability distribution for event start times.  We count the instances when a pair of event time windows overlap, and repeat for 20000 simulated spills. 

For the total mass of the ANNIE detector we find that the number of muon neutrino interactions per spill is low (less than 1) for the SciBooNE and NOvA ND locations and very high as expected for the MINOS ND (Table~\ref{rockeventnumber}). The rates  are below 1 Hz for the former locations and at 12 Hz for the latter. From this study is concluded that the on-axis location for the BNB beam and the slightly off-axis location for NuMI result in ideal manageable neutrino event interaction pileup rates whereas the on-axis location in NuMI has too high rates to be viable even if the in-detector interactions in the interesting energy range are comparable to the other more ideal sites. This study did not take into consideration cosmic ray pileup. A discussion of these rates as it affects the electronics design can found in Sect.~\ref{sec-electronics}.


\begin{table}[H]
\centering
\begin{tabular}{|c|c|c|c|} \hline
Location & $\nu_\mu$ events/POT/ton & $\nu_\mu$ events/spill & Avg. pileup/spill \\ \hline \hline
SciBooNE  & $2.80 * 10^{-16}$ &0.03 & $5.0\times10^{-5}$\\ \hline 
NOvA ND  &$6.04 * 10^{-16}$ &  0.65& 0.0045\\ \hline 
MINOS ND& $1.85 * 10^{-14}$& 20& 3.76\\ \hline  

\end{tabular}
\caption{Expected number of events per spill (rock and contained) and corresponding pileup rates for different detector locations.}
\label{rockeventnumber}
\end{table}

%

\section{Neutron Backgrounds}
\label{sec-neutron}

Several sources introduce neutron backgrounds to the ANNIE detector, which will require detailed understanding. A continuum of ambient neutrons from cosmic radiation and long-lived isotopes will be present, but can be largely suppressed by strict time cuts around the beam window. Somewhat more challenging are the correlated backgrounds tied to the time structure of the beam, these are:

\begin{itemize}
\item Dirt Neutrons: neutrons produced by interactions of beam neutrinos with the rock and dirt, upstream of the ANNIE detector.
\item Sky Shine: ambient neutrons from the beam dump, which travel through air and scatter into ANNIE Hall. 
\end{itemize}

\begin{figure}
	\begin{center}
		\includegraphics[width=0.65\linewidth]{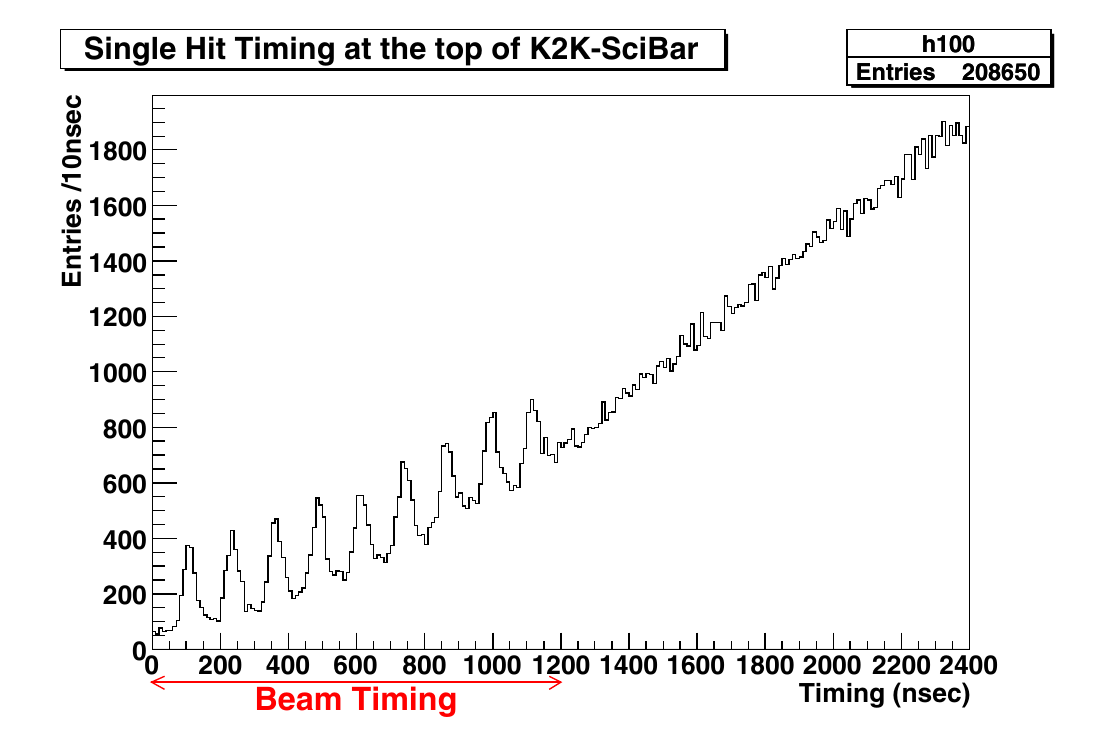}
	\end{center}
	\caption{A plot showing the fine time structure of single strip hits in the SciBar detector at KEK. Each successive bunch sits on top of an increasing pedestal of sky shine neutrons.}
	\label{SciBar-KEK}
\end{figure}

The general problem of sky shine has been discussed in several documents~\cite{{LeeSkyshine},{BartolNeutronMonitor}}. The SciBooNE collaboration examined the issue at the SciBooNE Hall~\cite{{SciBooNEskyshine},{Takeiskyshine}}. Figure~\ref{SciBar-KEK} helps illustrate this
phenomenon. The fine structure of single strip hits in the SciBar detector deployed at KEK is consistent
with successive bunches of signal on top of an increasing pedestal of background hits from sky shine.
Sky shine rates were also measured at KEK and a study was produced by the SciBooNE collaboration
in preparation for moving the SciBar detector to the BNB.

Neutrinos from the BNB can interact with dirt and rock upstream of the experimental hall, producing a correlated background. While this background may appear slow with respect to the prompt component of an event, it is fast on the time scale of Gd neutron captures, and will therefore present a problem for neutron counting.

\subsection{Understanding Neutron Backgrounds with ANNIE}

\begin{figure}
	\centering
	 \begin{tabular}{c c}
		\includegraphics[width=0.48\linewidth]{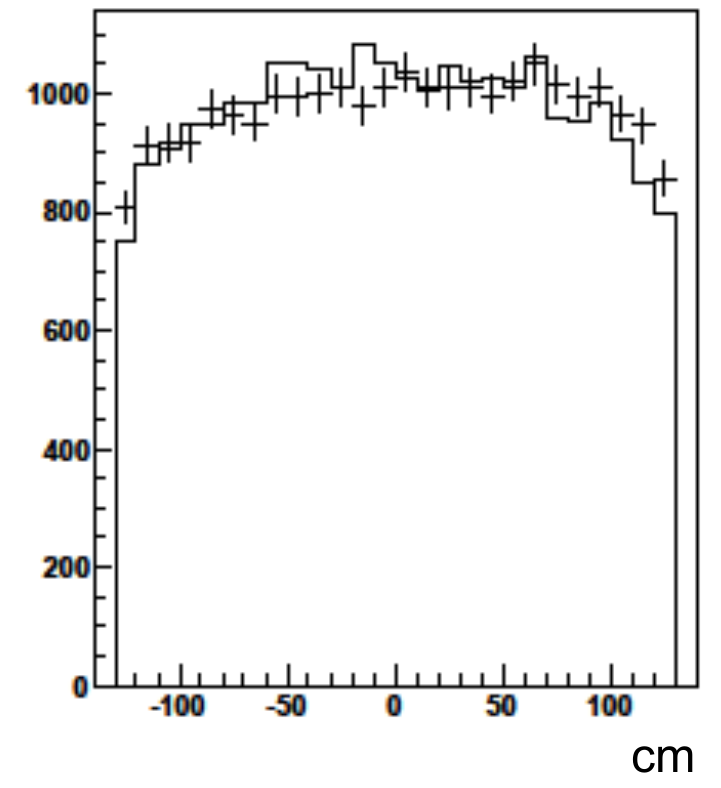} &
		\includegraphics[width=0.40\linewidth]{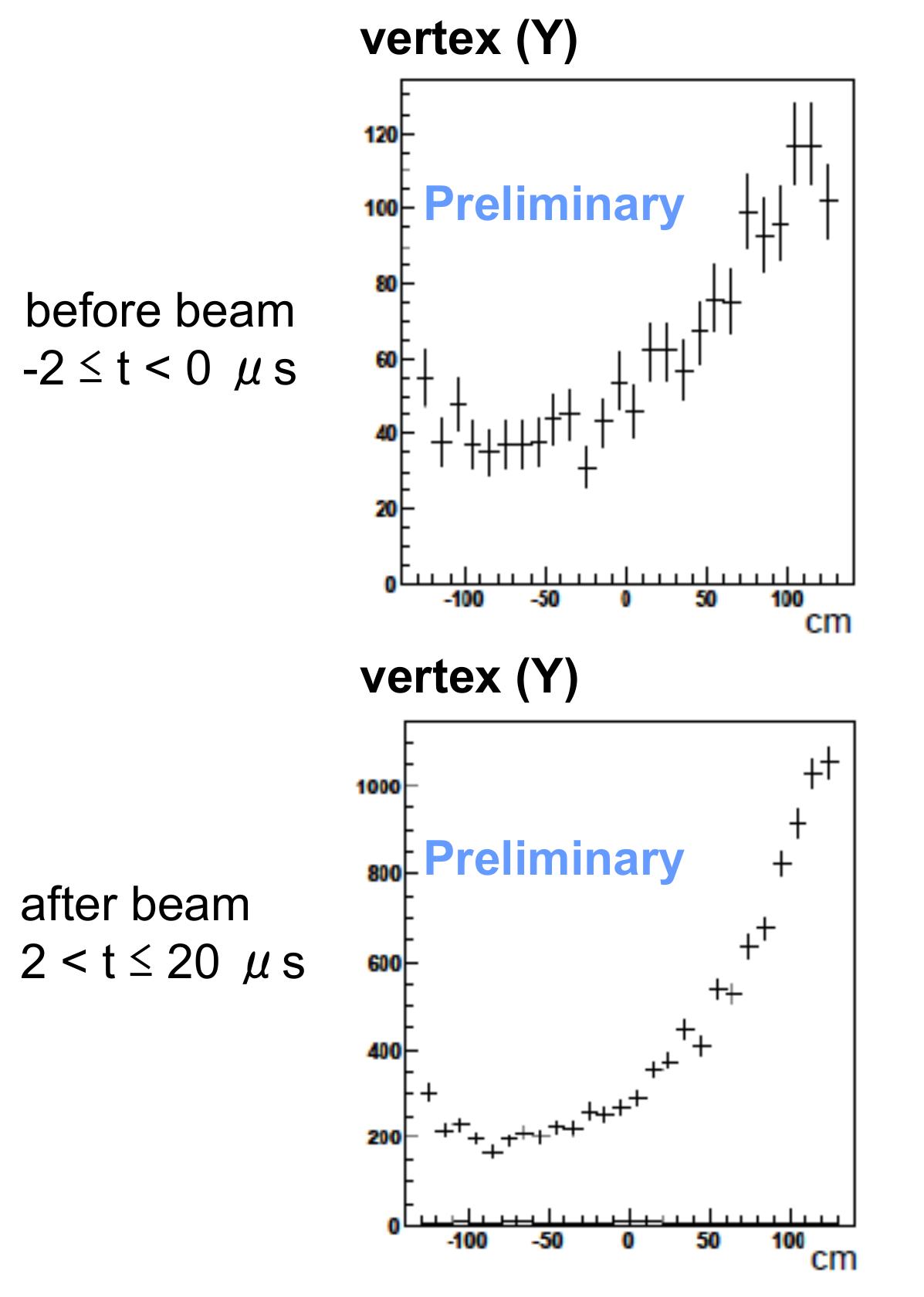}
	\end{tabular}

	\caption{The distribution of on-beamtime events in the upward-pointing, y direction in the SciBooNE detector (left), and the same distribution for hits before and after the beam window showing an excess towards the top of the detector.}
	\label{vertex-wrt-beamtime}
\end{figure}

It is thus important to carry out a detailed study of neutrons in the SciBooNE Hall. We must understand rates, energies, and stopping distances in various depths of water. For this purpose, we propose a series of studies using the proposed ANNIE water volume. Potentially we could also use a gas-phase TPC developed for use in the Double Chooz experiment, known as the Double Chooz Time Projection Chamber (DCTPC), built by a group at MIT. 

\begin{figure}
	\begin{center}
		\includegraphics[width=0.50\linewidth]{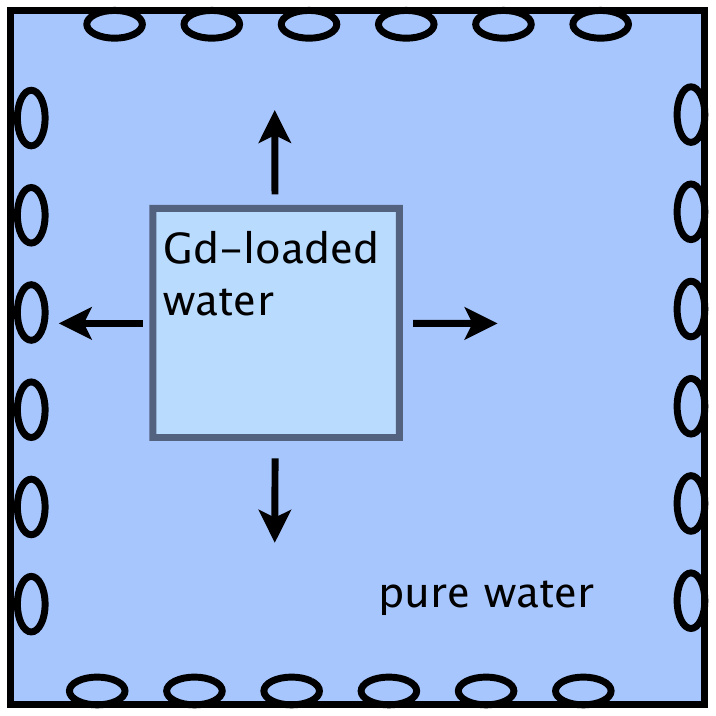}
	\end{center}
	\caption{A schematic showing a concept for ANNIE with a smaller, transparent Gd-loaded volume contained within a pure water volume. The inner volume can be moved around within the full tank to study how neutron capture rates very as a function of depth in the detector and distance from the beam.}
	\label{neutron-annie}
\end{figure}

We will seek to understand background rates directly in our target water volume. We propose to develop techniques similar to those used by the SciBooNE collaboration. Figure~\ref{vertex-wrt-beamtime} shows how non-uniformity in the vertex distributions of SciBooNE detector were used to identify the sky shine background. Selecting the number of events within the narrow beam window, vertices are reconstructed uniformly throughout the volume. However, when selecting a long window before and after the beam, a large number of vertices are reconstructed with a bias towards the top of the detector. This points to the specific background from sky shine. We can study non-uniformities in the neutron capture points, even without precision vertex reconstruction, by limiting the Gd-loaded water volume to a smaller portion of the total water volume. If this Gd-loaded target is transparent and movable, we can study how rates vary from top to bottom and in the beam direction, for beam-on events with no interaction and the water volume and for bunches with full-contained interactions (see Fig.~\ref{neutron-annie}).

\begin{figure}
	\begin{center}
		\includegraphics[width=0.50\linewidth]{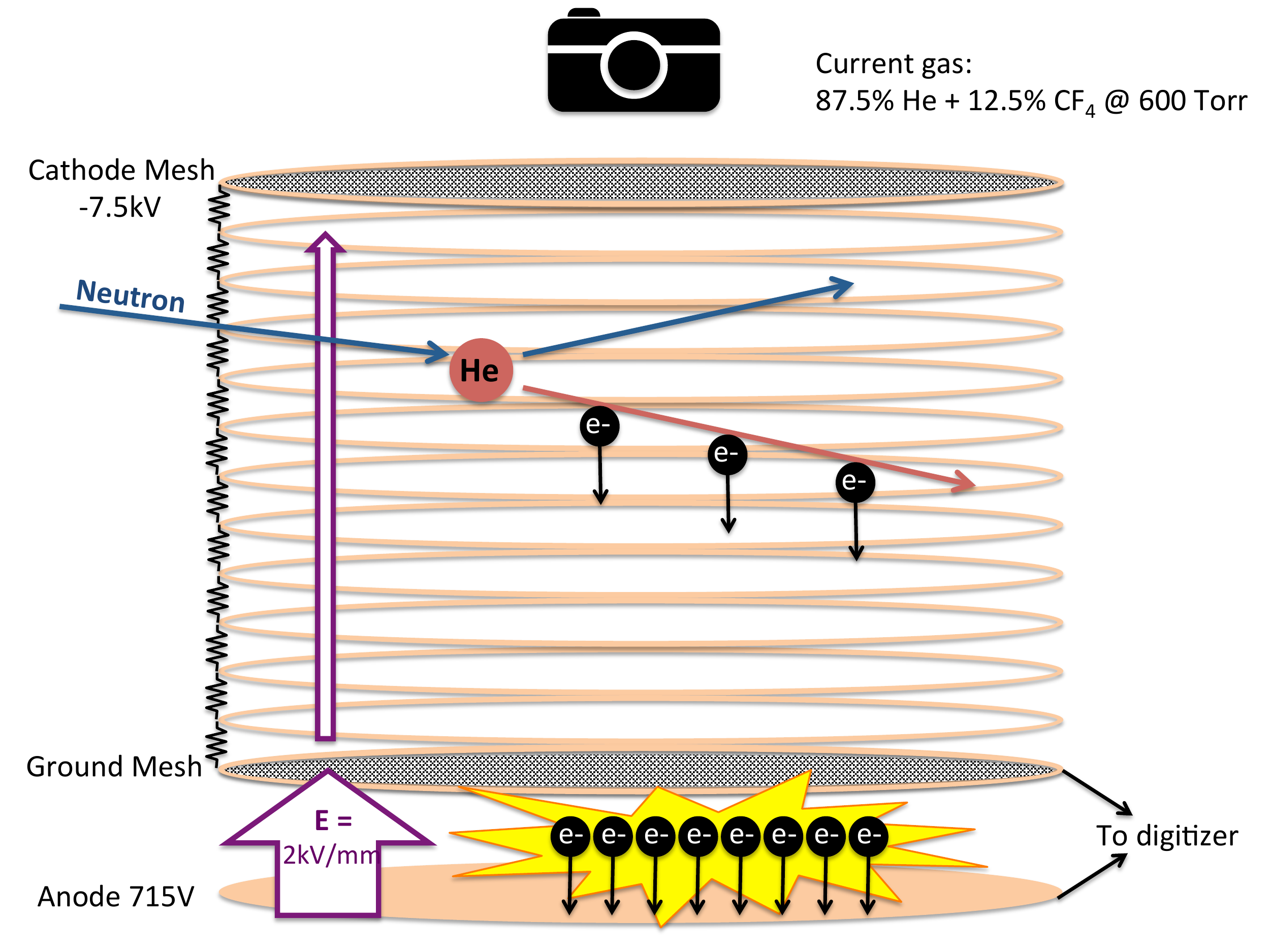}
	\end{center}
	\caption{A schematic showing the operating principles behind DCTPC.}
	\label{DCTPschametic}
\end{figure}

In addition, we could use the DCTPC which is sensitive to neutrons ranging from a few keV to tens of MeV (see Fig.\ref{DCTPschametic}). With a change in pressure and and target gas, the detector can be made sensitive up to 100 MeV. DCTPC has the advantage that it can reconstruct both the energy of the neutrons (from the recoil energy) and directionality. This could aid in pointing to where the expected backgrounds are coming from. Using the modified ANNIE volume and the DCTPC, it will be possible to compare {\it in situ} studies with direct {\it ex situ} measurements. If successful, the ANNIE collaboration may decide to build a neutron monitor based on DCTPC technology. The MIT group has expressed interest in collaborating in this aspect of ANNIE. 


\section{Mechanical Design of the Water Target}
\label{sec-mechanical}

\begin{figure}
	\begin{center}
		\includegraphics[width=0.7\linewidth]{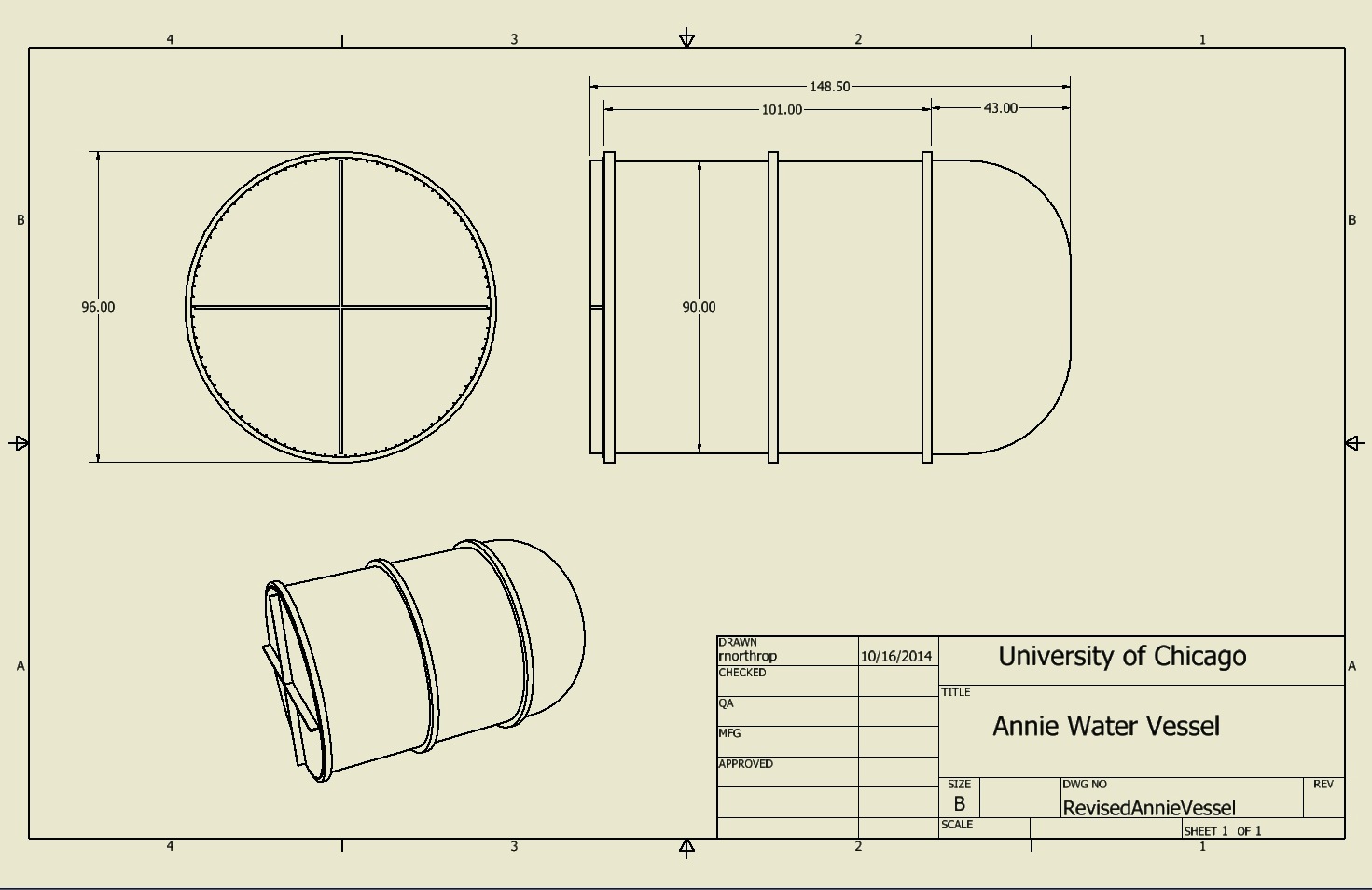}
	\end{center}
	\caption{Dimensions of the UChicago water tank to be used for the ANNIE target volume.}
	\label{ANNIEvessel}
\end{figure}

\begin{figure}
	\begin{center}
		\includegraphics[width=0.7\linewidth]{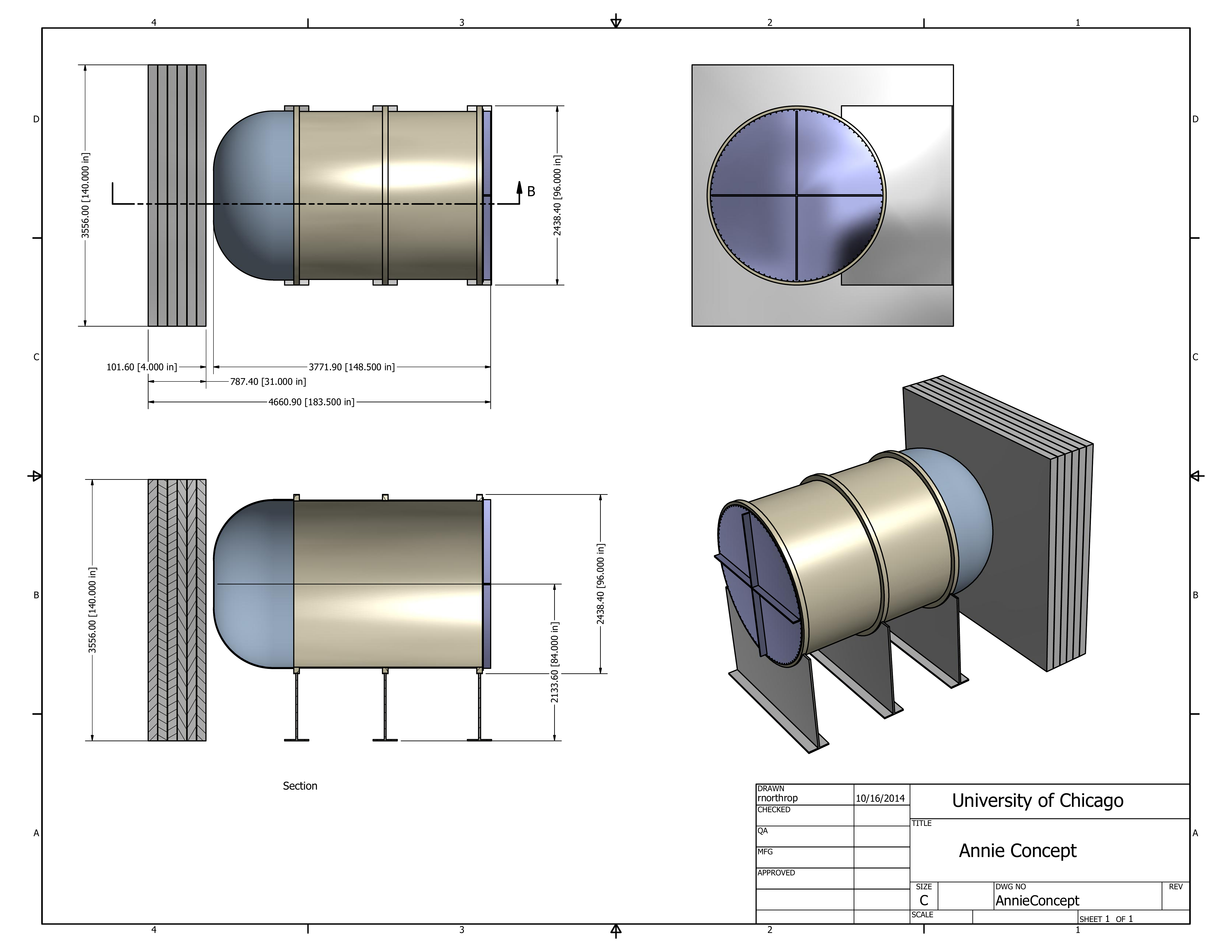}
	\end{center}
	\caption{The ANNIE tank, mounted on steel saddles, shown with the MRD.}
	\label{ANNIEdetector}
\end{figure}

\begin{figure}
	\centering

	 \begin{tabular}{c c}
		\includegraphics[width=0.45\linewidth]{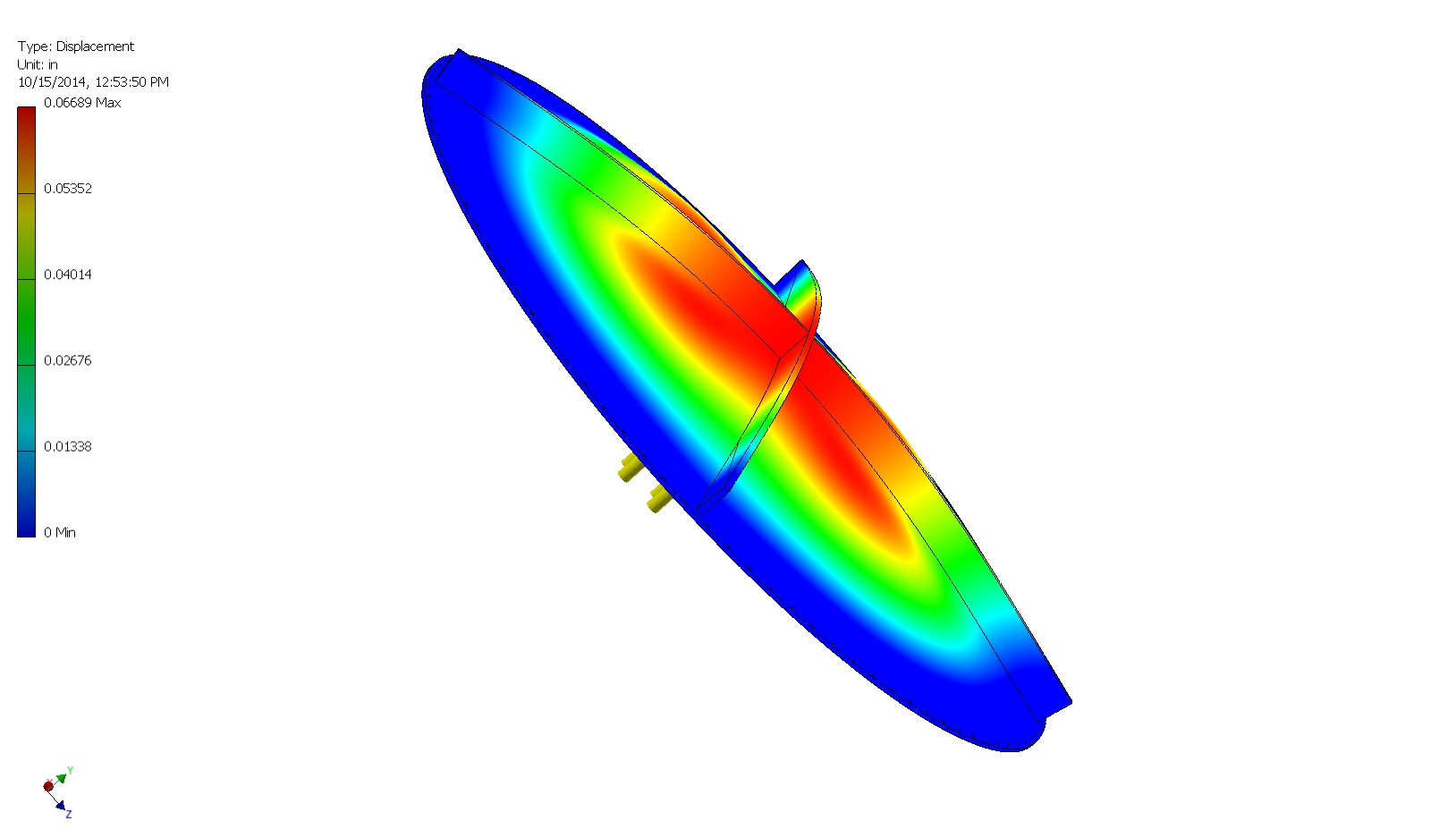} &
		\includegraphics[width=0.45\linewidth]{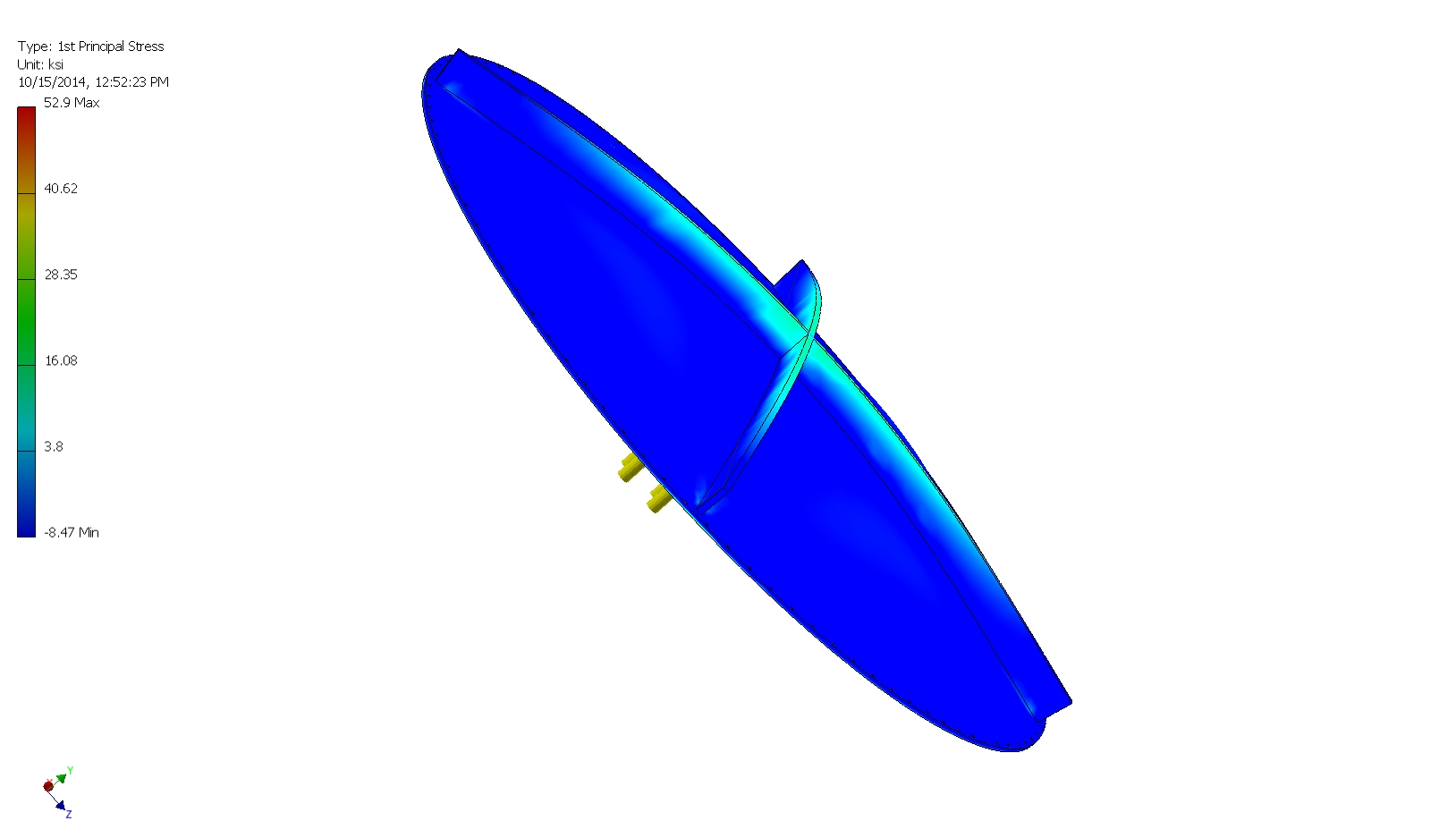}
	\end{tabular}

	\caption{Deflection (top) and stress (bottom) on the ANNIE endplate.}
	\label{ANNIEendplatestress}
\end{figure}

The baseline ANNIE design will make use of an existing pressure vessel, used for the cosmic-ray balloon experiment RICH, developed by Dietrich Mueller's group at the University of Chicago. The vessel is an aluminum tank, roughly 4.76 meters long and 2.29 meters wide, shown in Fig~\ref{ANNIEvessel}. Stress simulations show that the tank can withhold the stresses of a complete water fill. Plans are under way to perform a water-fill test and verify the integrity of the vessel. The tank is segmented, consisting of three parts: a central barrel and two dome-shaped end-caps. These three segments are joined by bolts at flanged ends with an O-ring between. Together, the barrel and two domes measure 4.76 meters (187.5 inches) long which will not fit in the SciBooNE hall. However, with a flat, reinforced steel flange to blank off one of the sides, the barrel plus one dome is an appropriate length of 3.77 meters (148.5 inches). This configuration will fit in the SciBooNE hall together with the MRD, as shown in Fig~\ref{ANNIEdetector}, leaving more than a foot of leeway. The design for the blank flange and stress studies are shown in Figure~\ref{ANNIEendplatestress}. The tank has one access port, already built-in, which will be used to feed through the cabling for our photosensors.


This 3.77 meter long and 2.29 meter diameter configuration will sit horizontally on a saddle consisting of three U-shaped steel yokes. Holding the central axis of the cylinder at 2.13 meters (84 inches) in line with the center of the BNB. The full detector system, including the water volume and MRD is shown in Figure~\ref{ANNIEdetector}.

The aluminum inner-surface of the tank raises concerns about water corrosion and reactivity with Gd salts. Two precautions will be taken to guard against this. First and foremost, the inner volume of the tank will be lined with a thermo-sealed plastic bag. Thickness of the plastic will be chosen to minimize the risk of puncture. Work is underway to determine how best to cover any sharp seem, rivets, and edges. We are also exploring glass epoxy coatings, which could be used to further prevent against contact between aluminum and water, should any punctures form.

The plastic liner will be held to shape by a plastic or steel support skeleton, which will double as the support structure for the photosensors. The skeleton may consist of two concentric cylinders, one encompassing the full water volume and a second, smaller structure surrounding a smaller volume, approaching the size of our target volume. The first run of ANNIE will likely make use of the smaller volume to instrument the fiducial mass with higher angular coverage, given a small batch of first LAPPDs. This inner structure may eventually be removed, the LAPPDs placed directly around the full volume, as prospects for high coverage improve. A second run of ANNIE will have 50-100 LAPPDs which would enable more efficient use of the target volume, as discussed in Section~\ref{sec-simulations}. One simple possibility for attaching photosensors to the skeleton is the use of water-proof velcro, rated to hold against weights much larger than those of the LAPPDs. The velcro fixtures would make the photosensor coverage easily adaptable and modular for various runs of ANNIE.

\section{Gadolinium Loading}
\label{sec-Gd}

\subsection{Neutron Capture in Water}

The fate of neutrons liberated in light water is first to be thermalized via collisions, primarily with free protons, then to be captured by a proton or oxygen nucleus. 
The cross sections for these capture reactions are 0.33 barns and 0.19 
millibarns, respectively, so to first approximation every thermal neutron is
captured on a free proton via the reaction $n + p \rightarrow d + \gamma$.
The resulting gamma has an energy of 2.2~MeV and makes very little detectable
light since the Compton scattered electron is close to Cherenkov threshold. 
The entire sequence from liberation to capture takes around 200 microseconds, with only a 
very small dependence (plus or minus a few $\mu$s) on initial neutron energy.

However, the situation is much improved by adding a water-soluble gadolinium (Gd) compound, 
gadolinium chloride,  GdCl$_3$, or the less reactive though also less soluble gadolinium sulfate, 
Gd$_2$(SO$_4$)$_3$, to the water. Naturally occurring gadolinium has a neutron capture  
cross section of 49,700 barns, and these captures produce an 8.0~MeV gamma cascade, easily 
seen in typical water Cherenkov detector configurations.  Due to the much larger cross section of 
gadolinium, adding just 0.2\% of one of these compounds (about 0.1\% Gd)
by weight is sufficient to cause 90\% of the neutrons to capture visibly on gadolinium 
instead of invisibly on hydrogen.  Following the addition of gadolinium the 
time between neutron liberation and capture is reduced by an order of magnitude to 
around 20~$\mu$s, greatly suppressing accidental backgrounds.

\subsection{Water Filtration System}

Starting with the very first large-scale Water Cherenkov detector -- the
Irvine Michigan Brookhaven (IMB) proton decay experiment, which began
taking data in the early 1980's, exceptional water clarity has been of
key importance for massive devices of this kind.  After all, there is
little benefit in making a very large detector unless the target mass
contained within the detector can be efficiently observed.  

The strategy employed to create extremely clear water has been to remove all 
suspended solids, dissolved gases, ions, and biologics from solution via a series of
filtration elements.  These include microfiltration filters, degasifiers
(vacuum and/or membrane type), reverse osmosis membranes (RO),
deionization resins (DI), and exposure to intense ultraviolet light (UV).  The water
in the detector must be continuously recirculated through the filtration system. This 
is necessary as transparency-impairing materials are steadily leaching into the 
chemically active ultra-pure water.  In addition, during the process of filtration 
the water is typically chilled to further impede biological growth, with the 
added benefit of simultaneously reducing PMT dark noise which is typically 
strongly temperature dependent.

When gadolinium loading is desired, the filtration situation becomes significantly 
more complicated: somehow the water must be continuously recirculated and cleaned of 
everything {\em except} gadolinium chloride or gadolinium sulfate.  Over the past 
decade there have been focused R\&D programs both in the US and Japan aimed at devising 
a method capable of maintaining this exceptional water transparency while at the same 
time maintaining the desired level of dissolved gadolinium in solution.  A solution has been 
developed, a novel, truly selective water filtration technology, now known as a 
``molecular band-pass filter''. 

However, the stringent requirements on water transparency in force at very large water Cherenkov 
detectors like Super-Kamiokande, where light can be expected to travel for tens 
of meters before it is collected by photomultiplier tubes or other technologies, will 
be significantly relaxed for ANNIE due to its modest physical dimensions. This also greatly 
simplifies the situation when Gd is added to the water. 

As an outgrowth of the aforementioned water filtration R\&D programs, the Gd-tolerance and 
efficacy of every component of a traditional water filtration design has been evaluated. 
This allows the straightforward design of a streamlined, Gd-tolerant water system tuned 
for the needs of ANNIE. An ANNIE water system would include 
microfiltration,  ultrafiltration, UV light, deionization (during the 
filling stage only), and membrane degasification, with temperature control 
via a chiller recommended but not strictly necessary.  Once the detector is 
full of pure water the deionization (DI) unit would be bypassed, and then the gadolinium 
compound -- roughly 40 kilograms in total -- would be added directly into the 
water stream; it will dissolve as the water is recirculated.  When it is ultimately 
time to remove the gadolinium from solution, simply turning the DI unit back on will 
do the trick.

Such a system has a variety of advantages: it is both cost-effective and easy to 
operate, it has a modest physical footprint in the experimental hall, and it will 
provide water of at least 15 meters attenuation length, perfectly adequate for a 
detector roughly three meters on a side.

\subsection{Gd Availability}

Gadolinium loading of ANNIE will come at a very modest cost.  Based upon years of experience building similar systems by ANNIE collaborators, the Gd-capable water filtration system for  ANNIE described can  be delivered and installed at the experimental site by collaborators at UC Irvine. 

The minimum amount of water soluble gadolinium compound needed by ANNIE is about 50 kilograms. 
This is enough for one test loading (4 kg) and one full loading (40 kg), with 6 kg left 
over for various other studies and tests. Over the last two decades the highly volatile 
rare-earth market has valued this quantity of Gd$_2$(SO$_4$)$_3$ as high as \$24,240.

As it happens, 50 kilograms of high-quality (99.99\% TREO) gadolinium sulfate is currently 
available as surplus from the Japanese R\&D program. It has been set aside for ANNIE by 
the director of the program (M. Vagins, an ANNIE collaborator) and would be supplied to 
Fermilab/UChicago free of charge.

\section{Photodetection}
\label{sec-photodetection}

In this section we describe the availability and development work regarding photosensors to be used in ANNIE. ANNIE will be a hybrid WCh, containing LAPPDs for detailed event reconstruction and a large number of conventional photomultiplier tubes to improve light collection for energy reconstruction and efficient neutron tagging.

\subsection{Photomultiplier Tubes}
\label{subsec-PMTs}

In order to maximize light collection, ANNIE will need to augment its LAPPD coverage with and existing stock of conventionalÄù 8-inch PMTs (manufactured by Hamamatsu) taken from the IMB and the Super-Kamiokande experiment. PMTs of type I (IMBÄù) have separate bases and require a water seal, while PMTs of type S (Super-Kamiokande) have integrated bases sealed by the manufacturer. Both PMTs are operated with positive high voltage between 1,500 V and 2,000 V. To select PMTs for ANNIE, the dark-noise rate of all PMTs was measured in a small dark box at various voltages. In each case, the PMT was also exposed to light from a wrapped piece of plastic scintillator attached to a light guide. Between the end of the light guide and the PMT there is a 10 cm air gap.  Inserting a plug into the light guide blocks the light from cosmic ray muons passing through the scintillator in order to facilitate the dark-noise measurements. Each PMT was measured at the same distance to the plugged or unplugged scintillator. All PMTs were triggered (by oscilloscope) with the same threshold. 

We have nearly completed basic quality assurance studies of the existing PMT stock. All of the PMTs were tested in a dark box, both for dark counts, and with light produced by cosmic rays passing through a scintillator bar. Performance of the PMTs was characterized by the metric given in Equation~\ref{eq:PMTquality}, where $L_i$ are the pulse heights for ``light" events produced by the scintillating bar, and $D_i$ are the pulse heights for "dark" events - spontaneous signals in a dark box.

\begin{equation}
Q = \cfrac{<L>-<D>}{\sqrt{(<L^2>-<L>^2)(<D^2>-<D>^2)} }
\label{eq:PMTquality}
\end{equation}

Each PMT was then categorized by Q value. The definitions of each classification can be found in Table~\ref{PMTclassification}. Among the PMTs, 63 type-S and 43 type-I tubes are of a quality 1 or higher, totally 106 (see Table~\ref{PMTinventory}).  For the 63 Type-S PMTs rated 1 or higher, no additional costs will be necessary for use in ANNIE.  Some work will be necessary to water proof  the usable stock of Type-I phototubes. Some tubes remain to be tested - 67 type-I tubes, and 7 type-S. However, based on the fractions measured so far, we expect that 60-120 usable PMTs will be available from this stock. 

There are thus two different coverage scenarios, one with 60 PMTs and a high coverage configuration with  more than 100 PMTs. The low coverage scenario is immediately possible with the ready and tested Super-K PMT stock. The 100 PMT scenario will still require work to attach water-proof HV bases to the IMB PMTs. The implications of these scenarios are discussed in Sec.~\ref{sec-PMTcovgsim}.

Once, basic QA testing is complete, we plan to more systematically study the PMTs. We of course will start with those most likely to be used in a first run of ANNIE. High voltages for the PMTs can be achieved using an existing supply possibly provided by the Super-Kamiokande collaboration. 

\begin{table*}
\begin{center}
\begin{tabular} {| c| c | c |}
\hline
-1 & ``bad" & nonfunctional \\
0 & ``okay" & $Mean+RMS<P_{max}<0$ \\
1 & ``good" & $Mean<P_{max}<Mean+RMS$ \\
2 & ``great" & $Mean-RMS<P_{max}<Mean$ \\
3 & ``excellent" & $P_{max}<Mean-RMS$ \\
\hline
\end{tabular}

  \end{center}
  \caption{Classification scheme for the ANNIE PMT-stock, based on QA testing.}
  \label{PMTclassification}
\end{table*}

\begin{table*}
\begin{center}
\begin{tabular} {| l | c | c | c | c | c | c | c |}
\hline
Rating &  -1 & 0 & 1 & 2 & 3 & $\ge$ 1 & untested \\
\hline
\hline
Type-S & 11 & 4 & 36 & 21 & 6 & 63 & 7 \\
Type-I & 76 & 2 & 18  & 16 & 9 & 43 & 67 \\
\hline
\end{tabular}

  \end{center}
  \caption{Inventory of the ANNIE PMT stock and their quality ratings. There are 63 Type-S PMTs rated 1 or higher, and 43 Type-I, with some remaining to be tested.}
  \label{PMTinventory}
\end{table*}

\subsection{Large Area Picosecond Photodetectors}
\label{subsec-LAPPD}

The Large Area Picosecond Photodetector (LAPPD) project was formed to develop new fabrication techniques for making large-area (8" x 8") MCP photodetectors using low-cost materials, well established industrial batch techniques, and advances in material science~\cite{LAPPD}. 

LAPPDs may be a transformative technology for WCh detectors. While conventional photomultipliers are single-pixel detectors, LAPPDs are imaging tubes, able to resolve the position and time of single incident photons within an individual sensor. This maximizes use of fiducial volume as it allows for reconstruction of events very close to the wall of the detector, where the light can only spread over a small area. The simultaneous time and spatial resolutions of the LAPPDs, at better than 100 picoseconds and 3mm for single photons, represent a major improvement over conventional PMTs. Preliminary Monte Carlo studies indicate that the measurement of Cherenkov photon arrival space-time points with resolutions of 1 cm and 100 psec will allow the detector to function as a tracking detector, with track and vertex reconstruction approaching size scales of just a few centimeters~\cite{fastneutrino}. Imaging detectors would enable photon counting by separating between the space and time coordinates of the individual hits, rather than simply using the total charge. This means truly digital photon counting and would translate directly into better energy resolution and better discrimination between dark noise and photons from neutron captures. Finally, at a thickness of less than 1.5 cm, LAPPDs maximize the use of the limited fiducial volume available to small detectors. 

\subsection{LAPPD Status and Availability}

As the LAPPD effort transitions into the next stage of the project, they will become available through the commercialization process. The DOE awarded  $\$3$M possible through the STTR program to Incom, Inc to develop a commercial line over the next several years. Incom has been a key institution in the LAPPD project from the beginning, as the company who fabricates the glass microcapillary arrays that serve as the gain stage central to LAPPDs. Over the years, they have greatly refined the process of making these substrates and are now scaling their process for higher volumes and better yields. They have also made significant progress in expanding the scope of their capabilities, from making the bare plates to building whole detectors. They have purchased an ALD-reactor system, necessary to coat and activate the plates. This reactor will be operated by an employee who has spent the last year working closely with the Argonne group who pioneered the coating process. They purchased an evaporation chamber for electroding the components. They are also building a large clean room space for full production line. The assembly of complete, sealed-tube detectors, will occur in a vacuum-transfer system designed based on LAPPD experts at Berkeley Space Science Laboratory. 

An industry standard for commercialization is around 3-4 years, but with the process already under way, the timeline for LAPPD tile commercialization has been refined. 
General availability of prototype LAPPDs is determined by the progress on the SBIR Phase II commercialization program being implemented by Incom Inc. In private communication with
Michael Minot of Incom Inc., he anticipates demonstrating pilot production of LAPPDs by Quarter III of 2015 followed by tile pilot production, providing limited numbers of LAPPD tiles to early adopters in the first half of 2016. Some small number of tiles may be available sooner. Given ANNIE's readiness as an early adopter, it is likely that we will be able to procure at least one such tile.

Production of LAPPDs at Incom will occur in small test batches, with volume increasing in time. Consequently, ANNIE will likely start with sparse LAPPD coverage: roughly five 20cm x 20cm LAPPDs to cover the entire 3m x 3m wall facing the beam. In the beginning, all ANNIE events will resemble those close to the wall of a much larger detector. In addressing these challenges, ANNIE provides an opportunity to understand the benefits of imaging photosensors.  ANNIE is an ideal first adopter of LAPPDs for neutrino detection, as light yields are low enough to comfortably fall within the rate handling abilities of the default LAPPD design, and because the time structure of the beam events and optical pile-up seems compatible with the electronics in their current, as discussed in Sec~\ref{sec-electronics}. Moreover, ANNIE is small enough that isotropic coverage is possible without huge volumes. 

Several other pathways may provide ANNIE with a stock of LAPPDs or smaller detectors based on similar technologies. Work at University of Chicago, parallel to that of Incom, explores a more radical and cost-effective fabrication method. This work is ongoing and if successful, could provide a new source for prototype LAPPDs as well as a path for higher volume production lines.

\begin{figure}
	\begin{center}
		\includegraphics[width=0.7\linewidth]{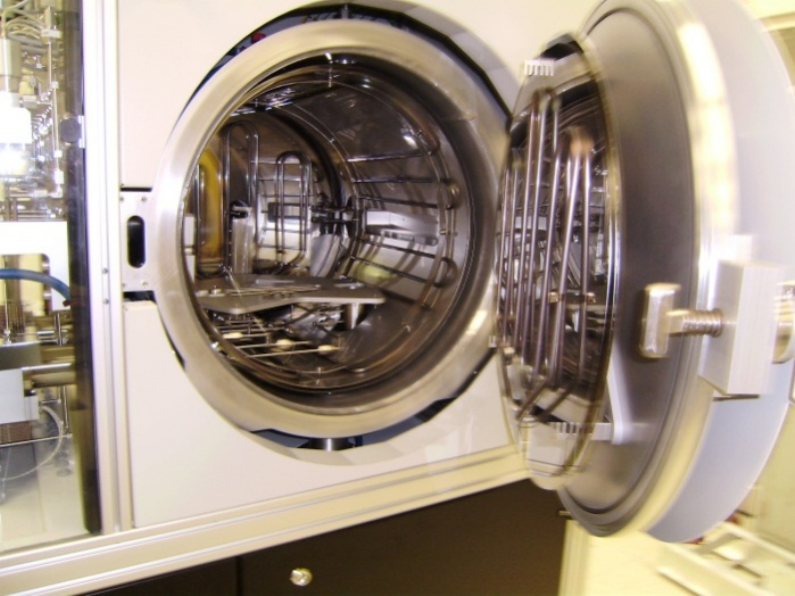}
	\end{center}
	\caption{The Beneq ALD reactor purchased for Incom's MCP fabrication facility.}
	\label{Beneq}
\end{figure}


\subsection{ANNIE Specific LAPPD Development Work}

Some application-specific detector development work will be necessary to ready LAPPDs for use in WCh detectors. In this section we discuss some of the steps necessary to ready these sensors for use in a WCh.

Once commercially available, early LAPPD prototypes will be thoroughly tested, first on a test bench and then in a scaled-down operational context. A large variety of testing facilities and fixtures are available for the task. In addition, there is a broader community of interested ``early adopters" with plans to use LAPPDs in similar experimental contexts. A facility at the Advanced Photon Source at ANL is designed to characterize the time-response of LAPPD detectors using a fast-pulsed laser~\cite{APS}. Fixtures exist to test complete, resealable LAPPD systems with electronics. These facilities can be used to benchmark the key resolutions of detector systems and study the effects of pileup from coincident pulses on the same anode strip. 

Tests of the readout and and DAQ can begin even before the first sealed commercial prototypes become available. Eric Oberla at University of Chicago is building a small, tubular WCh detector for timing-based tracking of cosmic muons using commercially available Planacon MCPs. The electronics are designed using the LAPPD front-end system, which will provide an excellent opportunity for developing operational experience and testing aspects of the front-end design. It may be possible to deploy this detector on a Fermilab test-beam. Finally, there are discussions of performing tests either with a radioactive calibration sample or a test beam on a small target mass coupled to a handful of LAPPDs. 

High voltage connections for LAPPDs will be made on the front window. Work will be carried out to design and test a scheme to make these connections and operate the front-end electronics in water. In addition to answering these specific technical challenges, the ANNIE effort will also provide critical feedback on the performance and long-term operation of LAPPDs, thus helping to expedite the commercialization process. 

Some funding has already become available through the WATCHMAN collaboration to work on these technical problems. The goals are to further develop the DAQ for multi-LAPPD systems and design and demonstrate a method for sealing the electronics and water in high voltage. Engineers at UChicago will begin work soon.

Once all of the WCh-specific development work is achieved, testing an LAPPD submerged in water will be the final demonstration of feasibility. Such tests could occur in the official ANNIE water volume as part of a ``Phase I" technical proposal, as described in Sec.~\ref{sec-timeline}. 


\section{Muon Range Detector}
\label{subsec-MRD}

\begin{figure}
	\begin{center}
		\includegraphics[width=0.55\linewidth]{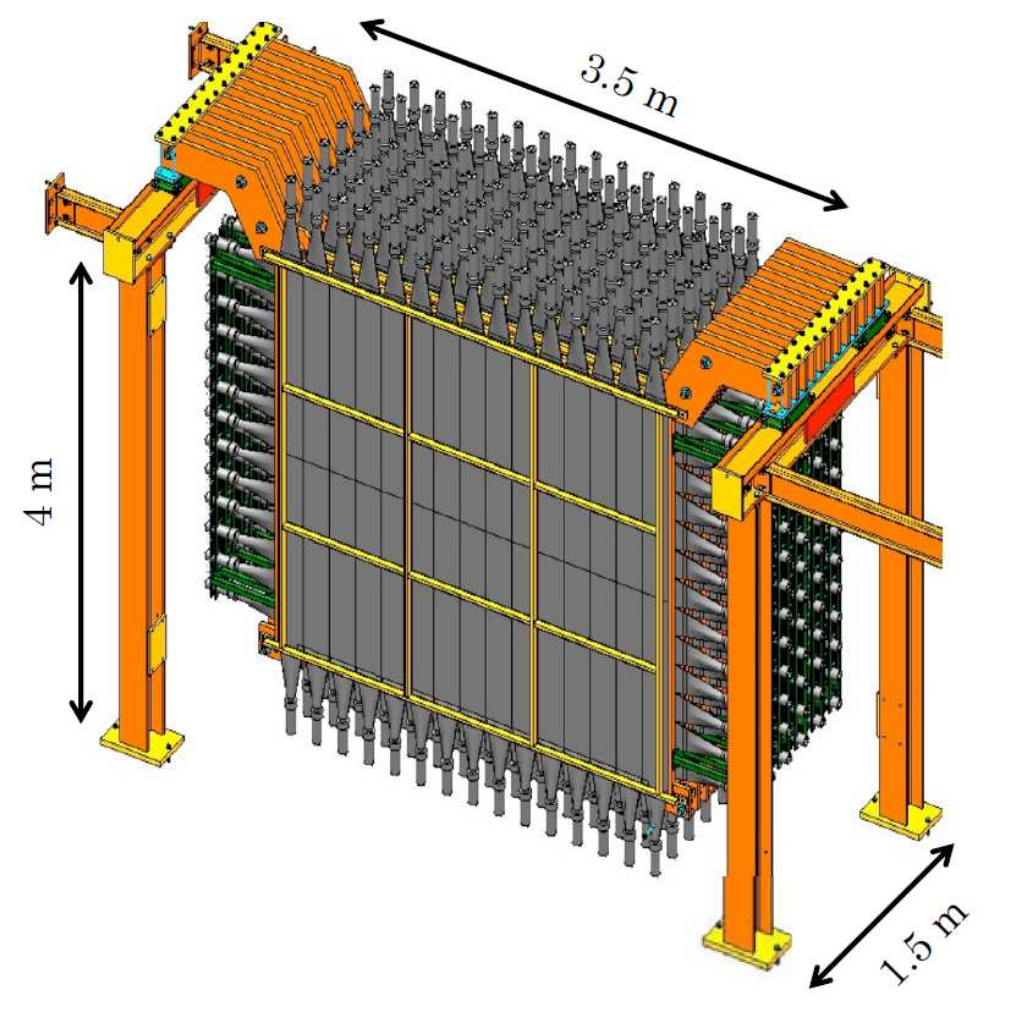}
	\end{center}
	\caption{A drawing of the SciBooNE Muon Range Detector.}
	\label{MRDdrawing}
\end{figure}

The Muon Range Detector (MRD), shown in Fig~\ref{MRDdrawing}, will be used to measure the momentum and total energy of muons leaving the ANNIE water volume with range-out energies up to 1.2 GeV/c. As a baseline, ANNIE will make use of the MRD as designed for the SciBooNE experiment, however there is potential for improvements. 

The SciBooNE MRD forms the core of the ANNIE muon range system. The SciBooNE MRD consists of twelve iron plates, each 2 inches thick, sandwiched between thirteen layers of scintillator. The scintillator planes were comprised of panels, each 20 cm wide and 0.6 cm thick, and were were arranged in alternating vertical and horizontal layers. The SciBooNE MRD was read out by 362 PMTs, each 2 inches in diameter. The iron plates cover an area of 274 $\times$� 305 cm$^2$.  

After SciBooNE ceased operations, various parts of the MRD have been claimed by other groups. In particular, the scintillator panels are owned by New Mexico State University and many of them may be used in the MicroBooNE veto detector. Thus, instead of relying on a steel-scintillator sandwich, ANNIE is planning to use a resistive plate chambers (RPCs) that are available from the digital calorimetry group at Argonne National Laboratory.  

The ANNIE MRD would keep the twelve steel plates intact and in their present configuration. These are currently owned by Dr. Morgan Wascko, a SciBooNE co-spokesperson, who is based at Imperial College London. We have initiated a property transfer to Dr. Matthew Malek, also at Imperial College and an ANNIE collaborator. Instead of thirteen layers of scintillator, the new design calls for only three layers, with the remaining ten layers comprised of RPCs from Argonne. 

The switch to RPCs can be considered a significant upgrade for the ANNIE MRD. The RPCs have centimeter-level precision at each layer, allowing for more precise reconstruction. Additionally, the RPCs can tolerate a magnetic field on the order of 1 Tesla. Applying such a field during a future phase of the experiment would allow for charge-sign reconstruction in the MRD. 



The RPCs will be provided by the digital calorimetry group at Argonne National Laboratory, and are ready now. Each RPC is layer is one square meter, with onboard readout pads, and has position resolution at the centimeter level for each layer. 

Unlike the SciBooNE MRD, which contained thirteen layers of scintillator, the ANNIE MRD could be arranged with three layers of scintillator and ten layers of RPCs in the following configuration: The first (i.e., most upstream) layer of the detector is to be composed of scintillator and used as a trigger. This will be followed by five layers of RPCs. A second scintillator layer will serve as a coincidence trigger for noise reduction, if necessary. During the test phase known as ANNIE Phase I, we will evaluate whether this coincidence trigger is required. Following the second scintillator layer, there will be five more layers of RPCs, ending with a final layer of scintillator. This last layer will tag exiting events, for which only a lower bound on energy can be set.

The Argonne RPCs have been tested with both a steel absorber (at Fermilab) and a tungsten absorber (at CERN). They are available now, with a working readout, and can be included in the ANNIE Phase I test. 

Additionally, the RPCs can tolerate a magnetic field on the order of 1 Tesla. Applying such a field during a future phase of the experiment would would greatly enhance the physics potential of the MRD by allowing charge-sign reconstruction. This capability would bring the MRD beyond simple calorimetry, enabling measurements of energy and charge-sign via curvature in the magnetic field. Even for the most energetic beam events, momentum measurements are then possible from curvature. The high position resolution of the RPCs enable precision measurements of the curvature and a strong particle identification (PID).

ANNIE could be the first experiment to use a magnetized MRD in conjunction with a Gd-loaded water Cherenkov detector. The neutron capture on gadolinium provides complementary information to the curvature in the magnetic field. Thus, using this information in a global reconstruction provides a high purity selection of lepton charge-sign and interaction type. 

\section{Electronics}
\label{sec-electronics}


The trigger and readout electronics will be developed jointly by UChicago, ISU and Queen Mary. The UChicago group together with UHawaii have been the main developers for the fast readout chip PSEC4.   The ISU group brings significant prior experience with data acquisition systems. The Queen Mary group also has significant expertise from their work in the T2K and SNO+ experiments. 

The electronics design for ANNIE needs to accommodate both fast sampling of the Cherenkov light from the muon track (100~ps resolution over $\sim$100 ns duration) and the delayed capture of the thermal neutrons, which occurs approximately 30 $\mu$s later. ANNIE will require an integrated system with both LAPPDs and conventional PMTs. Because of the near-surface operation and intense light from signal events spread over the small area of the detector walls, the front-end and DAQ systems will be designed to handle this optical pileup. 

A trigger rate of $\le$ 20 Hz is envisioned, which includes 7.5~Hz of beam spills from the Booster Neutrino Beam, each spill lasting approximately 1.6~$\mu$s.  The chance probability for event interaction pileup during a beam spill from  muon neutrino interactions is extremely small, so that one can assume less than 1 neutrino interaction per beam trigger.  This leaves ample room for calibration signals such as random triggers (which measure the detector in a quiescent state),  cosmic ray-induced muons passing vertically through the detector, injected light calibrations, etc.
As noted in section \ref{sec:crmuons}, an anti-coincidence shield will be required to moderate the rate of cosmic ray-induced muons and keep the overall trigger rate at a manageable level (tens of Hz) for the proposed electronics.

The event rate encompasses physics occurring on two vastly different time scales, which argues for a dual readout system. The fast signal from the muon track will be sampled with high time resolution using a fast readout system, described in section \ref{sec:fastreadout}.  The slow system requires only moderate bandwidth (62.5 MS/s - 250 MS/s) but must accommodate a readout window significantly greater than  30 $\mu$s  in order to capture the delayed component and to measure the background from sky shine. This system is described in section \ref{sec:slowreadout}. This dual scheme enables precision reconstruction of the prompt events with continual readout for the subsequent captures.

While the fast readout system was developed and optimized for the LAPPDs, we plan to record signals from the PMTs using both the slow and fast electronics.  This allows us to better use the PMTs to compensate for limited LAPPD coverage when measuring the fast component and it also has advantages for cross-calibration purposes.  Signal splitters developed for the CDF collaboration will be used to split the PMT signal for this purpose.


The large number of channels in the LAPPDs does not allow the use of the same dual readout scheme as the one used for the PMTs. However, even while ``dead'' the PSEC4 chips continue to write out prompt pulses.  This means that the PSEC trigger system can still detect, count, and time-tag pulses above the trigger threshold during the dead period. In this sense, a slow readout system could be designed in firmware to enable LAPPDs to detect neutron captures, even while the sampling electronics is frozen to read out the prompt event.


\subsection{Fast Readout System}\label{sec:fastreadout}

The fast readout system will be adapted from a complete front-end system for LAPPDs already developed and tested by groups at the University of Chicago and University of Hawaii.  The system is built around a new class of low power, custom designed waveform sampling chips~\cite{{UCelectronics2},{UHawaiielectronics},{UHawaiielectronics2},{UHawaiielectronics3},{UHawaiielectronics4}}. Some development work will be necessary to address the unique needs of the ANNIE detector. 

The default choice of electronics for ANNIE fast readout system is a suite of electronics built around the PSEC-4 chip~\cite{UCelectronics2}.  This chip is a CMOS-based sampling ASIC, sometimes described as a scope-on-a-chip.  The PSEC-4 allows full shape fitting of LAPPD pulses and consequently a more sophisticated analysis of the hits. The chip has a 256-sample circular buffer; the total duration of that buffer is set by the clock rate, which is adjustable over a range of 30-50 MHz.  The duration of a single sample is the length of one clock cycle divided by 256 and the sampling rate is the clock rate multiplied by 256. 

Nominal settings for the PSEC-4 puts the sample size at around 100 picoseconds, corresponding to a 25 nanosecond buffer and a clock rate of 40 MHz. Once a decision is made to read the buffer, that channel is dead for 2-4 microseconds. So, reading a chip can mean as much as 20 microseconds of dead-time.  

A proposed PSEC-5 chip would have a much deeper (order microseconds) circular buffer, almost capable of continuous operation at low rates. However, on the timescale for ANNIE, it is more realistic to design around the existing chip. Fortunately, given the length scales of the ANNIE detector, and the fact that we can trigger on the beam clock, it should be possible to cover the entire prompt event with the PSEC-4 buffer. 

%

The PSEC electronics has been integrated into a full readout system. Firmware already exists to operate 120 channels (2 LAPPDs) with self-triggering. Six PSEC chips are incorporated into an analog-to-digital card (ACDC). Two ACDCs are needed per LAPPD (one for each side). Four of these cards are integrated into an FPGA-based ``central card", although systems with more than 4 ACDCs will likely require a different architecture.

\subsection{Slow Readout}\label{sec:slowreadout}

The requirements of the slow readout system can be met by off-the-shelf commercial FADCs with buffer depth exceeding 100~$\mu$s and moderate bandwidth specification (62.5 MS/s - 250 MS/s).  Examples of these devices include the CAEN V1740 64-channel 12-bit digitizer and the CAEN 1720 8-channel digitizer, respectively. Both have memory buffers with adequate depth: 10MS/ch (CAEN 1720) and 1.5MS/ch (1740). Given a trigger rate of 20Hz envisioned for ANNIE, the data rate requirements for a 60~$\mu$s readout window ($\sim $ 2 times the minimally required window) results in 4 MB/s. Both CAEN modules accommodate data transfer rates of 60 MB/s - 160 MB/s using a VME interface.  Assuming the dead time of the modules is dominated by the data transfer, and given the trigger rate of 20~Hz planned for ANNIE, we expect a per trigger dead time of a few ms.

\subsection{Cosmic-Ray Induced Muons}\label{sec:crmuons}

The large flux from cosmic-ray induced muons in the Earth's atmosphere results in an integral flux of approximately 166~$\rm m^{-2} \, s^{-1} $ at sea level with a spectral peak energy of  $\sim$4~GeV.   For an accurate flux prediction at the SciBoonNE site, detailed simulations are required to take into account the fluctuations in the range of  muons passing through the surrounding rock.  However, we can treat the muon rate at the surface as an upper range for the actual rate, giving a rate of  $\sim$1.4~kHz using an effective collection area of 8.6 $ m^{2} $  for the ANNIE tank. In order to reduce these events to a rate of  less than 1~Hz,  an anti-coincidence shield with a high efficiency is required. While a large rate of cosmic ray muons is undesirable, muons passing through the detector volume along a well-defined path can provide an independent means of calibrating the ANNIE detector  throughput.  This could monitor the collective effects from changes in the optical throughput in the Gd-doped water, the phototube gains, the quantum efficiencies and any variations in the electronics system. For example, by forcing the electronics to trigger on a select range of vertical muons at a rate of 5~Hz, the detector response and changes therein could be monitored  at the few percent level over time scales of less than 30 minutes.


\subsection{Trigger Strategy}

We envision the following trigger strategy: beam spill events have the highest priority and an external trigger of the slow electronics readout occurs at each beam spill, irrespective of the status of the anti-coincidence to guard against a high rate of cosmic ray triggers.  A $60-100 {\mu}$s window is read out of the FADC buffers, starting at the time of the beam spill trigger.  
While the probability of a cosmic ray event passing through the detector is small ($10^{-4}$), a status bit, indicating the state of the anti-coincidence (and thus the presence or absence of cosmic ray-induced muons), will be stored along with the other event data.  It may also be desirable to force readout of the MRD with every triggered event. 

There is some additional subtlety required when triggering readout of the fast electronics system, due to that system's short ($\sim 30$ ns) buffer depth and high deadtime.  This means an additional trigger condition must be applied in addition to the beam spill requirement, based on the number of individual chips or channels that trigger.
This additional trigger requirement has two tiers.
First the buffers on each PSEC-4 chip will be frozen if triggered by simultaneous signals on more than one channel, preventing loss of data.
Second, a more aggressive requirement demanding a larger channel trigger multiplicity is imposed, ANDed with the beam spill trigger.  If no such condition is met, the hold is be taken off of the chip, immediately making it active again.  With this trigger scheme, the only dead time for non-events would be the few nanoseconds necessary to make a global trigger decision.  When a real event happens, then the relevant chips will be dead for several tens of microseconds. However, given the beam structure and expected event interaction pile-up, this still allows mostly active operation from bunch to bunch.

Calibration events  will be forced to fall outside the beam spill during which the electronics will be disabled for any other events.  Calibration events include vertical cosmic-ray muons.   Cosmic ray vertical muons are defined as muons passing through a yet to be determined set  of scintillator paddles, that allow one to select muons with a well defined vertical path length through the ANNIE detector volume.  The rate will be controlled by the acceptance (size) of the paddles.

\section{Simulations and Detector Requirements}
\label{sec-simulations}

Parallel with the progress in hardware, there is an existing effort to develop the simulations and reconstruction work necessary to guide the design of the detector and make full use of the LAPPDs. Groups at Iowa State University, University of Chicago, Queen Mary University, Imperial College, UC Irvine, and Ultralytics LLC will significantly contribute to this effort. The groups will use a variety of existing resources and their previous experience in this area. 

Critical capabilities of the ANNIE detector include:
\begin{itemize}
\item A detector with an appropriate form factor and fiducial volume to ensure efficient stopping of neutrons.
\item Sufficient PMT coverage to ensure efficient detection of the light from neutron captures.
\item Sufficient LAPPD and PMT coverage to discriminate between different event categories. At minimum, this means efficient separation of CCQE events from CC and NC events with pions and gammas. Ideally, we would also be able to categorize neutral current and multi-track events.
\item Accurate reconstruction of muon momentum and total visible energy.
\item Light yields and distributions compatible with the characteristics of the chosen photosensors.
\end{itemize}

Given the known physics of neutron scattering and capture in Gd-loaded water, we use GEANT simulations to answer these questions and thereby set the dimensions and determine the needed photocathode coverage of our detector. Simulations are also needed to answer questions regarding possible background from neutrino interactions from the rock behind the wall of the detector hall, as well as other cosmogenic backgrounds. 

Determining the rates and light yield of this event interaction pileup is necessary for the design of the trigger system and the LAPPD readout system. Also, light from signal events may be concentrated over small areas and thus the optical pileup of hits on individual anode channels must be understood. Optimizing pattern matching algorithms, and determining appropriate buffer sizes and tolerable dead times will also be important.

LAPPDs serve several purposes. First and foremost, we need to accurately determine the location of the interaction point within the detector. That way, it is possible to select events where the neutrinos start sufficiently far from the walls that the neutrons are unlikely to escape. LAPPDs are also critical in reconstructing details of the interaction: track parameters (especially for low energy muons and other particles that will not be reconstructed by the MRD) and possibly particle ID.

\subsection{Neutron Containment and Vertex Reconstruction}

\begin{figure}
       	\begin{center}
		\includegraphics[width=0.45\linewidth]{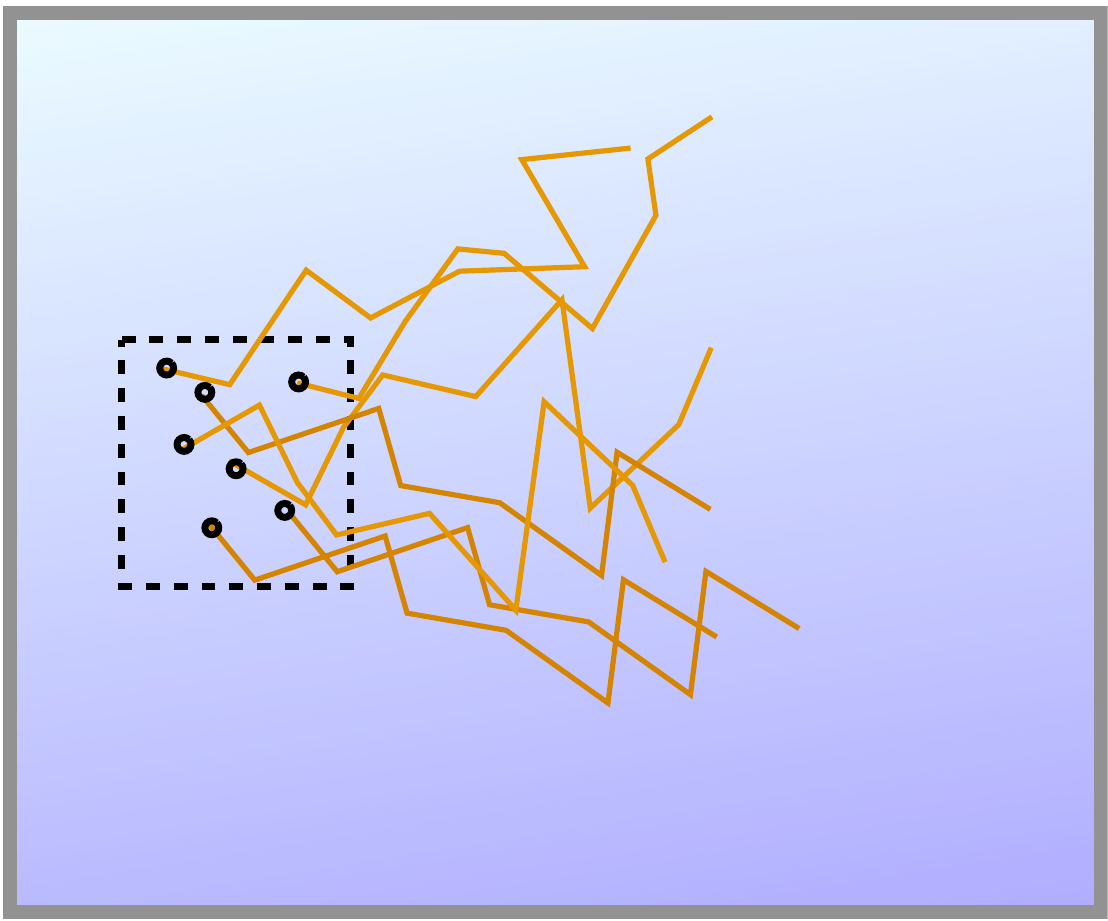}
	\end{center}
	\caption{A schematic showing the strategy for choosing a fiducial volume to ensure full containment of neutrons.}
	\label{neutronschematic}
\end{figure}

\begin{figure}
       	\begin{center}
		\includegraphics[width=0.85\linewidth]{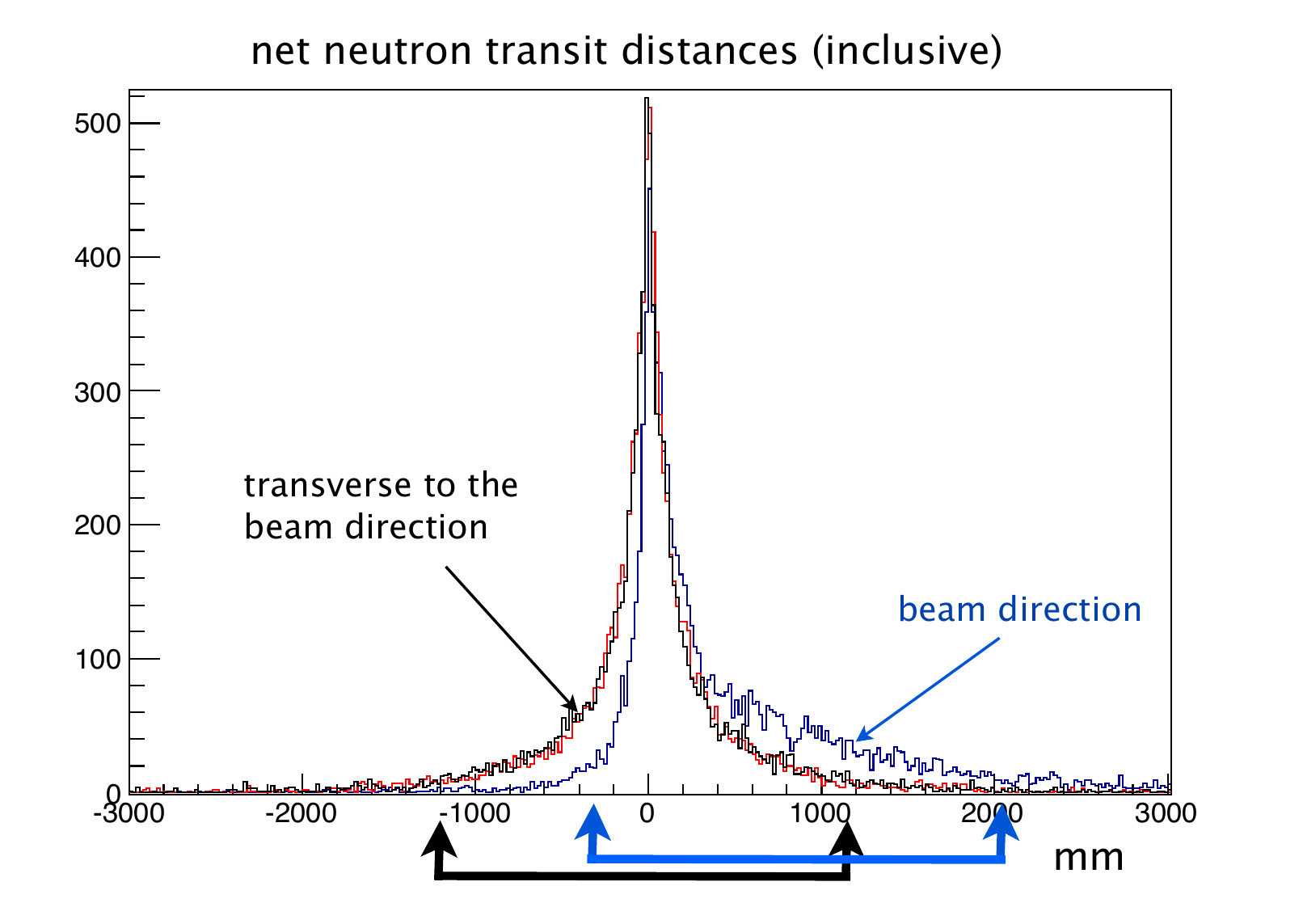}
	\end{center}
	\caption{Distributions showing the stopping distance of neutrons in ANNIE, in the two directions transverse to the beam (red and black) and along the beam axis (blue). In the transverse direction, 90$\%$ of all neutrons stop symmetrically $\pm$1.2 meters from their starting points (designated by the black arrows). In the beam direction, 90$\%$ of the neutrons fall in a window mostly in front of their starting position, ranging from -0.3 to 2.1 meters (designated by the blue arrows).}
	\label{neutronrangeout}
\end{figure}

High energy neutrons (10-100s of MeV) produced by neutrino interactions within the ANNIE volume will travel some distance before slowing to a stop. A few meters of water are sufficient to stop the vast majority of these neutrons. The ANNIE detector, with roughly 3 m length scales is sufficient for stopping neutrons produced sufficiently far from the walls. Thus, it is best to restrict the ANNIE analysis to a smaller fiducial volume, appropriately placed to allow range-out of neutrons. This is depicted schematically in Fig~\ref{neutronschematic}. Figure~\ref{neutronrangeout} shows the range-out locations of the neutrons with respect to their initial starting points. These stopping distances are symmetric about the starting point in the direction transverse to the neutrino beam, and forward of the starting point in the beam direction. Thus, the fiducial volume can be chosen in the center of the detector, 1.2 meters away from the wall in the transverse directions, and upstream of the detector center in the beam direction. Given a roughly 30 ton total volume, it is possible to chose such a fiducial region while still maintaining more than the 1 ton of target mass sufficient for a basic measurement. Moreover, if we can demonstrate sufficient understanding of the detector efficiencies, it may be possible to perform a more inclusive analysis, making use of much more target mass. Regardless, this analysis will require accurate reconstruction of the interaction points in the detector. 

\begin{figure}
	\centering
	 \begin{tabular}{c c}
		\includegraphics[width=0.48\linewidth]{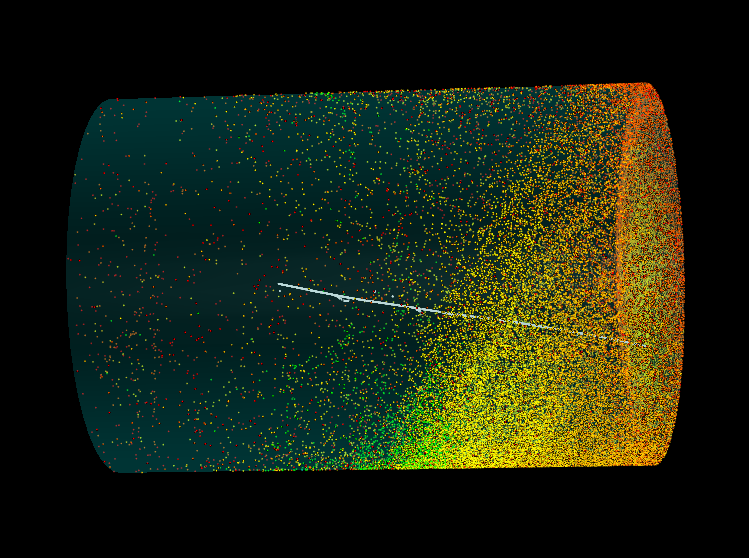} &
		\includegraphics[width=0.48\linewidth]{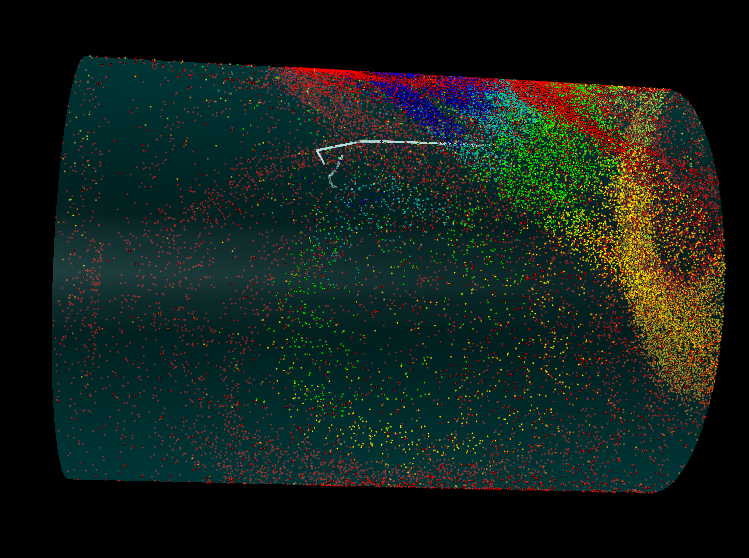} \\
		\includegraphics[width=0.48\linewidth]{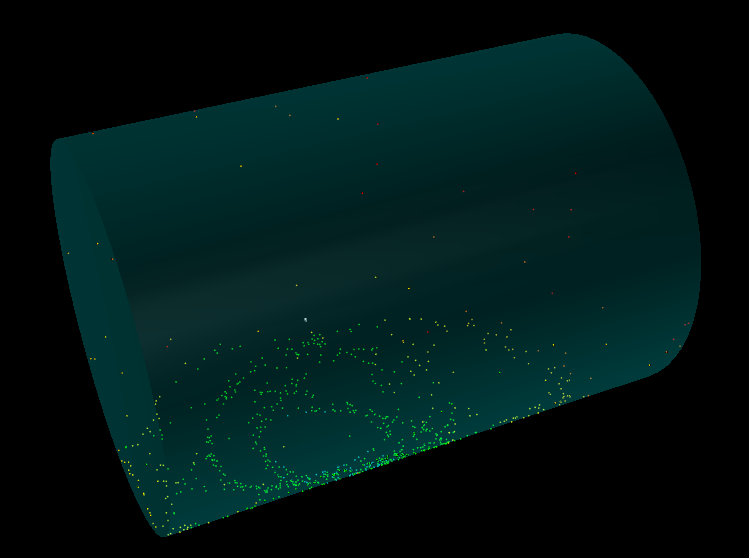}  &
		\includegraphics[width=0.48\linewidth]{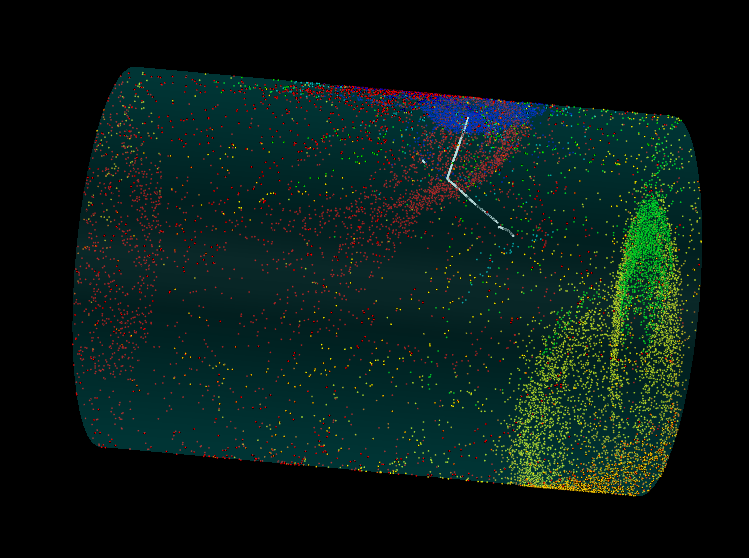}  
	\end{tabular}

	\caption{Four event displays showing examples of various interactions in the ANNIE detector. The colors represent time, with blue being earliest and red latest times. Upper left: CCQE interaction with a single muon exiting the volume. Upper right: CC resonant pion production with a fully contained muon and $\pi^0$ decaying to two gammas. Lower left: NC resonant pion production with a low energy $\pi^0$ decaying to two gammas. Lower right: A CC deep inelastic scattering interaction producing a muon, a $\pi^0$, and a $\gamma$. In all for plots, the beam direction is pointing along the cylinder axis to the right.}
	\label{evtdisplays}
\end{figure}

At the energies of the Booster Neutrino Beam, ANNIE will see a variety of neutrino interactions, ranging from Charged Current Quasi-Elastic (CCQE) and elastic Neutral Current (NC) interactions to resonant pion production (which turns on at around 1 GeV) and even some deeper, ineleastic scatters. Example event displays in a cylinder approximating the ANNIE form factor are shown in Fig~\ref{evtdisplays}.

Much of the light produced by these events will be in the forward region of the detector with respect to the beam direction. However, for increasingly inelastic interactions and high $q^2$ more tracks and light will be dispersed around 4$\pi$. 

\subsection{LAPPD Coverage and Fine Tracking}

Ultimately, we intend to operate ANNIE with more than 50 LAPPDs. However, given the expected time table for LAPPDs (see Sec.~\ref{subsec-LAPPD}), the current goal of our simulations and reconstruction work is to establish the minimum number of LAPPDs necessary to meet our minimum physics goals. We think it is possible to operate with as few as 12 LAPPDs.

One of the key requirements of the ANNIE detector, discussed above, is the ability to identify events with interaction points located in a fiducial volume offset sufficiently far from the walls of the detector. In pure CCQE events with only one track, the position of the interaction vertex along the direction track is not constrained by timing alone, since there is a degeneracy in time with the starting point and unknown $t_0$ of the interaction. The starting point of the track in the parallel direction is determined by finding the edge of the Cherenkov cone.  LAPPDs can help in the following way:

\begin{itemize}
\item Excellent directionality and reconstruction of the vertex in the plane transvers to the track direction. 
\item If a Cherenkov cone-edge crosses one or two LAPPDs, the ability to resolve the positions of hits within a single LAPPD module can enhance the accuracy for fitting the cone edge.
\item Timing on the LAPPDs can be used to separate between single-track and multi-track events, so long as light from more than one track hits an LAPPD. The causal inconsistency of the light with a single-track hypothesis would provide the relevant figure of merit.
\end{itemize}

The degeneracy between the unknown $t_0$ and starting point of the track is broken for multi-track events. Multi-track events have a single solution for the vertex, which can be determined by causality of the light arriving to the walls. Therefore, our strategy is to: (1) use LAPPDs to discriminate between single-track and multi-track events, (2) use timing on the LAPPDs to find the vertex in multi-track events, and (3) use the combination of PMT hits and MRD track reconstruction (at minimum) to find the cone-edge for single tracks.

\begin{figure}
       	\begin{center}
		\includegraphics[width=0.55\linewidth]{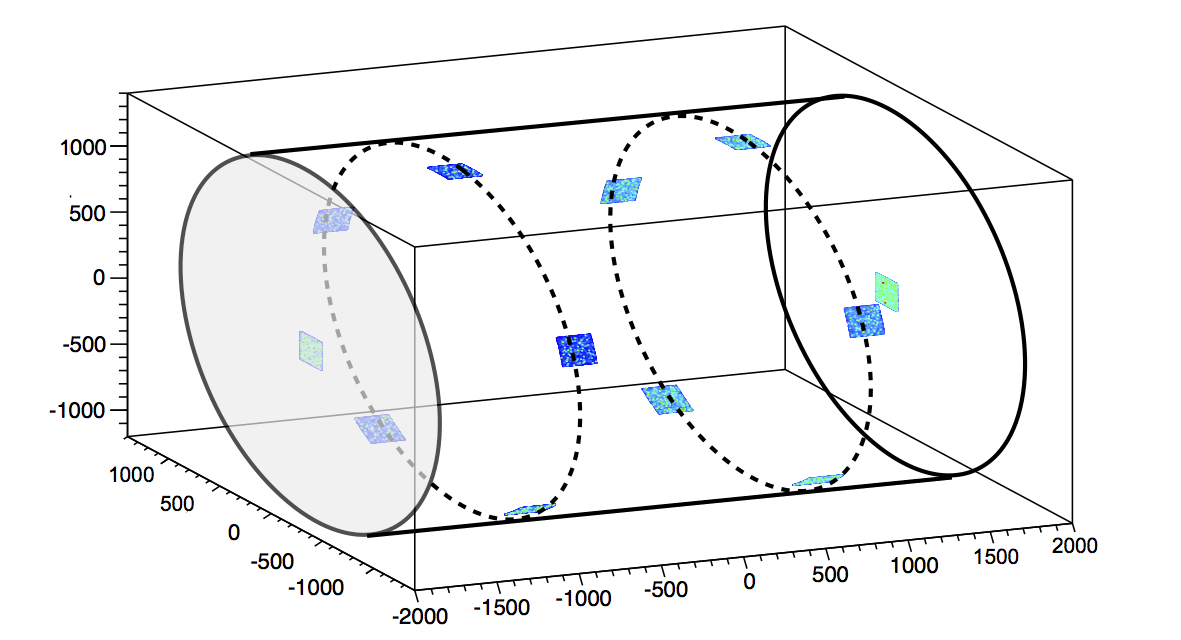}
	\end{center}
	\caption{A possible scenario for instrumenting the ANNIE volume with 12 LAPPDs.}
	\label{12lappds}
\end{figure}

Better reconstruction is possible as larger numbers of LAPPDs become available. Ideally, we would like 50-100 LAPPDs, enabling 5-10$\%$ coverage. The minimum number of LAPPDs is set by the requirement that some light from all tracks in the fiducial volume should hit at least one LAPPD. Since the fiducial volume is more than 1 meter away from the walls of the detector, this means that spacing of $\sim$2 meters between LAPPDs would suffice to meet this criteria. We are currently working on final optimization of that coverage, but we believe that a scheme involving $\sim$12 LAPPDs, such as the one shown in Fig~\ref{12lappds}, will  suffice.

\subsection{PMT Coverage and Neutron Tagging Efficiency}
\label{sec-PMTcovgsim}

\begin{figure}
       	\begin{center}
		\includegraphics[width=0.55\linewidth]{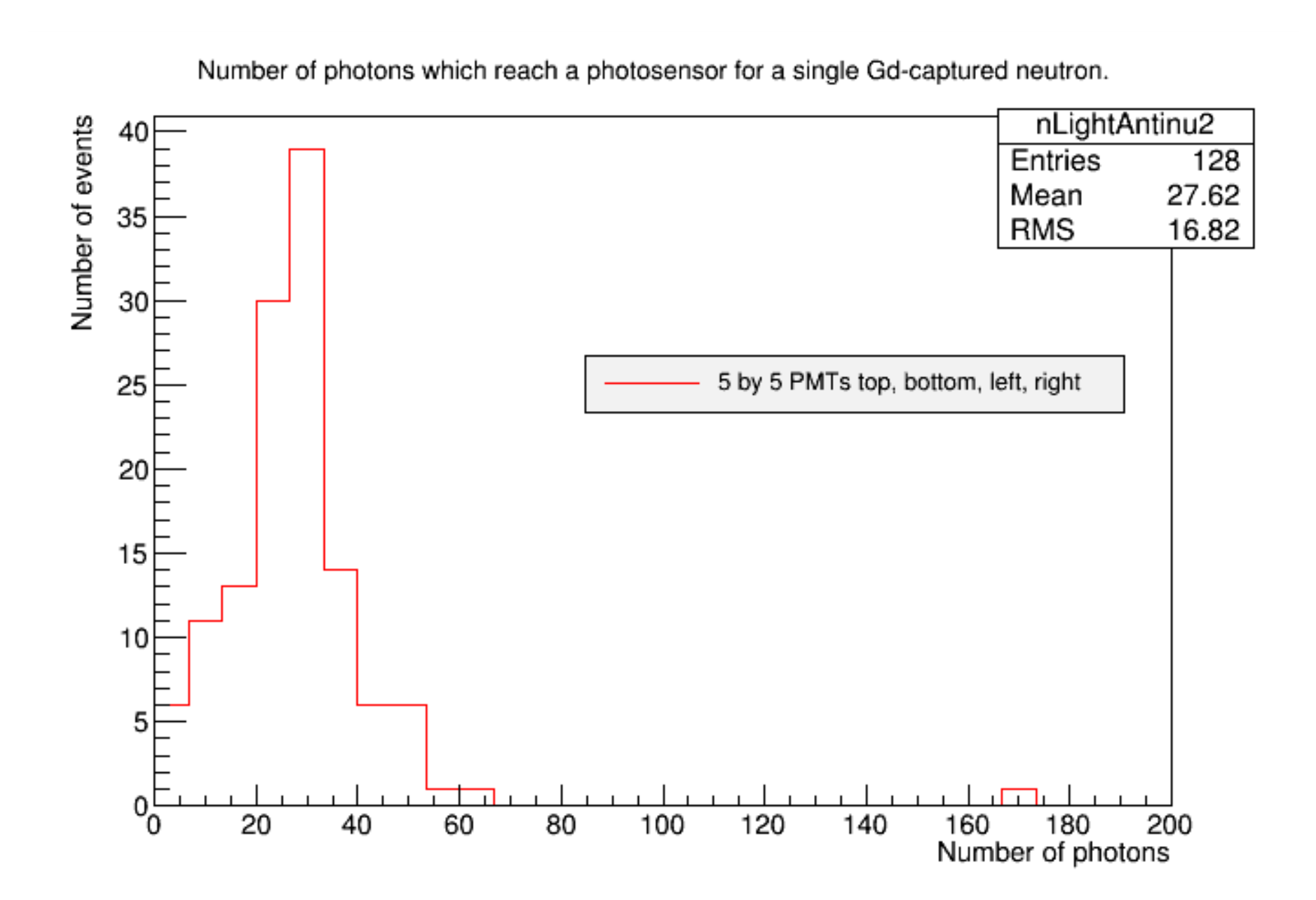}
	\end{center}
	\caption{The total number of hits detected by the PMT system for 100 PMTs in a cubic geometry.}
	\label{GdPhotHits}
\end{figure}

We are currently examining two coverage scenarios: a low coverage operation with 60 PMTs and a high coverage option with more than 100 PMTs. These two scenarios are chosen based on the known availability of PMTs for the detector: roughly 60 are immediately available for use, while more can be made usable with some effort to refurbish the base of the tubes (see Sec~\ref{sec-photodetection}). We have created simulations to study the effects of different numbers and arrangements of PMTs, with the goal of identifying how often we collect enough light to identify a Gd-capture. Conservatively assuming an effective quantum efficiency of 20$\%$, we look at the number of detected photons per capture. Figure~\ref{GdPhotHits} shows a simulated scenario with 100 PMTs distributed in a 3m cubic geometry, only on the top, bottom, and side walls (not the front and back). This study was conducted assuming an earlier cubic geometry and will soon be redone for the cylindrical case, with 60 and 100 PMTs arranged isotropically. Even with fewer PMTs, we expect higher efficiency with isotropic coverage given that fewer than $5\%$ of the events in the cubic simulation provide fewer than 5 hits.

\subsection{Simulating the LAPPD Response}
\label{sec-LAPPDcovgsim}

\begin{figure}
	\centering
	 \begin{tabular}{c c}
		\includegraphics[width=0.48\linewidth]{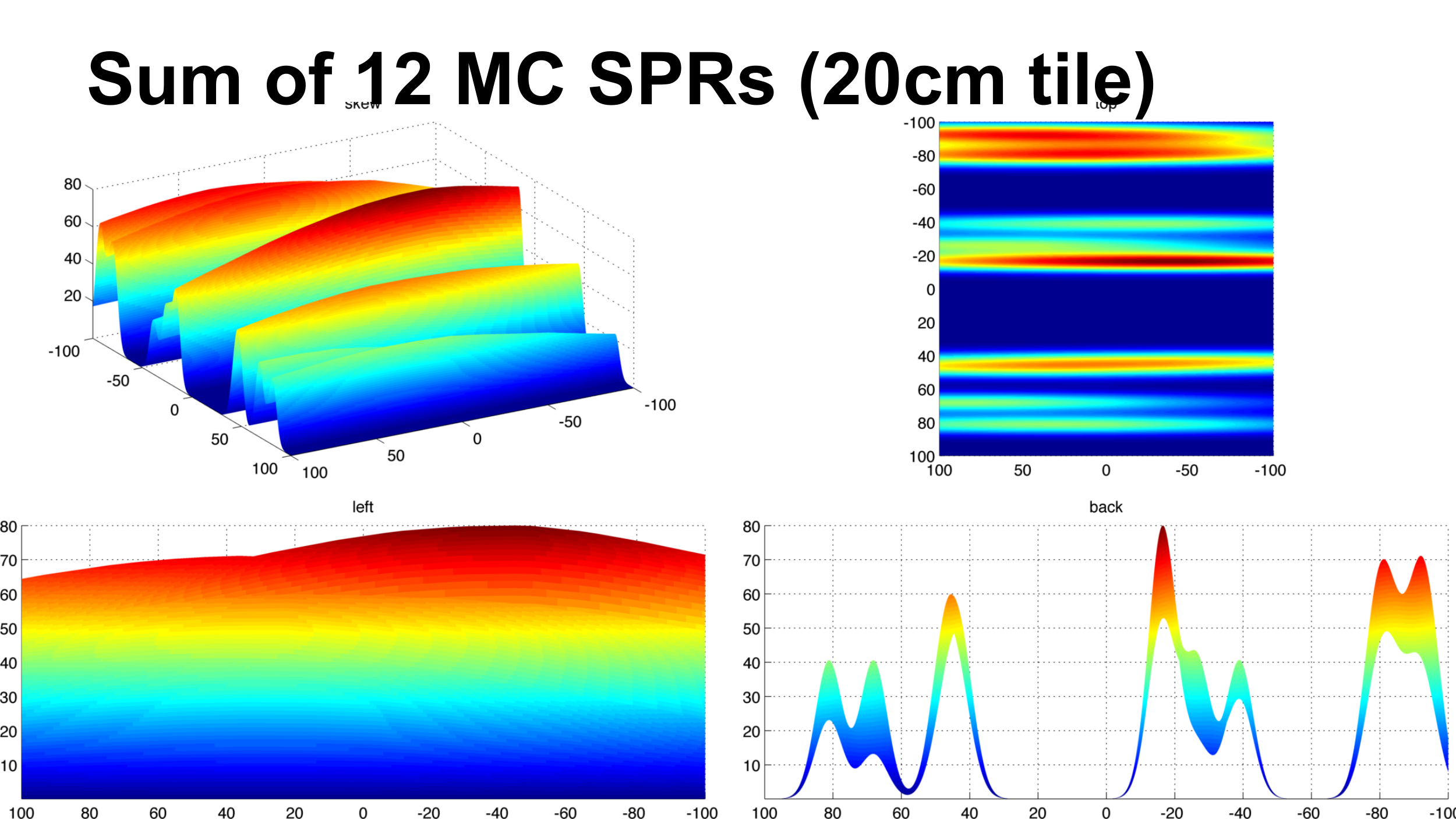} &
		\includegraphics[width=0.48\linewidth]{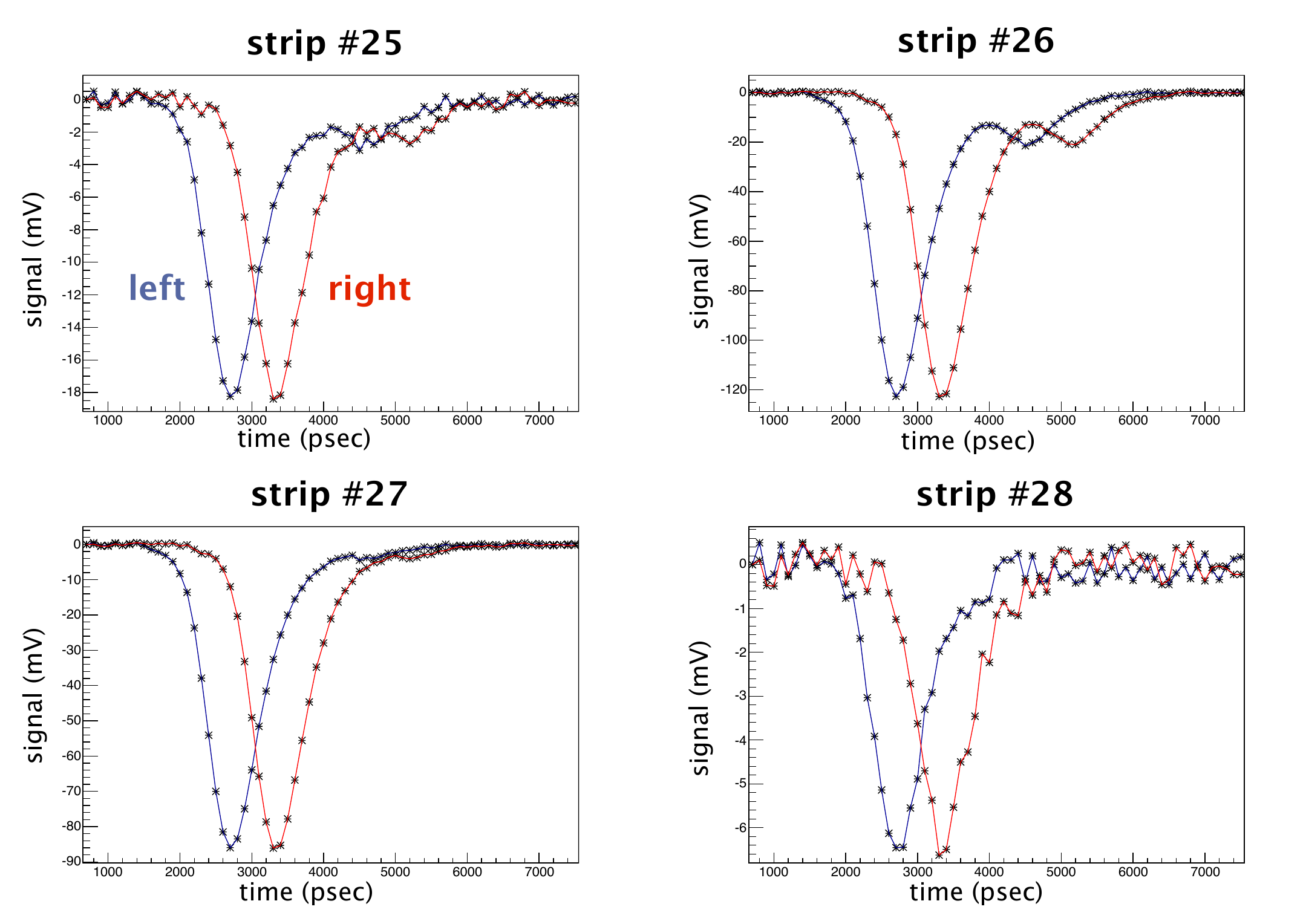}
	\end{tabular}
	\caption{Left: A temperature plot showing the time evolution of multiple photon hits on the striplines of an LAPPD. Right: A simulation of a four-stripline cluster struck by two photons, as would be seen in the output of PSEC electronics.}
	\label{anodesim}
\end{figure}

The stripline anode readout in the nominal LAPPD design presents reconstruction challenges when multiple photons hit the same strip, a phenomena we refer to as ``optical pileup". A variety of techniques can, in fact, be used to disambiguate multiple hits per strip. These are being studied in a real LAPPD test setup and used as the basis for a more sophisticated LAPPD response model than the parametric smearing currently used in our Monte Carlo.  Figure~\ref{anodesim} shows simulations of the LAPPD response.

\begin{figure}
	\centering
	 \begin{tabular}{c c}
		\includegraphics[width=0.45\linewidth]{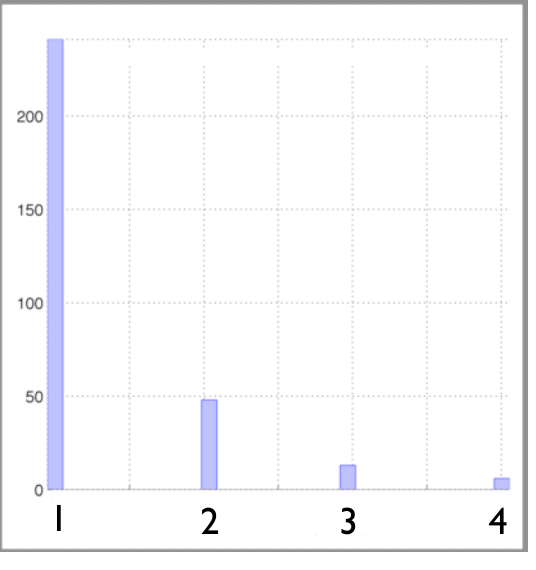} &
		\includegraphics[width=0.48\linewidth]{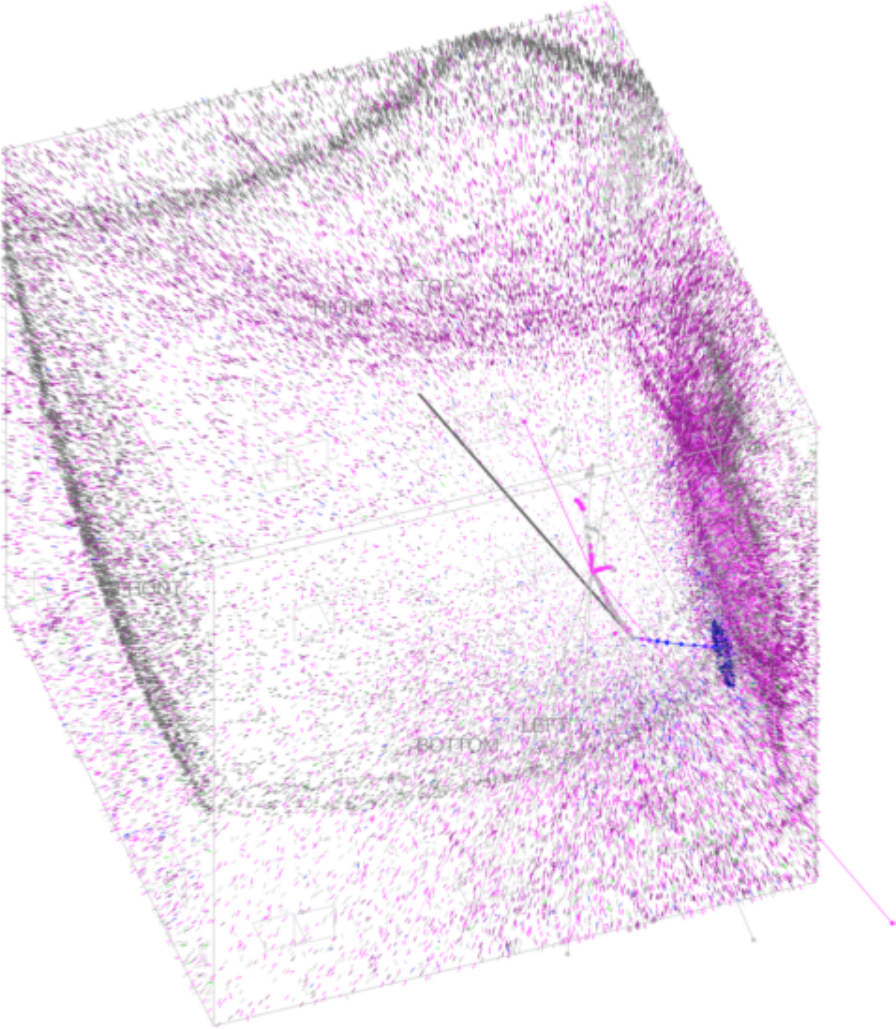}
	\end{tabular}
	\caption{Simulations of a particular event in a 3m cubic geometry. Left: histogram of the number of hits per stripline. Right: Event display.}
	\label{hitsperchannel}
\end{figure}

\begin{figure}
       	\begin{center}
		\includegraphics[width=0.55\linewidth]{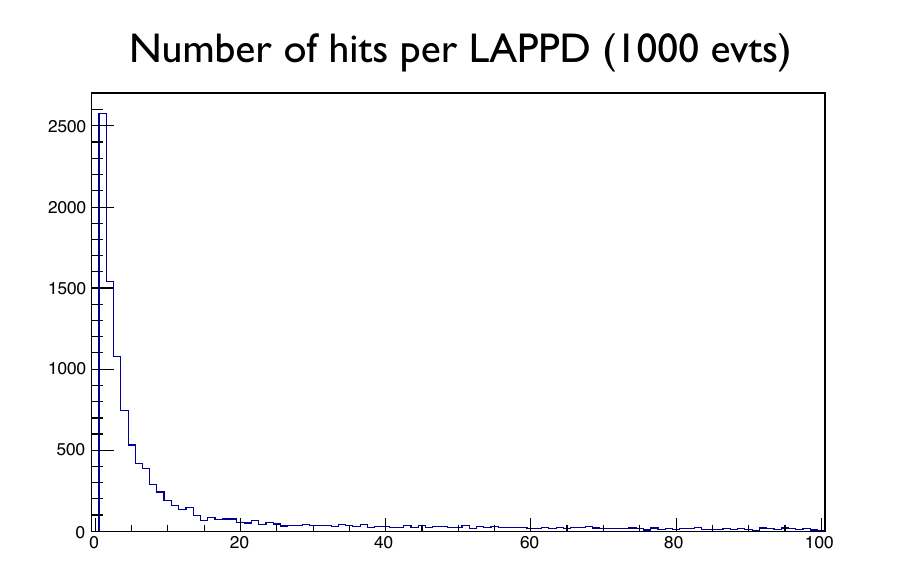}
	\end{center}
	\caption{The fraction of LAPPDs with more than 30 hits (some muti-photon optical pileup) per event.}
	\label{pileupfrac}
\end{figure}

ANNIE is an ideal context for an early application of LAPPDs precisely because the Chernkov light yield and area-to-volume of the detector minimize the likelihood of multiphoton pileup. Figure~\ref{hitsperchannel} shows a histogram of the number of hits per strip for a random ANNIE event. Figure~\ref{pileupfrac} shows what fraction of LAPPDs produce more than 30 photoelectrons per event, for 200 events. Since LAPPDs have 30 stripline anodes, this corresponds to the fraction of LAPPDs with at least some multi-photon pile-up. Typically, less than 15$\%$ of LAPPDs see more than 30 hits in any given event. The studies presented here were performed using an older cubic design and will need to be redone for the current cylindrical geometry. However, given comparable length scales we do not expect a substantial change in the results

\subsection{Future Simulation Work}

Studies presented in this section are ongoing and will continue, as we finalize the details of the ANNIE design. Most of the simulations and reconstruction work so far has relied on a simple Monte Carlo package called WChSandBox. Collaborators at Queen Mary, UChicago, and Ultralytics LLC are working on developing full Montecarlo, with digitization model and more realistic implementations of the photosensors. Reconstruction work will be performed by UChicago and Iowa State. This work will continue through next year.

\section{Proposed Timeline}
\label{sec-timeline}

We propose to separate the operation of the experiment into three phases. The first phase which is immediately relevant would focus on technical development and background characterization. The first physics run of ANNIE would occur in the second phase after having demonstrated key technical aspects of the experiment in its first phase of running. With a third phase allowing for a more precise measurement for different classes of interactions. 

\subsection{Phase I: Technical Development and Background Characterization}

(Begin installation Summer 2015, run Fall 2015-Spring 2016)

A variety of background neutrons, both those originating from rock, and those from sky shine, present a challenge for ANNIE.  Characterization of these background neutrons can occur before full LAPPD coverage is available. We thus propose to install and operate the detector with the components that are already available such as the water volume, Gd supply and stock of conventional PMTs.  This detector provides a useful setup for testing early LAPPD prototypes as they become available. In addition, the Double Chooz Time Projection Chamber (DCTPC), a small gas-phase TPC calibrated for neutron detection may be available to provide ex-situ measurements of neutron backgrounds to compare against neutron tagging in the water volume. This represents an opportunity to make relevant physics and technical measurements while testing and developing the technology for the full physics experiment. We therefore present ANNIE Phase I as a technical proposal.

The available components for ANNIE Phase I would be the ANNIE water tank instrumented with the 60 working Type-S PMTs, with a smaller, movable, plexiglass inner volume of Gd-loaded water. There will likely be at least one LAPPD prototype and one Argonne-made 6cm glass tile. The Gd-enhanced water volume would be moved around in the larger tank to study how measured rates of neutron detection rise and fall with distance from the walls in the horizontal and vertical direction, with and without in-volume beam events. These rates can be compared with predictions based on ex-situ measurements using DCTPC. They will provide critical information on beam-correlated dirt neutron production and sky shine neutrons, critical knowledge for the successful operation of ANNIE. They could also prove useful for understanding these backgrounds on LAr1ND.

Technical Objectives:
\begin{itemize}
\item Operation of the type-S PMTs in water
\item Operate basic electronics - ability to see beam structure, prompt event, and delayed
captures
\item In-situ optical calibration with pulsed LED or fiber laser
\item Test a single LAPPD (and/or small glass MCP), with HV connections,  in water
\end{itemize}


Support is required by FNAL engineering to estimate costs of installation, safety inspection and commissioning. Computing support for simulations and data acquisition will also be required. Also funds for the slow electronics and early LAPPD prototype tiles will also be needed.

\subsection{Phase II: ANNIE physics run I}

(Installation Summer 2016, run Fall 2016 - Spring 2017)

Phase II of ANNIE would represent the first full physics run. This phase will begin when we have acquired sufficient LAPPDs to operate our ``low-coverage" LAPPD scenario, as determined by simulation and reconstruction studies. As clarity is achieved regarding costs and availability of LAPPDs Phase II will be the target for submitting a complete physics proposal and soliciting operational budgets from various agencies. It is hoped that the upgrade and installation could occur over summer 2016 with full operation achieved by fall 2016.

The technical components of ANNIE Phase II will include a full, Gd-enhanced water volume with the 60 working Type-S PMTs, and a small but sufficient number of LAPPDs; the refurbished MRD; a full electronics scheme; and external neutron monitoring, either by DCTPC or a custom reproduction.

This first measurement will be an inclusive measurement of neutron yields as a function of reconstructed $q^2$ and $E_{vis}$, with the possibility of separating between Charged Current and Neutral Current events. We will also test out, with limited coverage, the ability to identify neutrino interactions that imitate proton decays.

Technical Objectives:
\begin{itemize}
\item Demonstration of full DAQ
\item Operation and successful tracking with water potted LAPPDs
\item Track/event reconstruction
\item Working MRD - compare MRD tracking with LAPPDs
\item Timing calibration
\end{itemize}

\subsection{Phase III: ANNIE physics run II}  

(Run Fall 2017 or on completion of Phase II - Fall 2018)

Phase III represents the full realization of the ANNIE detector. It will occur at such time as funding and availability allow moderate ($>$ 10\%) coverage, isotropically. We estimate that this will correspond to roughly 50-100 LAPPDs. With sufficient LAPPD coverage, detailed kinematic reconstruction and particle ID may become possible. In such a case, we will measure neutron yields for event classes separated by detailed inventory of final state particles. Phase III will be designed to explicitly identify PDK-background events. Detailed plans for this operation are in development and contingent on further simulations and reconstruction work, and validation against data collected in Phases I and II.

Technical objectives: complete event reconstruction - analysis using mostly active volume.

Physics goals:
\begin{itemize}
\item Neutron yield measurements with detailed event reconstruction
\item Explicit tagging of PDK backgrounds
\end{itemize}

\subsection{Additional ANNIE runs}

After successful operation of the two ANNIE physics runs to the satisfaction of the collaboration, additional runs may be possible with the goal of improving on the physics and testing new R\&D directions. These directions might include tests with water-based Liquid Scintillator.

\section{Budget and Funding Strategy}
\label{sec-budget}

The budget and funding strategy is currently being developed. A budget for phase I of ANNIE will be submitted in time for the PAC meeting in January 2015. In this phase the most significant costs will be related to engineering, including completing the design and installation of the water tank and PMTs as well as the electronics. Additional costs will include the acquisition of LAPPD prototype tiles and its readout. The already existing resources committed for ANNIE are shown in Table~\ref{freebies}. 

In terms of a funding strategy, we require PAC approval to be able to present a proposal to the agencies for the funding of manpower at the laboratory and universities in order to run the experiment. The Department of Energy has invested significant funds in the development of LAPPD in the past and thus there is interest in this experiment as a proof of principle HEP application. Also the physics of proton decay has been  recommended as one of the science drivers in the P5 report, under the category of Exploring the Unknown through Baryon Number Violation. Specifically the search for proton decay is cited in the report as ``an avenue to probe this [$10^{16}$~GeV] ultra-high energy scale". We thus expect both DOE and NSF to welcome ANNIE as a significant contribution to ``the planned program for neutrino experiments Hyper-Kamiokande and LBNE which offer the opportunity to increase the sensitivity to proton decay by an order of magnitude" as described in the report. 

%

\begin{table}
\begin{center}
\begin{tabular} {| l|  c | c |}
\hline
item& contributing institution&section of white paper\\
\hline
Water Tank  & U Chicago  &  Sec~\ref{sec-mechanical} \\
50 kg of Gd & UC Irvine  &Sec~\ref{sec-Gd} \\
63 SK-PMTs & UC Irvine &Sec~\ref{sec-photodetection} \\
43 IMB-PMTs & UC Irvine &Sec~\ref{sec-photodetection} \\
Steel and Scintillator MRD  & Imperial College/SciBooNE & Sec~\ref{subsec-MRD} \\
RPC-based MRD  & ANL & Sec~\ref{subsec-MRD} \\
\hline
\end{tabular}

  \end{center}
  \caption{Existing resources available for the ANNIE experiment.}
  \label{freebies}
\end{table}

\section{Conclusions}
\label{sec-conclusion}

The ANNIE experiment provides an opportunity to make an important measurement of the final-state neutron abundance from neutrino interactions with water nuclei as a function of momentum transfer. This measurement will have a significant impact on a variety of future physics analyses, including a nearly factor of five (or more) improvement in the sensitivity of WCh proton decay searches provided by efficient neutron tagging. ANNIE also provides a low-cost opportunity to maintain technological diversity and Water Cherenkov expertise in the US neutrino program. ANNIE will represent a first demonstration detector for WCh reconstruction using the newly developed, high resolution LAPPD imaging sensors. Much development work has already been done by the LAPPD collaboration, including a working readout system with capabilities well matched to the needs of ANNIE. The experiment-specific development mainly involves implementation, some amount of customization, and long-term systems testing in a real physics context. 

In this Letter of Intent we have addressed key elements of the design and planning for ANNIE. Namely, we have carried out a detailed site study for neutrino beams at Fermilab concluding that the SciBooNE hall is indeed optimal for this experiment. We have started to understand the neutron background at the site and propose to install a prototype that will allow us to characterize this background and at the same time allow us to demonstrate key technical aspects for a full physics run of the experiment. We have also made significant advances in the planning of the experiment, for example a mechanical design of the water target has been carried out and a potential vessel identified. Photomultiplier tubes have been tested and additional potential tubes have been committed for use in ANNIE. Potential substitute to the scintillator of the SciBooNE MRD has also been found. The requirements and potential design of the electronics system have been discussed in this document. Finally, significant progress has been made in the simulations and definition of the detector requirements to carry out the physics program of the experiment. 

As when we presented our first Expression of Interest,  Fermilab's support and interest plays a critical role in the success of ANNIE. The availability of the SciBooNE hall and access to the Booster Neutrino Beamline is critical. We are requesting approval and engineering support to plan the installation, safety inspections and commissioning of the phase I of ANNIE. We would also benefit from support in the form of computer resources and simulations expertise for this phase. This phase is crucial to enable the collaboration to seek funding to complete the experiment and achieve the physics goals of ANNIE.


\begin{thebibliography}{10}
\expandafter\ifx\csname url\endcsname\relax
  \def\url#1{\texttt{#1}}\fi
\expandafter\ifx\csname urlprefix\endcsname\relax\def\urlprefix{URL }\fi


\bibitem{eoi}
  I.~Anghel {\it et al.}  [ANNIE Collaboration],
  ``Expression of Interest: The Atmospheric Neutrino Neutron Interaction Experiment (ANNIE),''
  arXiv:1402.6411 [physics.ins-det].

\bibitem{FSneutrons}
J.Beacom and M.Vagins, ``Antineutrino spectroscopy with large water cherenkov detectors," Phys. Rev. Lett. 93 (2004) 171101.

\bibitem{Ejiri}
H. Ejiri, ``Nuclear deexitations of nucleon holes associated with nucleon decays in nuclei," Phys. Rev. C 48 (1993) 1442-1444.

\bibitem{Ncapturewater}
R. E. Meads {\it et al}, ``The capture cross section of thermal neutrons in water," Proc. Phys. Soc. A 69 (1956) 469.

\bibitem{NcaptureGd}
S. Dazeley {\it et al}, ``Observation of neutrons with a gadolinium doped water cherenkov detector," Nucl. Instrum. Methods Phys. Res. A 607 (2009) 616.
 
\bibitem{SKneutronyield} H. Zhang for the [Super-Kamiokande Collaboration], ``Neutron tagging and its physics application in Super-Kamiokande-IV," Proceedings of the 32 International Cosmic Ray Conference, Beijing (2011).




\bibitem{RabyNakamura}
See the review by S. Raby in K. Nakamura {\it et al}, [Particle Data Group] J. Phys. G 37 
(2010) 075021 and references therein.

\bibitem{SuperK2}
H. Nishino {\it et al} [Super-Kamiokande Collaboration], ``Search for proton decay via $p \rightarrow e^+ \pi^0$ and $p \rightarrow \mu^+ \pi^0$ in a large water cherenkov detector,"  Phys. Rev. Lett. 102 (2009) 141801.

\bibitem{SuperK3}
K. Abe {\it et al} [Super-Kamiokande Collaboration], ``Search for proton decay via $p \rightarrow \bar{\nu} K^+$ using 260 kiloton year data 
of Super-Kamiokande," Phys. Rev. D 90 (2014) 072005.





\bibitem{supernova-beacom}
R.~Laha and J.~F.~Beacom,
  ``Gadolinium in water Cherenkov detectors improves detection of supernova $\nu_e$,''
  Phys.\ Rev.\ D {\bf 89}, no. 6, 063007 (2014).

\bibitem{review-beacom}
 J.~F.~Beacom,
  ``The Diffuse Supernova Neutrino Background,''
  Ann.\ Rev.\ Nucl.\ Part.\ Sci.\  {\bf 60}, 439 (2010).

\bibitem{LBNESN}
See the discussion of SN detection in [LBNE Collaboration], ``The long baseline neutrino experiment (LBNE) water cherenkov detector (WCD) conceptual design report (CDR)" (2012); arXiv:1204.2295v1.

\bibitem{Raffelt}
G. G. Raffelt, ``Physics opportunities with supernova neutrinos," Progress in Part. and Nucl. Phys. 64 (2010) 393-399.



\bibitem{T2K2} 
K. Abe {\it et al} [T2K Collaboration], ``Observaton of electron neutrino appearance in a muon neutrino beam"; arXiv:1311.4750 [hep-ex].

\bibitem{MB1}
A.~A.~ Aguilar-Arevalo {\it et~al.}, ``First measurement of the muon neutrino charged current quasielastic double differential cross section," Phys.~Rev.~D 81, 092005 (2010).

\bibitem{MB2}
A.~A.~ Aguilar-Arevalo {\it et~al.}, ``First measurement of the muon antineutrino double-differential charged-current quasielastic cross section," Phys.~Rev.~D 88, 032001 (2013).

\bibitem{Benhar}
O.~Benhar, P.~Coletti, and D.~Meloni, ``Electroweak Nuclear Response in the Quasielastic Regime," Phy.~Rev.~Lett. 105, 132301 (2010).

\bibitem{Martini}
M.~Martini, M.~Ericson, G.~Chanfray and J.~Marteau, ``Neutrino and antineutrino quasielastic interactions with nuclei," Phys.\ Rev.\ C 81, 045502 (2010).

\bibitem{Nieves} 
J.~Nieves, I.~Ruiz Simo and M.~J.~Vicente Vacas, ``The nucleon axial mass and the MiniBooNE quasielastic neutrino–nucleus scattering problem," Phys.\ Lett.\ B 707, 72 (2012).

\bibitem{MBp} 
R.~Dharmapalan {\it et al.} [MiniBooNE+ Collaboration], ``A new investigation of electron neutrino appearance oscillations with improved sensitivity in the MiniBooNE+ experiment"; arXiv:1310.0076 [hep-ex].



\bibitem{LeeSkyshine} W. Lee {\it et al.}, Phys. Rev. Lett. 37, 186 (1976).

\bibitem{BartolNeutronMonitor} Bartol neutron monitor data, http://neutronm.bartol.udel.edu/. Neutron monitors of the Bartol Research Institute are supported by NSF grant ATM- 0000315.

\bibitem{SciBooNEskyshine}   A.~A.~Aguilar-Arevalo {\it et al.}  [SciBooNE Collaboration], ``Bringing the SciBar detector to the booster neutrino beam,'' arXiv:0601022 [hep-ex].

\bibitem{Takeiskyshine} H. Takei for the SciBooNE Collaboration, ``Identification of Recoil Proton Tracks for a Neutrino Neutral Current Elastic Scattering Cross-Section Measurement at SciBooNE", talk at the 2008 Meeting of the American Physical Society.








\bibitem{Sciboone}
K. Hiraide {\it et al} [SciBooNE Collaboration], ``Search for charged current coherent pion production on carbon in a few-GeV neutrino beam," Phys. Rev. D 78 (2008) 112004.

\bibitem{LAPPD}
The Large Area Picosecond Photodetector Collaboration: http://psec.uchicago.edu 


\bibitem{UCelectronics2}
E. Oberla {\it et al}, ``A 15 GSa/s, 1.5 GHz bandwidth waveform digitizing ASIC," Nucl. Instrum. Meth. A735 (2014) 452-461; \\ http://psec.uchicago.edu/library/doclib/documents/218 

\bibitem{UHawaiielectronics}
K. Bechtol {\it et al}, ``TARGET: A multi-channel digitizer chip for very-high-energy gamma-ray telescopes," Astropart.Phys. 36 (2012) 156-165; SLAC-PUB-14202 

\bibitem{UHawaiielectronics2}
L. Ruckman and G. Varner, ``Sub-10ps monolithic and low-power photodetector readout," Nucl. Instrum. Meth. A602 (2009) 438-445.

\bibitem{UHawaiielectronics3}
L. Ruckman, G. Varner, A. Wong, ``The first version buffered large analog bandwidth (BLAB1) ASIC for high luminosity collider and extensive radio neutrino detectors," Nucl. Instrum. Meth. A591 (2008) 534-545.

\bibitem{UHawaiielectronics4}
G.S. Varner {\it et al}, ``The large analog bandwidth recorder and digitizer with ordered readout (LABRADOR) ASIC," Nucl. Instrum. Meth. A583 (2007) 447-460.

\bibitem{APS}
B. Adams {\it et al}, ``A test-facility for large-area microchannel plate detector assemblies using a pulsed sub-picosecond laser", Rev. Sci. Instrum. 84 (2013) 061301;\\
http://psec.uchicago.edu/library/doclib/documents/214

\bibitem{miniboone}
R.B. Patterson {\it et al}, ``The extended-track reconstruction for MiniBooNE", Nuclear Inst. and Methods in Physics Research, A 608 (2009), 206-224; arXiv:0902.2222.

\bibitem{chroma}
S. Siebert and A. LaTorre, ``Fast optical monte carlo simulation with surface-based geometries using chroma"; http://chroma.bitbucket.org

\bibitem{fastneutrino}
I. Anghel presenting work by I. Anghel, E. Catano-Mur, M.C. Sanchez, M. Wetstein, and T. Xin at DPF 2013; arXiv:1310.2654 [physics.ins-det].

\bibitem{OpticalTPC}
Work based on ideas by H. Frisch and developed by  I. Anghel, E. Catano-Mur, A. Elagin, H. Nicholson, E. Oberla, M.C. Sanchez, M. Wetstein, and T. Xin. For example, see M.~Demarteau, R.~Lipton, H.~Nicholson, I.~Shipsey, D.~Akerib, A.~Albayrak-Yetkin, J.~Alexander and J.~Anderson {\it et al.}, ``Instrumentation frontier report", pp. 8.

\end{thebibliography}

\end{document}